\documentclass[fleqn,usenatbib]{mnras}

\usepackage{newtxtext,newtxmath}

\usepackage[T1]{fontenc}
\usepackage{ae,aecompl}
\usepackage{ulem}

\usepackage{tikz}
\usetikzlibrary{tikzmark}
\usetikzlibrary{arrows,shapes,backgrounds}

\usepackage{graphicx}	
\usepackage{amsmath}	
\usepackage{amssymb}	

\usepackage{enumitem}
\newlist{legal}{enumerate}{10}
\setlist[legal]{label*=\arabic*.}
\usepackage{epsfig}
\usepackage{color}
\usepackage{natbib}
\usepackage{float}
\usepackage{dblfloatfix}
\usepackage{csvsimple}
\usepackage{natbib,twoopt}
\usepackage{csquotes}

\usepackage{pifont} 

\usepackage{rotating}
\usepackage{setspace}
\usepackage{etoolbox}
\usepackage{threeparttable}
\usepackage{mathtools,amsmath}

\title[SED selected AGN in the VIPERS sample]{An obscured AGN population hidden in the VIPERS galaxies: identification through spectral energy distribution decomposition\footnote{This paper uses data from the VIMOS Public Extragalactic Redshift Survey (VIPERS). VIPERS has been performed using the ESO Very Large Telescope, under the "Large Programme" 182.A-0886. The participating institutions and funding agencies are listed at http://VIPERS.inaf.it.}}

\author[Pouliasis et al.]{E.~Pouliasis$^{1,2}$\thanks{E-mail: epouliasis@astro.noa.gr}, 
G.~Mountrichas$^{3,1}$,
I.~Georgantopoulos$^{1}$,
A.~Ruiz$^{1}$,
M.~Yang$^{1}$, 
\newauthor
A.~Z.~Bonanos$^{1}$
\\
$^{1}$IAASARS, National Observatory of Athens, 15236 Penteli, Greece\\
$^{2}$Department of Astrophysics, Astronomy \& Mechanics, Faculty of Physics, University of Athens, Zografos, 15783
Athens, Greece\\
$^{3}$Instituto de Fisica de Cantabria (CSIC-Universidad de Cantabria), Avenida de los Castros, 39005 Santander, Spain\\
}

\date{Accepted XXX. Received YYY; in original form ZZZ}

\pubyear{2019}

\let\ACMmaketitle=\maketitle
\renewcommand{\maketitle}{\begingroup\let\footnote=\thanks \ACMmaketitle\endgroup}
\begin{document}
\label{firstpage}
\pagerange{\pageref{firstpage}--\pageref{lastpage}}

\maketitle
\begin{abstract}
The detection of X-ray emission constitutes a reliable and efficient tool for the selection of Active Galactic Nuclei (AGNs), although it may be biased against the most heavily absorbed AGNs. Simple mid-IR broad-band selection criteria identify a large number of luminous and absorbed AGNs, yet again host contamination could lead to non-uniform and incomplete samples. Spectral Energy Distribution (SED) decomposition is able to decouple the emission from the AGN versus that from star-forming regions, revealing weaker AGN components. We aim to identify the obscured AGN population in the VIPERS survey in the CFHTLS W1 field through SED modelling. We construct SEDs for 6,860 sources and identify 160 AGNs at a high confidence level using a Bayesian approach. Using optical spectroscopy, we confirm the nature of $\sim$85\% of the AGNs. Our AGN sample is highly complete ($\sim$92\%) compared to mid-IR colour selected AGNs, including a significant number of galaxy-dominated systems with lower luminosities. In addition to the lack of X-ray emission (80\%), the SED fitting results suggest that the majority of the sources are obscured. We use a number of diagnostic criteria in the optical, infrared and X-ray regime to verify these results. Interestingly, only 35\% of the most luminous mid-IR selected AGNs have X-ray counterparts suggesting strong absorption. Our work emphasizes the importance of using SED decomposition techniques to select a population of type II AGNs, which may remain undetected by either X-ray or IR colour surveys.
\end{abstract}




\section{Introduction}\label{intro}

Recent studies show that almost all galaxies in the Local Universe host a super-massive black hole (SMBH) in their centre \citep{magorrian1998,kormendy2004,filippenko2003,barth2004,greene2004,greene2007,dong2007,greene2008}. Their mass ranges between $10^5$ and $10^{10}$ solar masses. When matter from the galaxies starts to accrete into the SMBH, an enormous amount of energy is released across the electromagnetic spectrum (from radio emission up to X- and $\gamma$- rays). This constitutes the characteristic signature of Active Galactic Nuclei (AGNs). In many cases, the power of a single AGN with a similar size to our Solar System is higher than the power emitted by the stellar population of its host galaxy. The detection and study of AGNs is one of the most active fields of extra-galactic astrophysics today. Their demographics (e.g. space density), their distribution on the cosmic web (e.g. two-point auto-correlation function) and their physical properties (e.g. luminosity, black hole mass, accretion rate, absorption) play an important role in understanding the evolutionary models of black holes and whether AGN affect their host galaxy properties (e.g. star formation rate). Recent studies suggest a close interaction between the creation and evolution of galaxies and that of SMBHs \citep[e.g.][]{silk1998,granato2004,dimatteo2005,croton2006,hopkins2006,hopkins2008,menci2008}, though the physical processes governing this relationship are not yet fully understood.

To investigate any parallel evolution between galaxies and AGN, as well as to examine whether and how AGN feedback affects the evolution of galaxies, observations in infrared (IR), optical wavelengths and X-rays are required \citep{hopkins2006,bower2006}. The properties of galaxies hosting a black hole can be studied using optical and near-IR observations, while the absorbed AGN properties can be investigated in both X-rays and the mid-IR regime \citep[ and references therein]{hickox2018}. When the radiation of AGN dominates, X-ray emission is capable of penetrating large amounts of dust and gas without being absorbed. Consequently, the detection of X-rays is one of the most efficient methods of identifying AGNs and is essentially independent of absorption \citep{luo2008}. However, even hard X-rays can be absorbed in the presence of huge amounts of dust and gas \citep{gilli2007,fiore2008,treister2009,akylas2012}.

Mid-IR selection techniques offer a powerful tool to separate AGNs from stars and galaxies. This is achieved by separating the (approximately) power law AGN spectrum from the black body stellar spectrum of galaxies. Thus, another way of detecting AGNs is through mid-IR observations, which has the advantage of being affected less by extinction. Mid-IR broad-band colour selection criteria have been proven extremely useful in revealing the presence of an AGN \citep{lacy2007,stern2012,mateos2012,donley2012,assef2013} based on observations made by the \textit{Wide-field Infrared Survey Explorer} \citep[\textit{WISE};][]{wright2010} and the \textit{Spitzer} Space telescope \citep{werner2004}. Furthermore, many studies used the IR selection criteria in combination with X-ray or optical photometric and spectroscopic data to uncover absorbed AGN \citep[e.g.][]{rovilos2014, hainline2014, assef2015, hviding2018, glikman2019}. However, IR selection techniques are biased against low-luminosity AGNs \citep{barmby2006,georgantopoulos2008}. SED decomposition can alleviate this problem by utilizing large wavelength range and properly disentangle accretion from star formation \citep[e.g.][]{Ciesla2015}. SED decomposition can therefore provide a complementary tool to the X-ray and IR selection techniques, revealing weaker AGN systems with lower luminosities by reliably quantifying the host galaxy contribution to mid-IR colours.

This work focuses on selecting AGNs through SED modelling and fitting techniques in a sample of galaxies from the VIPERS survey \citep{guzzo2014,garilli2014}. We aim to find an AGN population with intermediate/high obscuration, which would be missed from current X-ray surveys and simple mid-IR colour cuts. The optical, near-IR and mid-IR data used for the SEDs along with ancillary data (spectroscopic data, X-ray catalogues, etc.) are presented in Section~\ref{data}, while in Section~\ref{methods}, we describe the methods we used to construct and model the SEDs and also a Bayesian approach to select AGNs. In Section~\ref{results}, we explore the optical spectra, demonstrate the properties of the SED selected AGNs in the mid-IR and X-ray regimes and we focus on the obscuration of these sources by applying different diagnostic tests. In Section~\ref{discussion}, we discuss these properties along with the differences between SED, mid-IR and X-ray selected AGNs, while Section~\ref{summary} presents the summary of the results and conclusions. Throughout the paper, we assumed a $\Lambda$CDM cosmology with H\textsubscript{o}=75 km s\textsuperscript{-1} Mpc\textsuperscript{-1}, $\Omega$\textsubscript{M}=0.3 and $\Omega$\textsubscript{$\Lambda$}=0.7.


\section{Data}\label{data}
The selection of obscured AGNs through SED fitting analysis requires secure redshifts and available optical and IR photometry for the SED construction. For that purpose, we used data from the VIMOS Public Extragalactic Redshift Survey (VIPERS) in the Canada-France-Hawaii Legacy Survey (CFHTLS) W1 field that contains rich multi-wavelength data from X-rays to the mid-IR bands.

\subsection{VIPERS}\label{vipers}

The VIPERS survey used the VIsible MultiObject Spectrograph \citep[VIMOS]{fevre2013} to perform deep optical spectroscopy within the CFHTLS W1 field. Follow-up spectroscopic targets were selected to the magnitude limit i'$=22.5$ from the CFHTLS catalogues \citep{guzzo2014}, while an optical colour pre-selection excluded galaxies at z$<0.5$ \citep{lefevre2013}. Comparing with a sample of VVDS-Deep and VVDS-Wide surveys, the selection criteria yielded $100\%$ completeness for z$>0.6$ \citep[Fig.4,][]{scodeggio2016}. In our analysis, we use the Public Data Release 2 \citep[PDR-2,][]{scodeggio2016} of the VIPERS survey, which consists of 86,775 galaxies with available spectra and their corresponding optical photometric data in the u, g, r, i and z bands from the CFHTLS T0007 data release \citep{hudelot2012}. Each spectrum is assigned a quality flag. In this work, we use sources with flags higher than 2 (confidence level of redshift measurement higher than 90\% based on more than one spectral features) that are considered as the most reliable \citep{garilli2014,scodeggio2016}. In total, 45,180 galaxies meet this criterion within $0.5<z<1.2$.

\subsection{VISTA-VHS}\label{vhsdata}

The VISTA Hemisphere Survey \citep[VHS,][]{mcmahon2013} is one of the six large surveys that are coordinated by the Visible and Infrared Survey Telescope for Astronomy \citep[VISTA,][]{emerson2006} to observe the entire sky in the southern hemisphere, covering 20,000 sq. degrees. The data used in this work are from the Data Release 6 that produced by the VISTA Science Archive (VSA). VSA handles all the data products generated by the VISTA Infrared CAMera (VIRCAM). The depth of the VHS observations is higher than other near infrared surveys, such as UKIDSS-LAS \citep{lawrence2007UKIDDS}, 2MASS \citep{skrutskie2006} or DENIS \citep{epchtein1994}, and the magnitude limits are 20.6, 19.8 and 18.5 mag (Vega) for the J, H and Ks band, respectively.

\subsection{AIIWISE}\label{wisedata}

The usage of mid-IR photometry, in our analysis, is twofold: First, its inclusion in the SED fitting process allow us to identify AGN candidates (see Section 3). Second, mid-IR photometry has been proven very efficient in detecting AGNs, since it is less affected by extinction. We, thus, compare our SED derived AGN candidates with those selected by IR colours (see Section 4). Launched by NASA, \textit{WISE} mapped the whole sky in the mid-IR regime with its four band-passes: W1=3.4 $\mu$m, W2=4.6 $\mu$m, W3=12 $\mu$m and W4=22$ \mu$m and reached 5$\sigma$ depths of 16.5, 15.5, 11.2, and 7.9 mag in Vega system, respectively. The AllWISE source catalogue \citep{cutri2013} consists of more than 700 million objects with signal-to-noise ratio higher than 5 in at least one band in the combined images.

\subsection{XMM-XXL}
XMM-XXL field \citep{pierre2017} covers an area of $\sim$50\,deg$^2$ with an exposure time of about 10\,ks per \textit{XMM-Newton} pointing. The survey is split into two approximately equal fields. In the analysis, we use data from the equatorial sub-region of the field (XMM-XXL-North; XXL-N) that overlaps with the CFHTLS W1 field and covers an area of about 25\,deg$^2$. We make use of the most recent X-ray catalogue of \citet{chiappetti2018}, which consists of 14,168 X-ray sources in the northern XMM-XXL field.


\subsection{Final sample}\label{finalsample}
\begin{figure}
   \begin{tabular}{c}
    \includegraphics[width=0.47\textwidth]{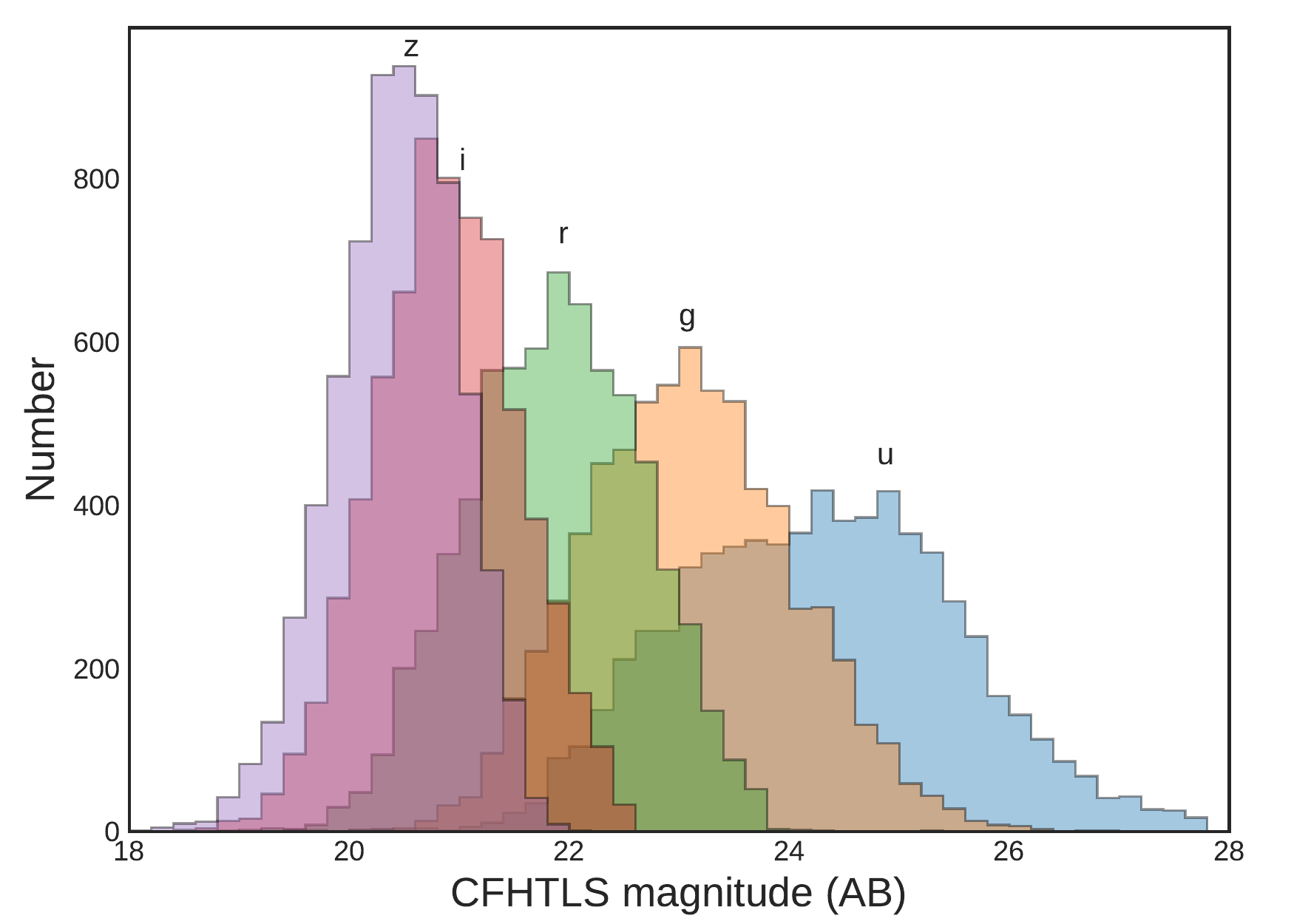} \\
    \includegraphics[width=0.47\textwidth]{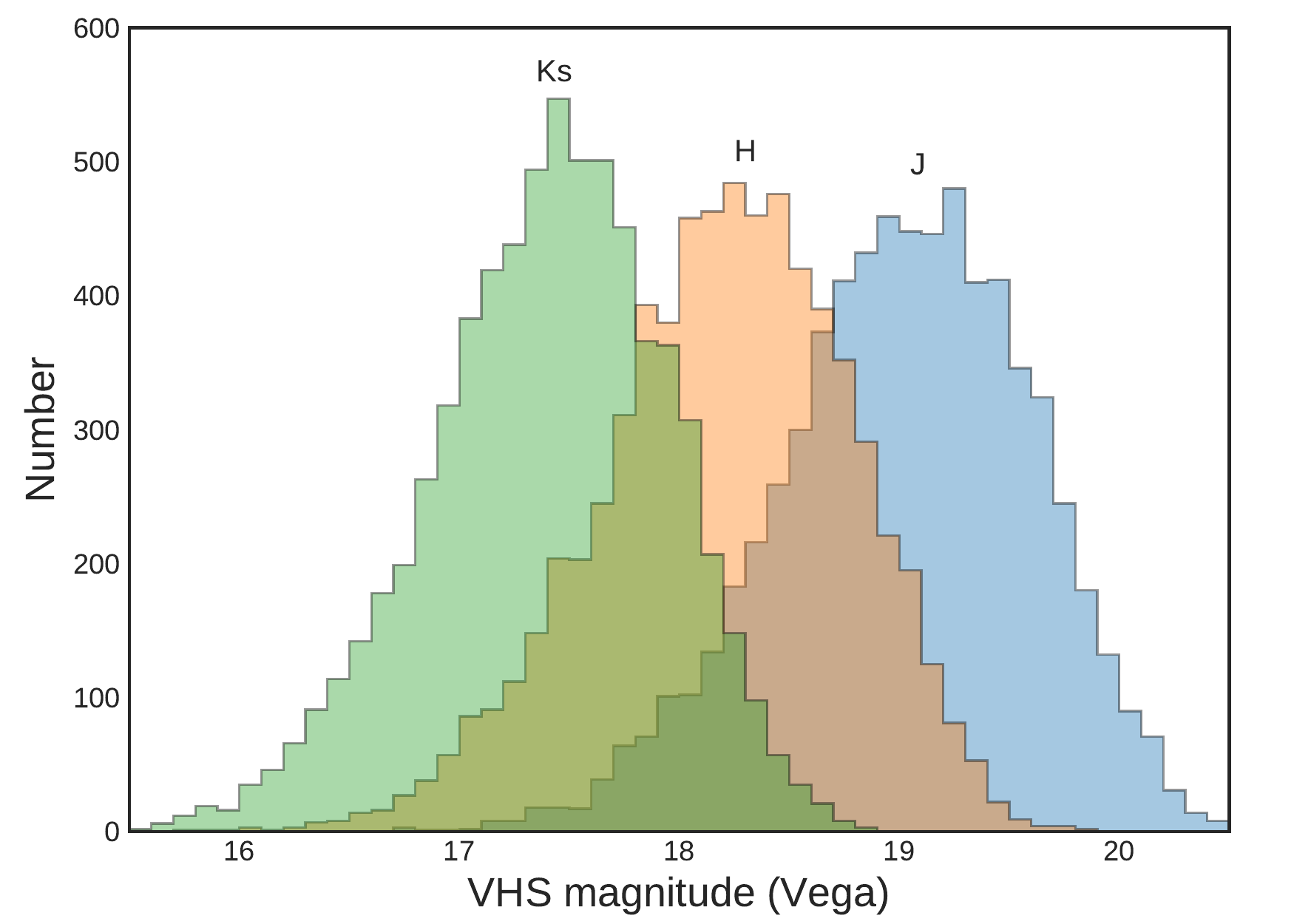} \\
    \includegraphics[width=0.47\textwidth]{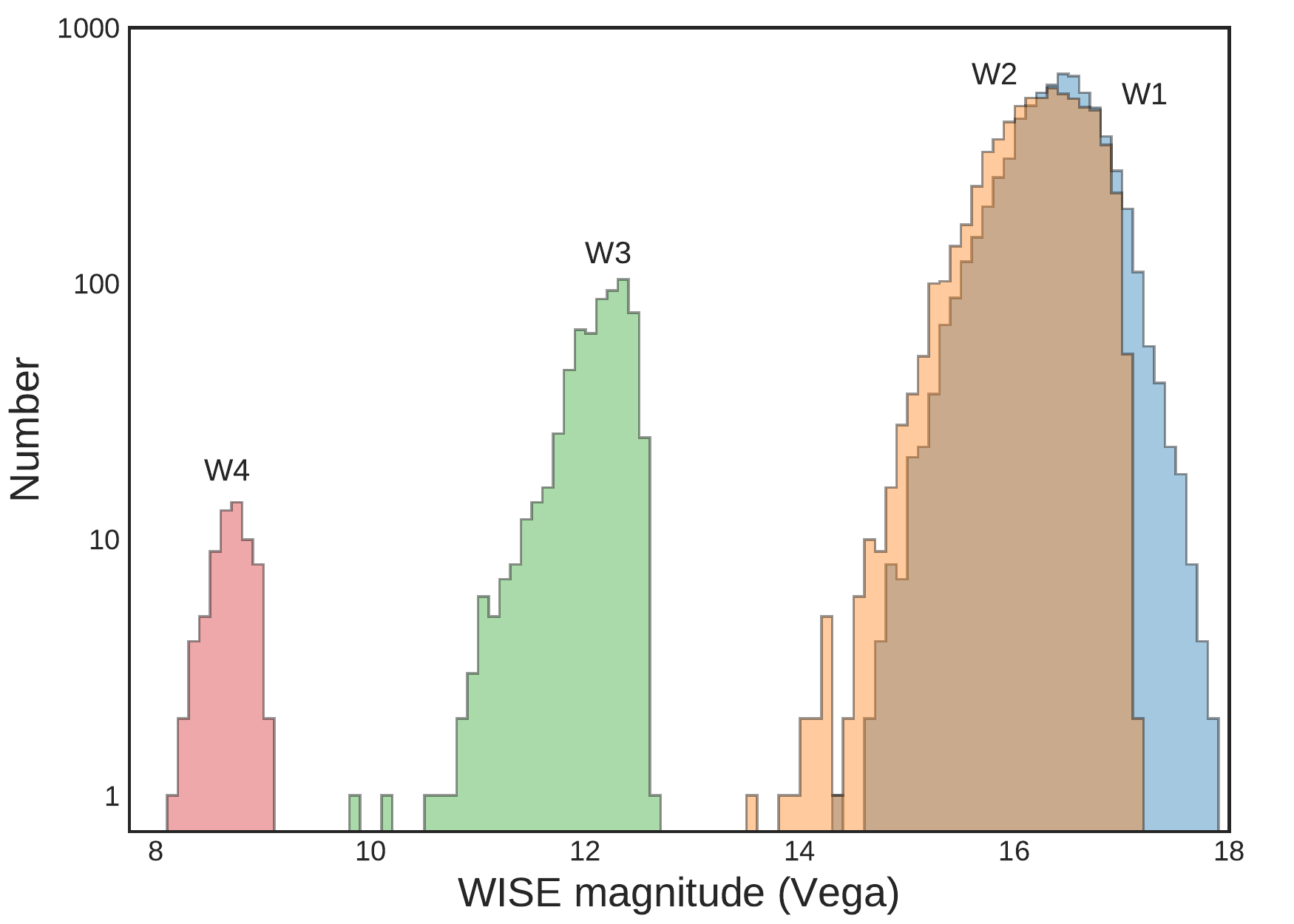}
    \end{tabular}
\caption{The magnitude distributions of our initial sample of 6,860 sources for the CFHTLS optical bands u, g, r, i and z (top), the VHS near-IR bands J, H and Ks (middle) and the \textit{WISE} mid-IR bands W1, W2, W3 and W4 in Vega system (bottom) with signal-to-noise ratio higher than three.} \label{magHist}
\end{figure}

The VIPERS catalogue with spectroscopic redshfits and optical photometry was cross-matched with the list of mid-IR AllWISE and the near-IR VHS sources at the same time using the \texttt{xmatch} tool from the \texttt{astromatch}\footnote{\url{https://github.com/ruizca/astromatch}} package that uses different statistical methods for cross-matching of astronomical catalogues. This tool matches symmetrically a set of catalogues and gives the Bayesian probabilities of the associations or non-associations \citep{pineau2017}. To properly perform \texttt{xmatch}, all the cross-matched catalogues must cover the same footprint. As AllWISE is an all-sky survey and the VHS data cover the entire VIPERS field, we selected only the IR sources that lie in the footprint of the VIPERS survey. After the cross-match with \texttt{xmatch}, we kept only sources with a high probability of association (>68\%). When one source was associated with several counterparts, we selected the association with the highest probability. We then filtered out sources with signal-to-noise ratio lower than three in the near-IR and mid-IR bands to better constrain the SED fitting in the IR regime. The resulted catalogue of 6,860 sources was used as the parent sample (hereafter as VIPERS sample).

\begin{figure}
    \includegraphics[width=0.475\textwidth]{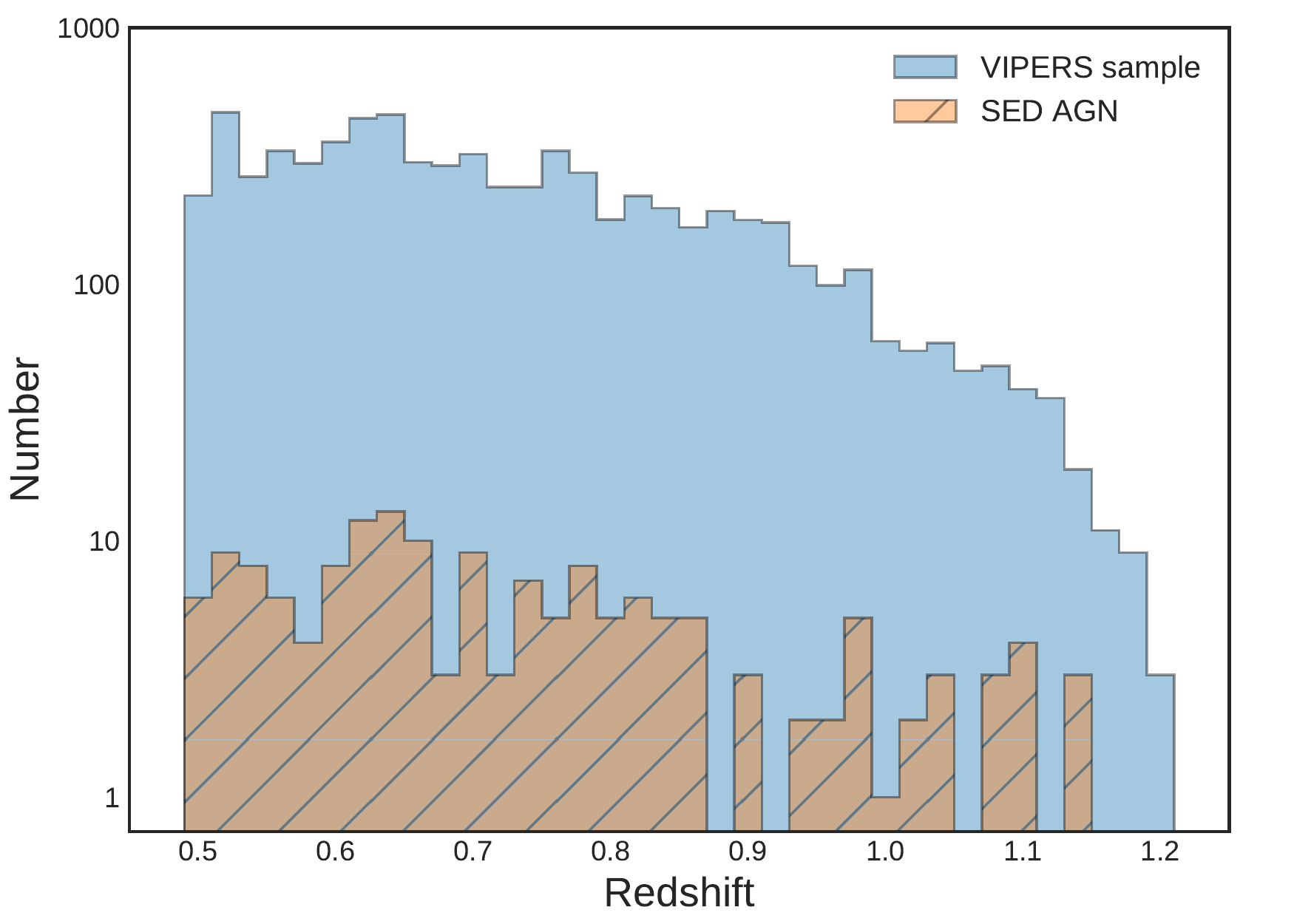}
\caption{Redshift distribution for all sources in the initial sample (blue) and for the 160 high-confidence SED selected AGNs (orange-hatched).}\label{Zhist}
\end{figure}

In Figure~\ref{magHist}, we show the optical, near-IR and mid-IR (from top to bottom) magnitude distributions of our final sources with signal-to-noise ratio higher than three, while in Figure~\ref{Zhist} we present their corresponding redshifts (blue histogram). All the mid-IR (\textit{WISE}) magnitudes were measured with profile-fitting photometry, since all the sources are point-like (\textit{ext}\_\textit{flg}=0) in the mid-IR images, which may be due to the low angular resolution of \textit{WISE} telescope (e.g. $\sim$6.1" for W1 band). The shape of the magnitude distributions in the optical and near-IR bands (Fig.~\ref{magHist}) is Gaussian-like, while the mean magnitude value becomes brighter from the u band (24.3 mag) to Ks band (17.4 mag). This is consistent with the expected magnitudes of the underlying host galaxy SEDs. Finally in the mid-IR bands, the number of W3 and W4 detections is lower compared to the shorter-wavelength mid-IR bands. This could be attributed to the fact that the AllWISE W3 and W4 bands have much lower sensitivity compared to W1 and W2 bands.

\section{Analysis}\label{methods}

In this Section, we present the templates and parameter space used to fit the SEDs of the sources and the Bayesian methodology applied to select AGNs among these sources. We performed SED analysis using all the available optical (CFHTLS), near-IR (VHS) and mid-IR photometry (\textit{WISE}), using \texttt{X-CIGALE} \citep{yang2020} that is the latest version of the Code Investigating GALaxy Emission \citep[\texttt{CIGALE},][]{noll2009,Ciesla2015,boquien2019}. This allows us to estimate the physical parameters of the sources. In particular, we are interested in the fraction of the IR luminosity originating from the AGN to the total IR luminosity of the galaxy, which is used as a proxy of the AGN activity \citep{Ciesla2015}. The SEDs were first fitted with a) a galaxy template and b) a galaxy plus an AGN component. For each case, we obtained the best-fitting solution provided by \texttt{X-CIGALE}, and used the Bayesian Information Crirerion \citep[BIC,][]{schwarz1978} to quantify which template (galaxy/galaxy+AGN) provides the best fit for each source and the highest probability to host an AGN. We describe the analysis steps in detail below.


\begin{table*}
\caption{Models and the values for their free parameters used by \texttt{X-CIGALE} for the SED fitting of our initial sample (6,860 sources).}
\begin{threeparttable}
\begin{tabular}{ l c r }
\hline
\multicolumn{1}{|l|}{Parameter} &  & Value \\ \hline \hline
\multicolumn{3}{|c|}{Star formation history: double-exponentially-decreasing ($\tau$-dec) model}\\
Age of the main stellar population && 0.5, 1.0, 3.0, 5.0, 7.0, 9.0, 11.0, 13.0 \\
$\tau_{main}$ && 0.1, 0.5, 1.0, 3.0, 5.0, 10.0, 20.0 \\
$age_{burst}$ && 0.1, 0.2, 0.4, 0.5  \\
\hline

\multicolumn{3}{|c|}{Stellar population synthesis model}\\
Single Stellar Population Library&&\citet{bruzual2003}\\
Initial Mass Function&& Salpeter \\
Metallicity && 0.02 (Solar) \\
\hline

\multicolumn{3}{|c|}{Nebular emission}\\
Ionization parameter ($logU$)&& -2.0 \\
Fraction of Lyman continuum escaping the galaxy ($f_{esc}$)&& 0.0 \\
Fraction of Lyman continuum absorbed by dust ($f_{dust}$)&& 0.0 \\
Line width in km/s&& 300.0 \\
\hline

\multicolumn{3}{|c|}{Dust attenuation: \citet{calzetti2000}} \\
Colour excess of stellar continuum light for young stars E(B-V) && 0.05, 0.1, 0.2, 0.3, 0.35, 0.4, 0.5, 0.6 \\
Reduction factor for the E(B-V) of the old stars compared to the young ones && 0.44\\
\hline

\multicolumn{3}{|c|}{Dust template: \citet{dale2014}}\\
IR power-law slope && 2.0 \\
\hline

\multicolumn{3}{|c|}{AGN models from \citet{fritz2006}}\\
 Ratio between outer and inner radius of the torus ($r_{ratio}$) && 60.0  \\
 Optical depth at 9.7 $\mu$m ($\tau$) && 0.1, 1.0, 6.0, 10.0 \\
 Parameter linked to the radial dust distribution in the torus ($\beta$) && -0.5 \\
 Parameter linked to the angular dust distribution in the torus ($\gamma$)& & 4.0 \\
 Angular opening angle of the torus ($\theta$) && 100.0 \\
 Angle with line of sight ($\psi$) && 0.001, 50.100, 89.990  \\
 AGN fraction && 0.1, 0.2, 0.3, 0.5, 0.7, 0.9 \\
 Extinction in polar direction, E(B-V) && 0.0, 0.05, 0.1, 0.15, 0.2, 0.3, 1.0\\
 Emissivity of the polar dust && 1.6 \\
 Temperature of the polar dust (K) && 100.0\\
\hline
\end{tabular}
\begin{tablenotes}
\item \textbf{Note.} -- $\tau_{main}$ is the e-folding time of the main stellar population model in Gyr, age is the age of the main stellar population in the galaxy in Gyr (the precision is 1 Myr), and $age_{burst}$ is the age of the late burst in Gyr (the precision is 1 Myr). $\beta$ and $\gamma$ are the parameters used to define the law for the spatial behaviour of the torus density. The functional form of the latter is $\rho(r, \theta)\propto r\beta e^{-\gamma|cos\theta|}$, where r and $\theta$ are the radial distance and the polar distance, respectively. $\theta$ is the opening angle and $\psi$ the viewing angle of the torus. Type-2 AGNs have $\psi$ = 0.001 and Type-1 AGNs have $\psi$ = 89.990, while values equal to $\psi$ = 50.100 are for intermediate type of AGN. The extinction in polar direction, E(B-V), included in the AGN module, accounts for the possible extincion in type-1 AGN, due to polar dust. The AGN fraction is measured as the AGN emission relative to IR luminosity (1--1000 $\mu$m). 
\end{tablenotes}
\end{threeparttable}
\label{proposal}  
\end{table*}

The \texttt{X-CIGALE} algorithm fits the observational multi-wavelength data with a grid of theoretical models and returns the best-fitted values for the physical parameters. The results are based on the energy balance, i.e., the energy absorbed by dust in UV/optical is re-emitted after heating at longer wavelengths, such as the mid-IR and far-IR. In this work, we built a grid of models including different stellar populations, dust attenuation properties, dust emission, star formation history and AGN emission. In particular, the models and the parameter space covered by these components are described below:

\begin{legal}[leftmargin=*]
\item To convolve the star formation histories of our sample, we used the double-exponentially-decreasing model (2$\tau$-dec). This model, provides the best stellar mass and star formation rates of the sources, at the expense of unrealistic galaxy ages \citep{Ciesla2015}. Using a different model, e.g., 1$\tau$-dec, delayed SFR, does not affect our measurements, as shown in \citet{mountrichas2019}.

\item For the simple stellar population, we used the synthesis models of \citet{bruzual2003}, assuming the Initial Mass Function by Salpeter. We adopted a constant, solar metallicity of Z\,=\,0.02, and a separation age between the young and the old stellar populations within the range of 1.5-1000 Myrs. A constant metallicity for all sources prevents long time consuming calculations, but also it does not affect significantly the shape of the SEDs compared to the observed ones and the derived properties \citep{yuan2018,hunt2019}. 

\item For the attenuation due to absorption and scatter of the stellar and nebular emission by interstellar dust, we utilized the attenuation law by \citet{calzetti2000}.

\item The emission by dust in the IR regime was modelled by the semi-empirical \citet{dale2014} templates. The parameter that describes them is the power-law slope of the dust mass distribution over heating intensity, $\alpha$. These templates are strongly correlated to the adopted attenuation models through the dust luminosity that is the outcome of the energy balance that \texttt{X-CIGALE} is based on.

\item The AGN emission was modelled using the \citet{fritz2006} template that includes both the emission from the central AGN and also the re-emitted radiation from the dusty torus heated by the central engine at longer wavelengths. Additionally, this latest version of the \texttt{X-CIGALE} code introduces polar-dust extinction to account for the possible extinction in type-1 AGN \citep{yang2020}.

\end{legal}

\noindent Table \ref{proposal} presents the models and the values of their free parameters, used in the fitting process.


\begin{table*}
\caption{Explanation of Bayes Factor and $\Delta$BIC values according to \citet{kass1995} and selection of high-confidence SED AGNs compared to mid-IR and X-ray selected samples.}
\centering
\scalebox{0.90}{
\begin{threeparttable}
\label{tableXrays}  
  \begin{tabular}{c c c c c c c c }
\hline
 $\Delta$BIC & BF & Evidence in favor & AGN Probability & Number of & mid-IR & X-ray & X-ray \& mid-IR\\ 
 (2*log(BF)) &  & of  model m & (\%) & SEDs (6,860) & AGN (52/35) & AGN (116) & AGN (17/14)\\ 
 \hline\hline
 >10 & >150 & very strong & 0.08 &           2,346 & 0/0 &       5 & 0/0 \\
 (6-10)& (3,20) & strong & 2.50  &           2,053  & 1/0&      29 & 0/0 \\
 (2-6)& (1,3) & little & 11.35  &            1,894 &5/3 &       40 & 2/2 \\
 \hline
 (-2-2)& (<1) & equal for both & 43.00  &    407 & 2/0&      8 & 0/0 \\
 \hline
 (-6,-2)& (3,20) & little & 84.30   &         87 & 7/2 &      6 & 1/0 \\
 (-10,-6)& (20,150) & strong & 97.62  &       17 &7/5 &       4 & 3/2 \\
 <-10& >150 & very strong & 99.89 &         56 & 30/25&      24 & 11/10 \\
\hline
 \textbf{<-2}& \textbf{>3} & \textbf{strong} & \textbf{91.21} &         \textbf{160} & \textbf{44/32}&      \textbf{34} & \textbf{15/12} \\
\hline
\end{tabular}
\begin{tablenotes}
\item \textbf{Note.} -- For positive values of 2log(BF) the best model, m, is assumed to be this with only galaxy templates, while for negative values m represents the model when an AGN component is included. Thus, the evidence in each case refers to model m. The AGN probabilities are the average of each bin and refer to the model with AGN templates: high values indicate AGN activity. The number of SEDs refers to the number of sources in each $\Delta$BIC bin. The last three columns give the number of mid-IR, X-ray and both mid-IR and X-ray selected AGNs in these bins. In mid-IR AGN and the last columns, the two numbers correspond to selection criteria of \citet{assef2013} with reliability of 75\% and 90\%, respectively. The last row corresponds to our AGN selection criteria with $\Delta$BIC$<-2$.
\end{tablenotes}
\end{threeparttable}}
\label{tableKassRaft}
\end{table*}

In Bayesian statistics, the choice of the model that fits best to the data is achieved through the Bayes Factor, BF. In Appendix A, we describe the procedure of calculating and evaluating the BF. The BF requires the calculation of the posteriori and the posteriori complementary probabilities. Alternative methods are required when the BF is calculated with not enough information for the a priori probabilities. Information Criteria, such as the Akaike Information Criterion \citep[AIC]{akaike1974} or the Bayesian Information criterion \citep[BIC]{schwarz1978}, are applied, since they are capable of calculating an approximation of the BF in the absence of priori distributions. This is achieved, by taking into account the complexity of the models, in addition to the goodness of fit. In this work, we used BIC, since it evaluates the true model among all possible and alternative hypotheses, while it is more conservative in the parameter impact on the selection compared to AIC. It favors models with small number of parameters, while by increasing the parameter space, it penalizes more the model. The values of BIC for a model are given by:
\begin{equation}\label{equation 1}
BIC=-2*ln(L)+2*p*ln(N),
\end{equation}

\noindent where p and N are the number of parameters and the number of observations, respectively, while L is the maximum likelihood of the model. To compare two models and select the best one, we calculate the difference of the information of the two models:
\begin{equation}\label{equation 2}
\Delta BIC=-2*ln(L_1/L_2) -(p_2-p_1)*ln(N),
\end{equation}

\noindent where the first term of this equation gives the ratio of the likelihoods of the two models and (p\textsubscript{2}-p\textsubscript{1}) is the difference of the parameters used in each model. It can be shown that the difference in BIC values for two models, $\Delta$BIC, are related to BF, and in this case the latter can be calculated approximately by:

\begin{equation}
BF= exp(-\Delta BIC/2). 
\end{equation}

\noindent We further consider the Schwarz weights \citep{burnham2002}:
\begin{equation}
weight=\frac{exp(-\Delta BIC_j/2)}{\sum_{n=1}^{2}exp(-\Delta BIC_n/2)}.
\end{equation}

\noindent These weights indicate the relative preference between two candidate models (n=1,2) and also provide a method to combine a parameter when using multiple model averaging. In other words, they express a probability that favours one model (in this case j) against the other. In Table~\ref{tableKassRaft}, we give the interpretation for each value of $\Delta$BIC and we select the best model according to $\Delta$BIC values and the posteriori probability of each model.


\section{Results}\label{results}

\subsection{SED selected AGNs}\label{4.1}

\begin{figure}
    \includegraphics[width=0.48\textwidth]{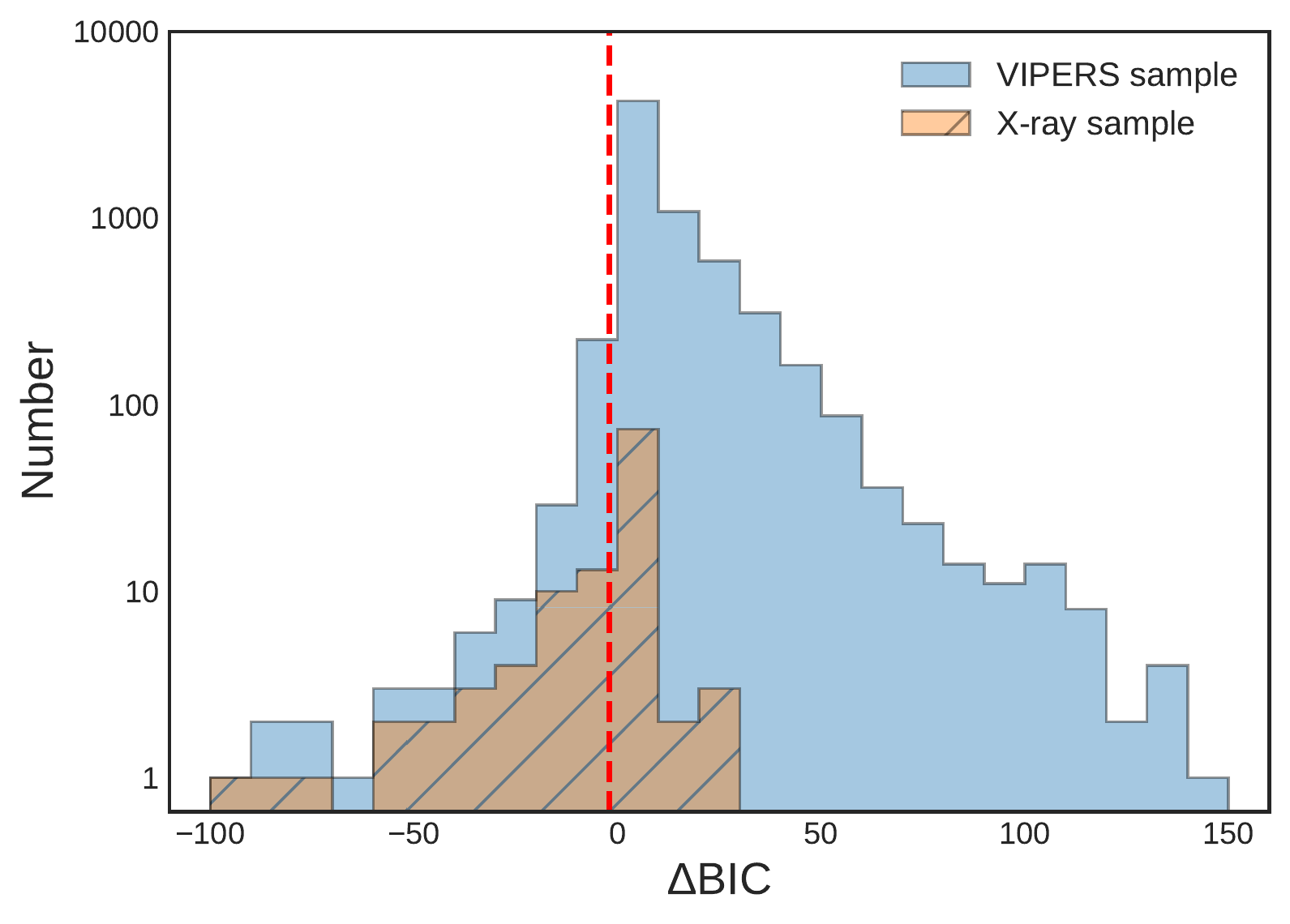}
\caption{The distribution of $\Delta$BIC for the VIPERS sample (blue). The orange-hatched histogram represents the distribution only for the X-ray sample. The vertical dashed line represents the threshold used in this work adopted by \citet{delmoro2017} corresponding to evidence that favors the model when including the AGN template.}\label{cumul}
\end{figure}

To identify AGN candidates, we constructed and modelled the SEDs of all the 6,860 sources in the VIPERS sample. Each SED was fitted twice. In the first run, we used only galaxy model templates. In the second run, we used both galaxy and AGN templates. For each case, we obtained the best fitting model and we calculated the BIC values (Eq.~\ref{equation 1}). Based on the difference, $\Delta$BIC (Eq.~\ref{equation 2}), and its interpretation (Table~\ref{tableKassRaft}), we classified the sources into samples with different evidence ratio (or AGN probability). Following \citet{delmoro2016}, we adopted a threshold of $\Delta$BIC$<-2$ to select AGN candidates. This results to 160 galaxies ($\sim$2.3\% of the initial sample) with probabilities hosting an AGN higher than 73\% (the mean value is equal to 91.20\%) and evidence that favors the model when including the AGN templates. These sources also have lower values of the reduced $\chi^2$ in the model fitting when adding an AGN component. Figure~\ref{cumul} shows the $\Delta$BIC distribution for the VIPERS sample (blue histogram). The vertical dashed line denotes the threshold used in this work.

 \begin{figure}
   \begin{tabular}{c}
    \includegraphics[width=0.5\textwidth]{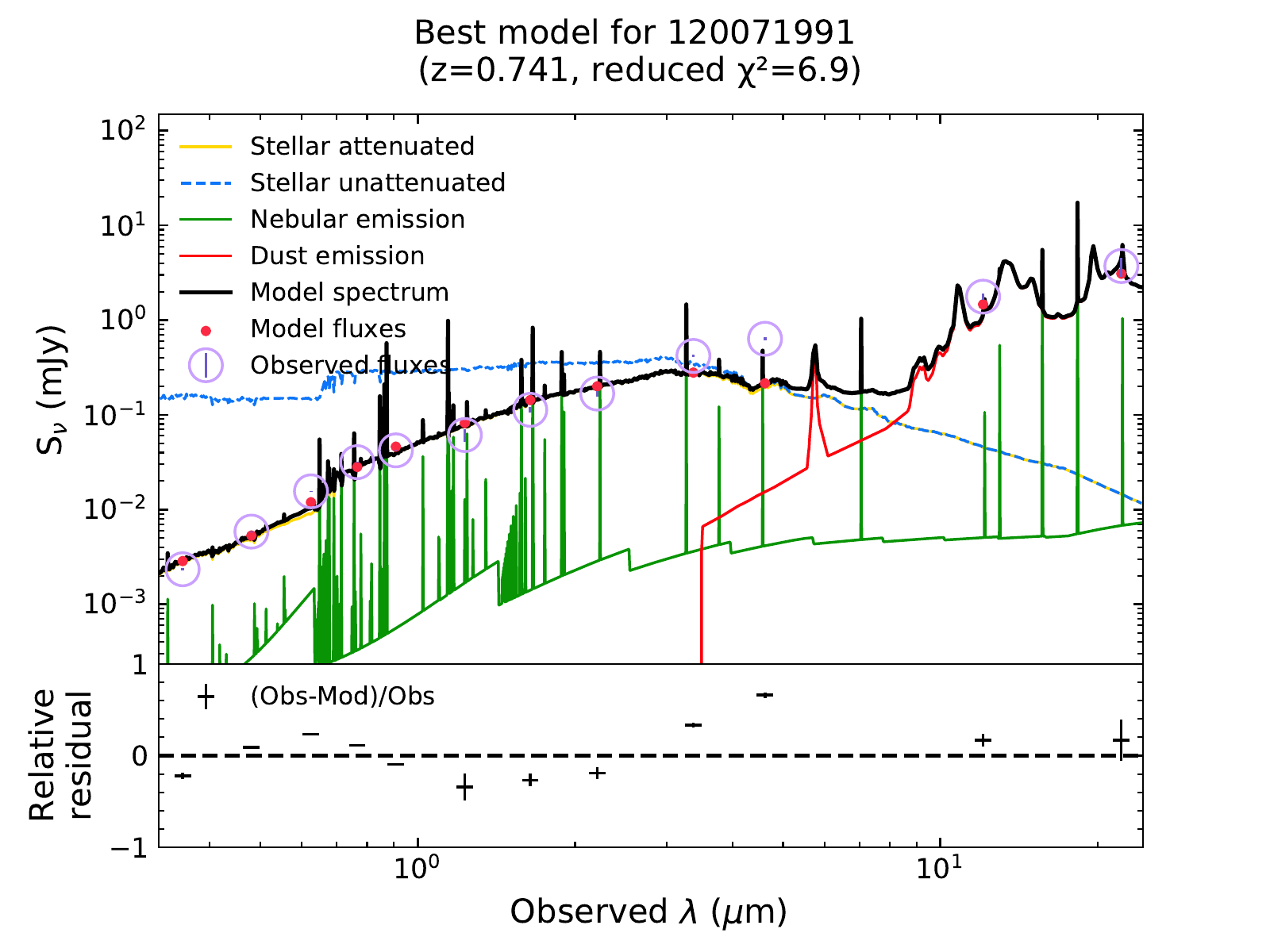} \\
    \includegraphics[width=0.5\textwidth]{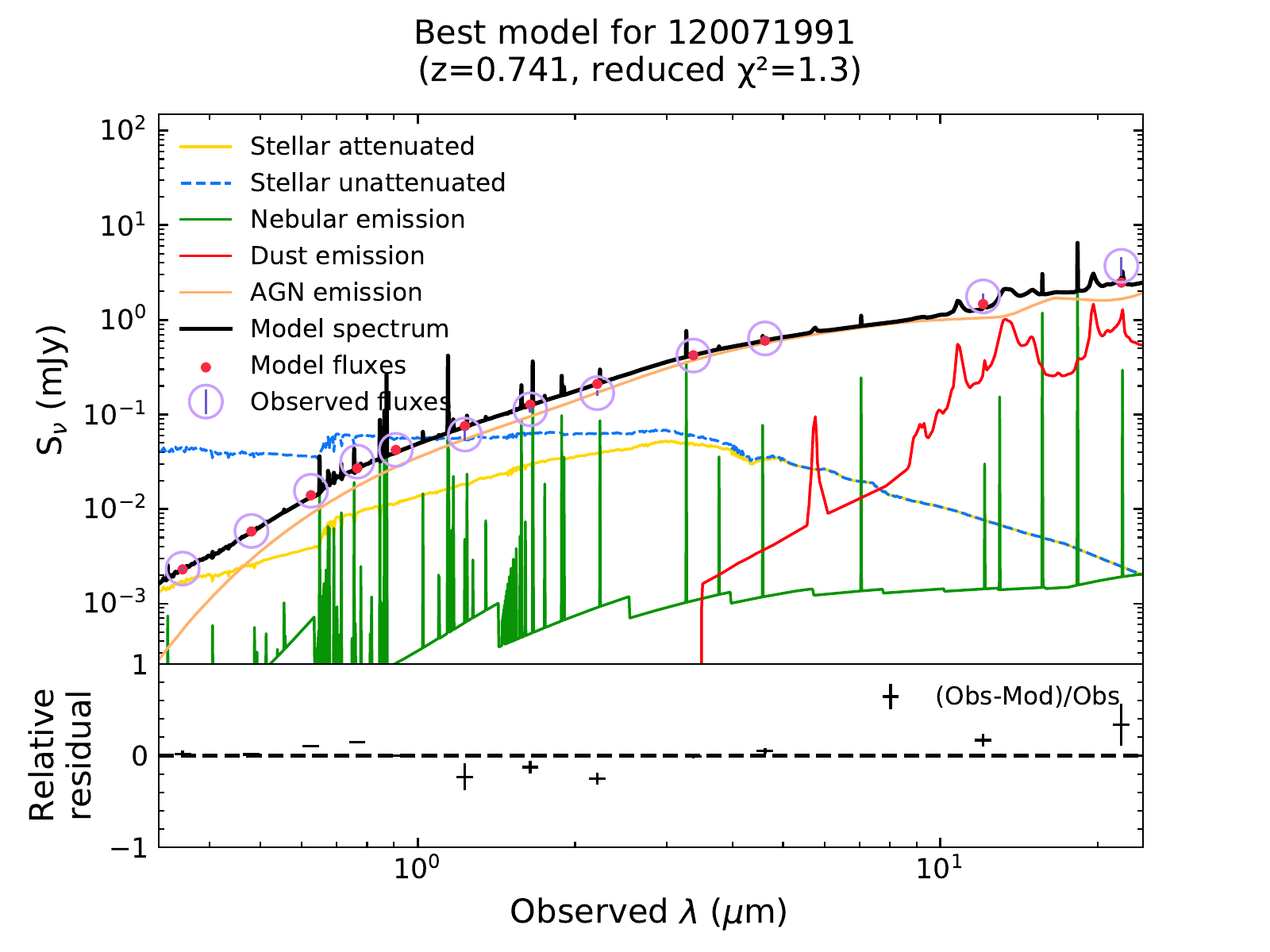}
    \end{tabular}
\caption{SED fitting example of a source classified as AGN with probability 99.99\%. The upper and lower SEDs correspond to models without and with AGN templates, respectively. The dust emission is plotted in red, the AGN component in green, the attenuated (unattenuated) stellar component is shown with the yellow (blue) solid (dashed) line, while the orange lines shows the nebular emission. The total flux is represented with black colour. Below each SED, we plot the relative residual fluxes versus the wavelength.}\label{sedshigh}
\end{figure}

In Figure~\ref{sedshigh}, we present the SEDs with and without AGN templates (lower and upper panels, respectively) for a source classified as AGN, with probability of 99.99\%. When an AGN component is added to the fitting process, the fit is significantly improved and the relative residual fluxes are minimized, as shown at the bottom of each panel. In Table~\ref{tableALL} in the Appendix, we list the properties of the 160 SED selected AGNs, while Figure~\ref{Zhist} shows their redshift distribution (orange-hatched histogram). We, further, explored (Appendix~\ref{farIR}) whether the availability of photometry at different wavelengths affects the AGN selection method. Our analysis showed that near-IR photometry (in addition to optical and mid-IR) is crucial for the reliability of the methodology. However, absence of far-IR or mid-IR photometry at longer wavelengths (W3, W4 photometric bands) does not affect the selection of AGN candidates.

The inclination angle, $\psi$, is defined as the angle between the equatorial AGN axis and the line of sight. Although, (X-) CIGALE cannot constrain the exact value of $\psi$ this parameter can be used as an indication for classifying AGN into type-1 and type-2 \citep{Ciesla2015,yang2020}. Specifically, $\psi =90^{o}$ denotes type-1 AGN, whereas $\psi \leq 50^{o}$ indicates intermediate or type-2 AGN. According to the SED fitting results, $71\%$ of the sources have inclination angle values $\leq 50^{o}$. In particular, 46/160 (29\%) and 81/160 (51\%) have $\psi =50^{o}$ and $\psi =0^{o}$, respectively. Thus, the vast majority (70\%) of our AGN candidates seem to present some level of obscuration. To further investigate the nature of the sources, in the next subsections we examine their optical spectra, their X-ray and mid-IR properties and we apply different obscuration diagnostics using optical and IR colours.

 
\subsection{X-ray detections and upper limits}\label{xray}

In this section, we explore the X-ray properties of our 160 SED selected AGN. We use the most recent available X-ray catalogue to search for counterparts, while for non detections we derive upper limits by constructing the X-ray mosaic using all available images in the field. Out of 14,168 X-ray sources in the northern XMM-XXL field \citep{chiappetti2018}, 10,029 sources ($\sim$70\%) fall inside the field of view of the VIPERS pointings. Cross-matching this sample with the whole VIPERS catalogue using the \texttt{xmatch} software (see Section~\ref{finalsample}), there are 4,736 sources with high probability of association (68\%). 359 of them have good quality spectroscopic redshifts (see Section~\ref{finalsample}) and are within the redshift range used in this study $(0.5<z<1.2)$. $116/359$ have optical, near-IR and mid-IR counterparts with a signal-to-noise ratio greater than three in the W1 and W2 \textit{WISE} bands. The $\Delta$BIC distribution of the 116 X-ray sources is shown in Figure~\ref{cumul}. Out of these, 34 sources ($\sim$30\%) are also SED selected AGNs ($\Delta$BIC$>$-2). This means that the remaining 70\% is not selected with the SED fitting technique. These sources should be AGNs with small amount of dust in their torus (or dust-free) and/or low IR contribution to the total of their hosts making them difficult to identify in the mid-IR wavelengths. We examine the nature of the latter in the next sections.

For the 126 SED selected AGNs not detected in X-rays, we derived the upper limits in the soft [0.5-2 keV] and hard [2-8 keV] bands. First, we constructed the mosaics taking into account all the publicly available \textit{XMM-Newton} observations \citep{jansen2001} covering the XMM-XXL northern field. The detailed imaging procedure is described in Ruiz \& Georgakakis (in prep.). Briefly, we retrieved and processed the overlapping observations in the field from all detectors by following the standard data reduction tasks of the \textit{XMM-Newton} Scientific Analysis Software \citep[SAS]{gabriel2004} and accounting for all the latest calibration files. \texttt{EPIC-pn} \citep{struder2001} and both \texttt{MOS-1} and \texttt{MOS-2} \citep{turner2001} detectors were operated in full frame mode using the thin filter. The event files from all detectors were cleaned from hot pixels/columns and pixels at the edges of the cameras and were screened to remove high particle background and soft proton flares by setting {\texttt{FLAG}=0} and selecting \texttt{pn} and \texttt{MOS} single events with {0<\texttt{PATTERN}<4} and {\texttt{PATTERN}<=12}, respectively. In a similar way, we built the combined mosaic with the exposure maps. The background mosaics were produced by masking the areas around the X-ray detections.

For the upper limits, we extracted the total counts around the sources within a circular region of radius equal to 15". For the background, we used a circular region centered on the target position with a radius of 30" and we normalized this to the area of the sources. Then, we used a Bayesian approach of 99.7\% confidence level to derive upper limits \citep{kraft1991}. To find the final count rates, we divided the upper limits with the exposure time and the encircled energy fraction ($eef$): $count\, rate=upper\, limit/eef/exposure\, time$. The exposure time was taken as the average value of the pixels in a circular region of 15", while we adopted $eef$=0.7 that corresponds to the radius used for the sources extraction counts. The count rates were converted to fluxes, using an energy conversion factor ($ecf$) calculated for each band. For the $ecf$ calculation, we used the \texttt{webbPIMMS} site by assuming a power-law model with photon index of $\Gamma=1.7$ \citep{nandra2005,tozzi2006} and galactic absorption $N_{H}=2.6*10^{20}$cm\textsuperscript{-2}. We calculated separately the $ecf$ for the \texttt{pn} and \texttt{MOS} thin filters and we took the average of them all. The final $ecf$ used are equal to $2.58\times10^{11}$ and $6.82\times10^{10}$ ergs photon\textsuperscript{-1} for the soft and hard band, respectively. The luminosities were calculated using the following equation:
\begin{equation}
L_X=4*\pi *D_L^{2}*F_X*(1+z)^{\Gamma -2},
\end{equation}
\noindent where L\textsubscript{X} and F\textsubscript{X} are the flux and luminosity in the hard band, respectively, D\textsubscript{L} is the luminosity distance, z is the redshift and $\Gamma$ is the photon index \citep{alexander2003,xue2011}. The units of L\textsubscript{X} are given in erg s\textsuperscript{-1}.


\subsection{Mid-IR selected AGNs}\label{mid-IR}

In addition to studying their X-ray properties, we explore whether the 160 SED selected AGNs could be characterized as AGN via simple mid-IR colour selection criteria. These criteria are based on the power-law that appears in the mid-IR bands (5-10 $\mu$m) when AGN luminosity is at least comparable with that of its host. A number of diagnostics have been proposed using \textit{WISE} data that provides imaging at 3.4, 4.6, 12 and 22 $\mu$m for a large sample of galaxies. These include a simple colour cut-off (W1--W2$\geq$0.8) defined by \citet{stern2012} for bright sources (W2$\leq$15.05 mag), a more refined magnitude-dependent cut-off taking into account faint sources in the W2 band by \citet{assef2013} and two wedges in W1--W2 vs. W2--W3 and W1--W2 vs. W3--W4 colour-colour diagrams suggested by \citet{mateos2012} using three and four \textit{WISE} bands, respectively.
Similar methods have also been proposed for sources observed with the IRAC instrument \citep{fazio2004} on board the \textit{Spitzer Space Telescope} \citep{werner2004} with observations at 3.6, 4.5, 5.8 and 8.0 $\mu$m filters. These are the "Lacy wedge" \citep{lacy2004,lacy2007,sajina2005}, the "Stern wedge" \citep{stern2005} and, more recently, the "Donley wedge" \citep{donley2007,donley2012}.
\begin{figure}
     \includegraphics[width=0.48\textwidth,height=0.48\textwidth]{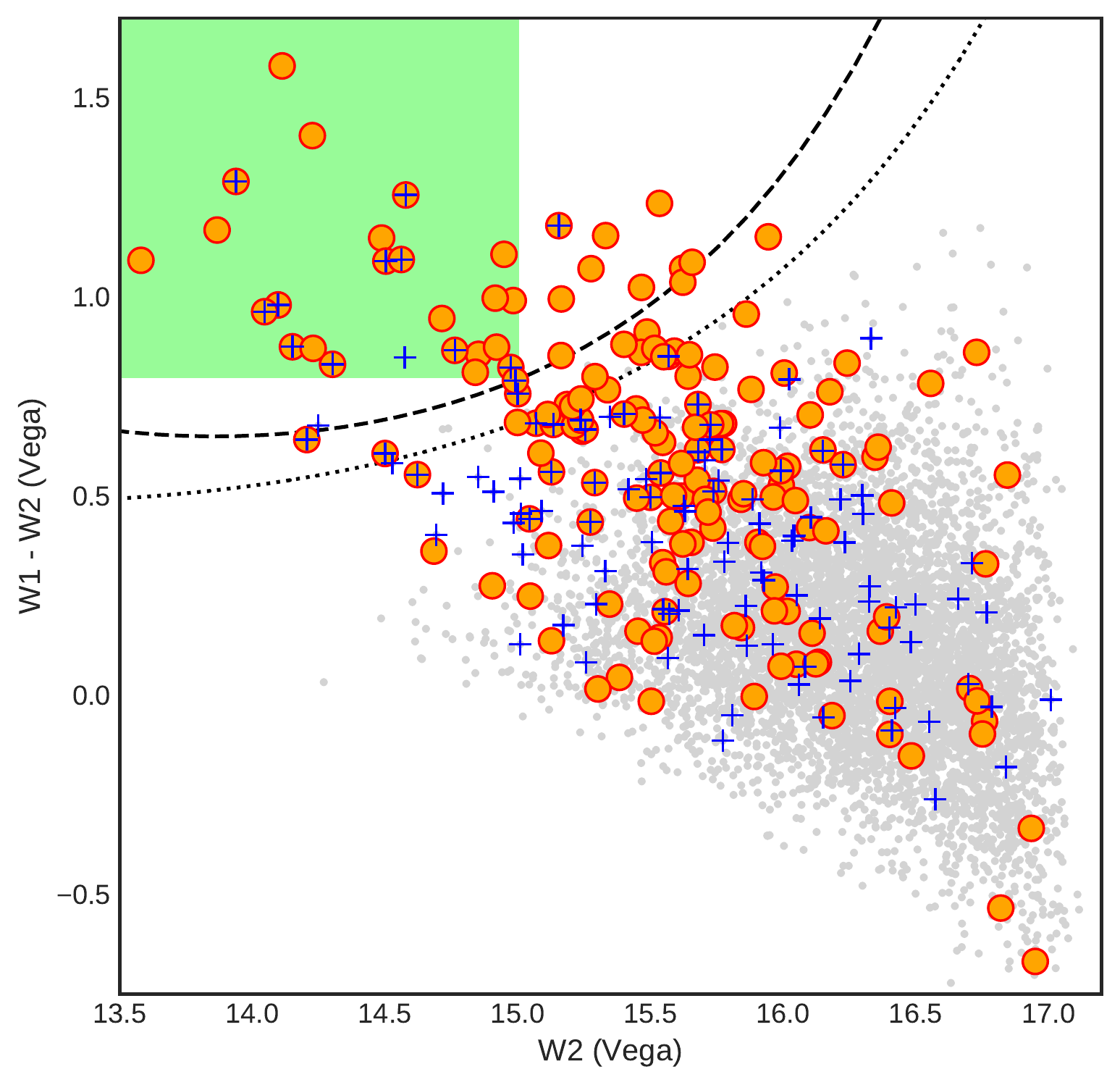}
\caption{\textit{WISE} magnitude-colour (W2, W1--W2) diagram for all 6,860 VIPERS sources (gray points). The dashed (dotted) line represents the \citet{assef2013} selection threshold with reliability of 90\% (75\%). The orange circles and blue crosses represent the SED and X-ray selected AGNs, respectively. The shaded area indicates the AGN selection criteria by \citet{stern2012}}\label{assef}
\end{figure}

\begin{figure}
    \includegraphics[width=0.48\textwidth,height=0.48\textwidth]{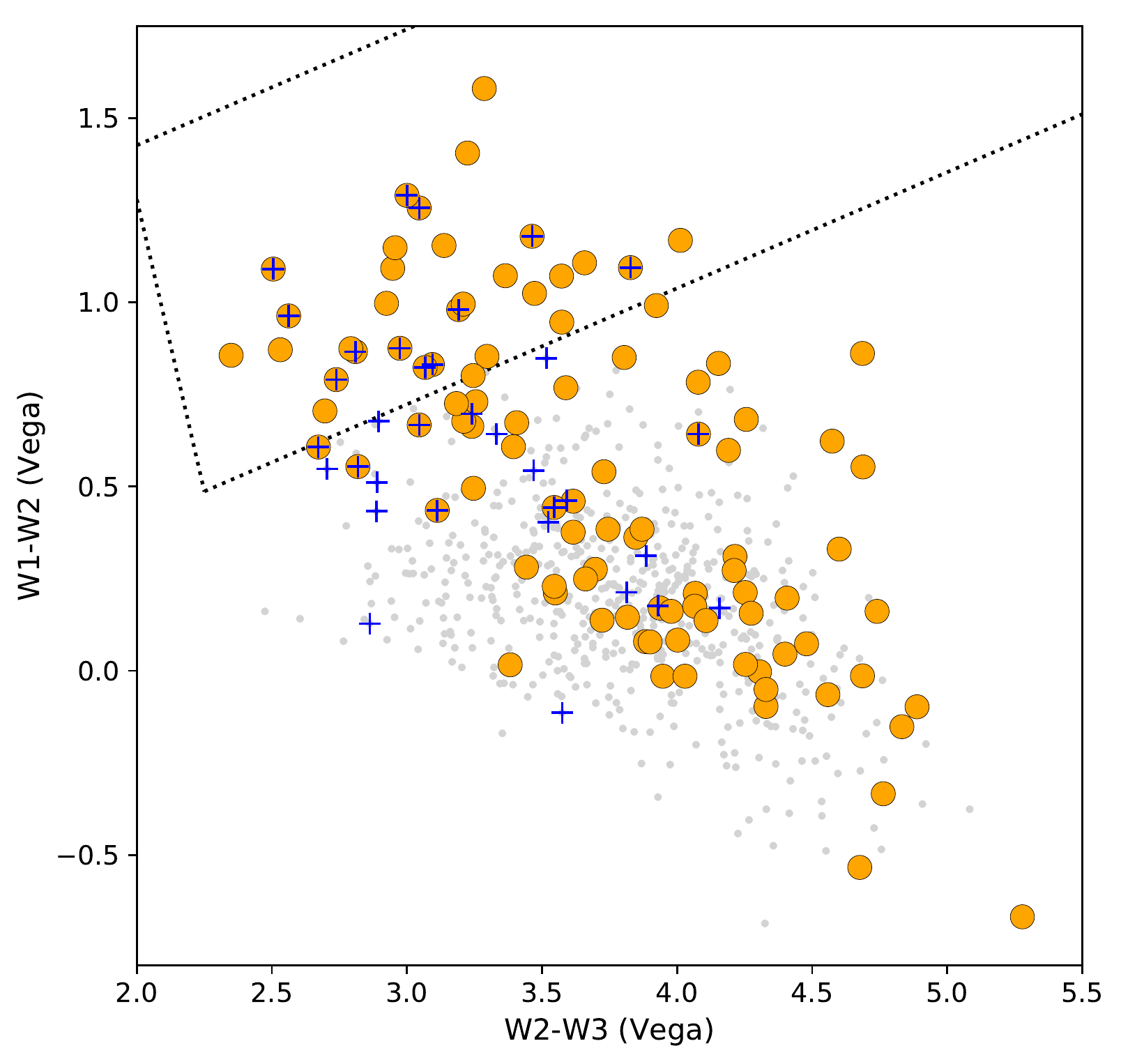}
\caption{\textit{WISE} colour-colour (W2--W3, W1--W2) diagram of the \textit{VIPERS} sample  with detections in all three bands (gray points) with the \citet{mateos2012} selection wedge (dotted lines). The orange circles and blue crosses represent the SED and X-ray selected AGNs, respectively, with detections in W1, W2 and W3 bands.}\label{mateos}
\end{figure}

First, we examined the VIPERS sample using the diagnostics of \citet{stern2012} and \citet{assef2013}. We used the AllWISE data mentioned in Section~\ref{wisedata}. This sample is photometrically complete at 16.8 mag (Vega), where there is a turn-over in the number density plot of the fluxes in the W2 band. Using the \citet{stern2012} criteria, we selected 25 AGN with magnitude brighter than W2=15.05 mag. Since this method is not reliable for fainter objects, we utilised the AGN selection criteria by \citet{assef2013}:

\begin{equation}    
{\rm{y> \alpha_r*exp(\beta_r*(x-\gamma_r)^2), x>\gamma_r \quad }}
\end{equation}
\begin{equation}
{\rm{y> \alpha_r, x\leq\gamma_r,}}
\end{equation}

\noindent where $x=W2$ and $y=W1-W2$ and the revised constant values of ($\alpha_r,\beta_r,\gamma_r$) given by \citet{assef2018} are equal to ($0.650,0.153,13.86$) and ($0.486,0.092,13.07$) for the 90\% and 75\% reliability levels, respectively. We found 35 sources at 90\% and 52 sources at 75\% reliability. Figure~\ref{assef} shows the \textit{WISE} colour-magnitude (W1--W2 versus W2) plot for the VIPERS sample. The lines represent the wedges as defined in \citet{assef2013} for both 90\% and 75\% reliability. We also over-plotted the 160 SED ($\Delta$BIC<=-2) and X-ray selected AGN samples. The shaded green area represents the \citet{stern2012} threshold for AGN selection.

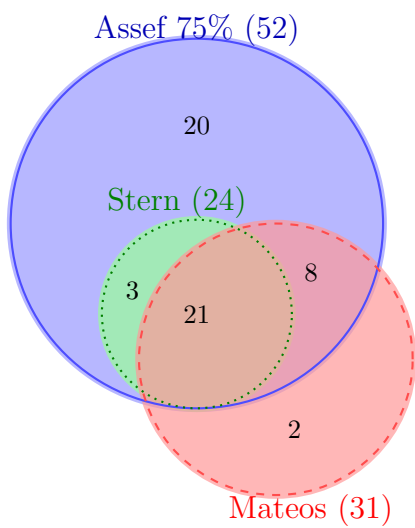
\begin{figure}
\centering
\begin{tikzpicture}[thick,scale=1, every node/.style={transform shape}]
  \begin{scope}[opacity=0.7]
    \fill[blue!40!white]   ( 90:1.2) circle (2.5);
    \fill[green!40!white] (200:0.) circle (1.30);
    \fill[red!40!white]  (330:1.2) circle (1.85);
 \end{scope}
 \def\firstcircle{(90:1.2) circle (2.45)}
 \draw[solid,white!30!blue] \firstcircle node[below] {};
  \def\firstcircle{(200:0.) circle (1.25)}
 \draw[dotted,black!50!green] \firstcircle node[below] {};
  \def\firstcircle{(330:1.2) circle (1.8)}
 \draw[dashed,white!30!red] \firstcircle node[below] {};

  \node at ( 90:3.77)  [font=\Large][text=black!30!blue]   [align=center]{Assef 75\% (52)};
  \node at (100:1.5)  [font=\Large][text=black!50!green]  [align=center] {Stern (24)};
  \node at (300:3.0)   [font=\Large] [text=red][align=center] {Mateos (31)};
 
  \node at (90:2.5) [font=\large] {20};
  \node at (310:2) [font=\large] {2};

  \node at (270:0.0) [font=\large] {21};
  \node at (20:1.6) [font=\large] {8};
  \node at (160:0.9) [font=\large] {3};

\end{tikzpicture}
\caption{Venn diagram of the AGN samples selected through the mid-IR colour selection criteria defined by \citet{mateos2012} (red-dashed), \citet{assef2013} (blue-solid) and \citet{stern2012} (green-dotted).}\label{vennMIR}
\end{figure}

Additionally, we selected mid-IR AGNs through \citet{mateos2012} colour selection criteria by using the colours (W2--W3) and (W1--W2). We do not apply the second diagnostic of \citet{mateos2012} that utilizes four \textit{WISE} bands as the inclusion of the W4 band with high signal-to-noise ratio reduces our sample to only 53 sources. There are 668/6,860 sources in our sample that have detections in all three bands (W1, W2 and W3) and signal-to-noise ratio greater than three. Following \citet{mateos2012}'s criteria:  
\begin{align}
W1-W2<0.3150\times(W2-W3)+0.796\\
W1-W2>-3.172\times(W2-W3)+7.624\\
W1-W2>0.3150\times(W2-W3)-0.222
\end{align}

\noindent we selected 31 mid-IR AGNs. All of them are SED selected AGN by our analysis, while 29 sources are also selected by the Assef criterion with 75\% reliability. In Figure~\ref{mateos}, we plot the (W2--W3) versus (W1--W2) colours. The dotted lines define the \citet{mateos2012} wedge. For reference, we plot the SED (orange circles) and X-ray (blue crosses) selected AGN samples. As illustrated in Figure~\ref{vennMIR} shows the Venn diagram between the different mid-IR colour selection methods. They are consistent with each other, resulting in 54 mid-IR AGNs using at least one of the aforementioned criteria.

\subsection{Optical spectroscopy}

We further inspected the spectra of all the SED selected AGNs to check for any AGN signature in their emission lines. In particular, we searched for the [NeV] forbidden emission line at $\lambda$=3426 \r{A} that is often used as a diagnostic tool to distinguish between AGN and star-forming galaxies \citep[e.g.][]{schmitt1998}. Furthermore, we looked whether broad emission lines are present. Additionally, we used two optical emission line diagnostics to separate the AGN and star-forming populations: the Mass Excitation diagram \citep[MEx,][]{juneau2011,juneau2014} and the colour excitation diagram \citep[TBT,][]{trouille2011}, that use the ([OIII], Hb) and ([NeIII], [OII]) emission line flux ratios, respectively. For this part of our analysis, we used the latest version of the \texttt{specutils}\footnote{\url{https://specutils.readthedocs.io/en/stable/index.html}} packages in \texttt{PYTHON}.

\subsubsection{[NeV] emitters and broad lines}

Based on the information provided in the VIPERS spectroscopic catalogue, 27 out of the 160 SED selected AGN, present broad emission lines in their spectra. As already mentioned, the [NeV] emission line at $\lambda$=3426 \r{A} is a good indicator of AGN activity \citep{schmitt1998,gilli2010}. Its high reliability is based on the fact that the energy needed to ionize [NeV] is 97 eV and may only come from high energy sources as opposed to, for example, stellar emission, since in the latter case the maximum emitted energy is lower than 55 eV \citep{haehnelt2001}. The [NeV] at $\lambda$=3346 \r{A} can also be used to identify AGN. However, it has less diagnostic power, as its intensity is significantly lower compared to [NeV] $\lambda$3426 \r{A} \citep[see e.g. Fig.~1 in][]{Maddox2018}. Many previous studies have used the [NeV] emission to select a large number of AGNs. \citet{mignoli2013} identified 94 type 2 [NeV] emitters and compared them to X-ray selected AGN and those from line ratio diagnostics. They concluded that the [NeV] emitters can identify low-luminosity and heavily absorbed AGNs with increasing fraction to higher stellar masses. More recently, \citet{vergani2018} studied the properties of the hosts of [NeV] AGNs, such as stellar masses, ages and colours. For the optical spectral coverage of VIPERS, the [NeV] line is accessible only for sources that lie at redshift higher than $\rm{z}>0.62$. There are 114 SED AGNs above this redshift limit. After removing objects with artefacts in their spectra and/or those with very low quality, we ended up with 42 [NeV] emitters. The vast majority of them have signal-to-noise ratio higher than five.

\subsubsection{MEx and TBT diagram}
\begin{figure}
    \includegraphics[width=0.48\textwidth]{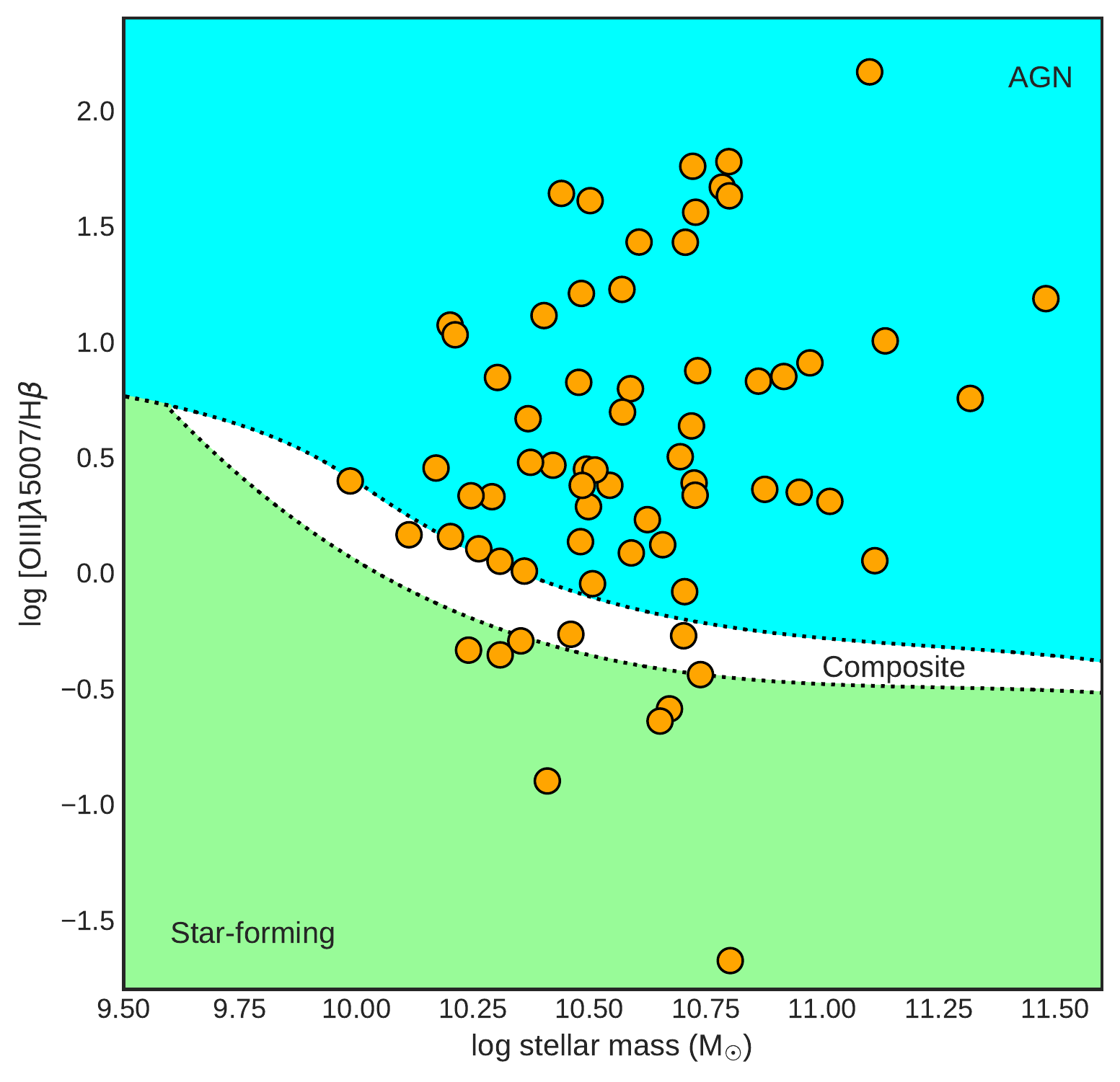}
    \caption{The Mass Excitation (MEx) diagram for the 66 SED selected AGNs that have significant both [OIII] and H$\beta$ emission lines. The dotted lines separate the star-forming galaxies from the AGNs. AGNs are found above the upper line (blue region), while under the lowest line the star-forming galaxies lie (green shaded area). In between these lines, there is a composite population that consists of both AGN and star-forming systems.}\label{MeXplot}
\end{figure}
For low redshifts $\rm{z}<0.5$, the classic emission line ratio diagnostic diagram usually used is the Baldwin, Phillips, \& Terlevich diagram \citep[BPT,][]{baldwin1981}. In our case, the standard emission lines used in the BPT diagram lie outside the wavelength coverage of the optical spectrographs. For example, for redshifts higher than $\rm{z}>0.5$ the [NII] and H$\alpha$ lines are redshifted to the observed near-IR regime. Thus, alternative indicators have been proposed in the literature, such as the rest-frame Bessel U-B galaxy colour \citep[Colour Excitation diagram]{yan2011}, H band absolute magnitude \citep{weiner2006}, [OII]/H$\beta$ \citep{lamareille2010}, Dn4000 break \citep{marocco2011}, the stellar mass (MEx diagram) or the rest-frame colour (TBT diagram). All these replacements of the [NII]/H${\alpha}$ ratio are based on the correlation between the galaxy stellar mass and the gas phase metallicity or the rest-frame colours \citep[depended on stellar mass;][]{kauffmann2003}. In this work, we used the MEx and TBT diagrams.

MEx diagram avoids blending of lines in low-resolution spectra, by utilizing stellar mass and only two spectral lines, i.e., [OIII]5007 and H$\beta$. In our analysis, we used the calibrated diagram of \citet{juneau2014} that is reliable to higher redshift (up to z$\sim$1). Using the flux ratio of [OIII] and H$\beta$ and the stellar mass of a galaxy, we utilized the mass excitation diagram to classify a source as star-forming galaxy, AGN or composite galaxy (i.e., both star-formation and AGN emission). To have both of the aforementioned lines present, we limited our sample to sources with $\rm{z}\leq0.9$, due to the VIPERS spectral coverage. Among our 160 SED selected AGNs, 66 sources have both [OIII] and H$\beta$ lines with good quality spectra. In Figure~\ref{MeXplot}, we present the MEx diagnostic plot, including the two empirically determined dividing lines of \citet{juneau2014} that separate the pure star-forming (under), the galaxies with significant AGN emission (above) and the composite (in between) galaxies. The masses were derived from the SED fitting technique described in the previous sections. The MEx diagnostic was calibrated by using the \citet{chabrier2003} IMF, thus we corrected the stellar masses of our sources as we assumed initially the \citet{salpeter1995} IMF when we modelled the SEDs. In particular, we multiplied our stellar masses by a factor of 0.62 as found in \citet{zahid2012}. At the end, according to MEx diagnostic, 53 sources (88\%) lie in the AGN area, six sources in the composite region, while seven sources in the star-forming region.

\begin{figure}
\centering
    \includegraphics[width=0.48\textwidth]{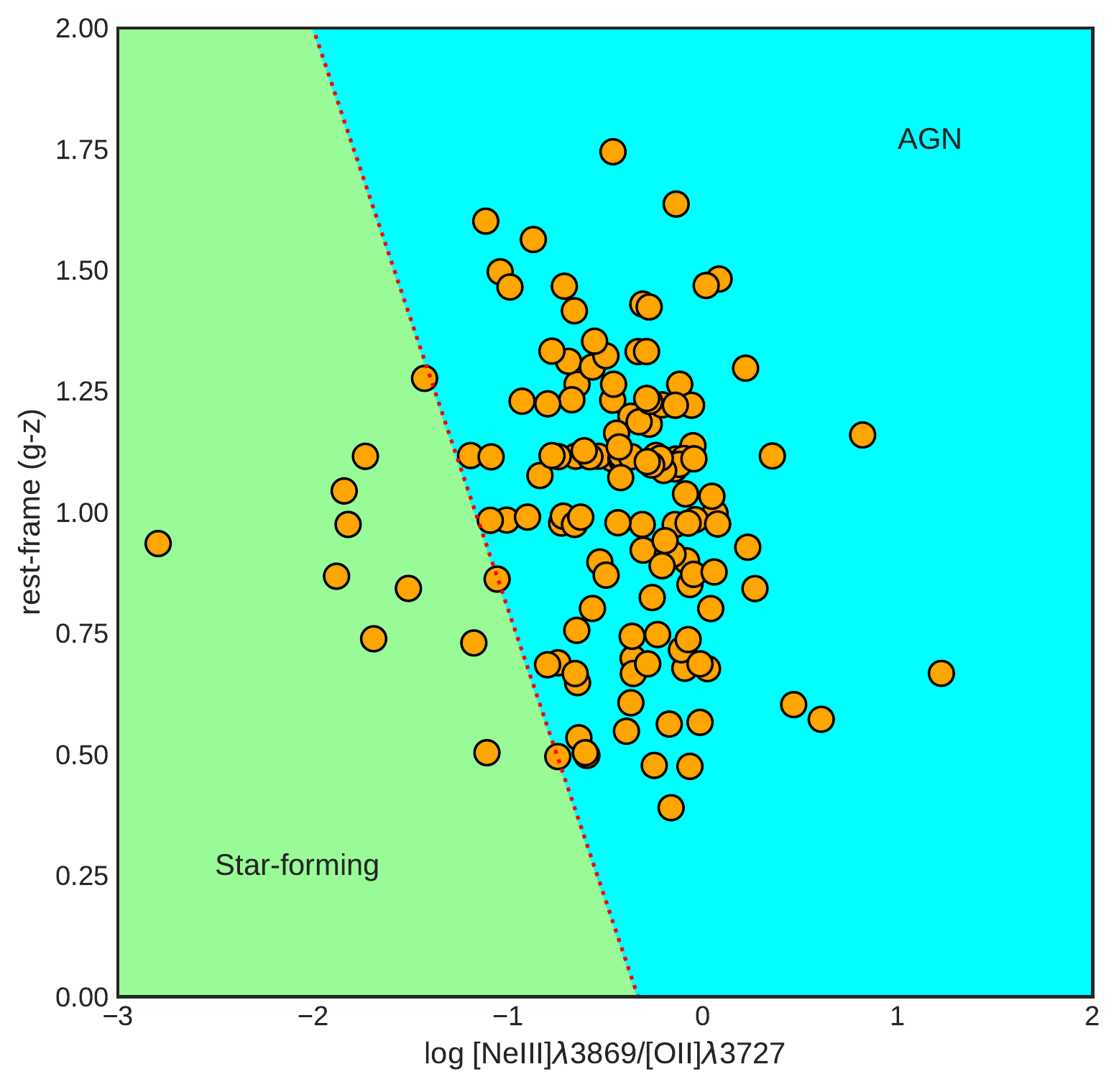}
    \caption{The TBT diagram for the 138 SED selected AGNs that have significant [NeIII] and/or [OII] emission lines. The dotted line separates the star-forming galaxies from the AGNs. The latter are found on the right part of this diagram (blue shaded region), while on the left side the star-forming galaxies exist (green shaded area).}\label{TBTplot}
\end{figure}

The TBT diagram uses the rest frame colour \textsuperscript{0.0}(g-z) as a function of the ratio of the emission lines [NeIII] and [OII]. This method relies on the assumption that AGNs are hosted by massive, fast-rotating galaxies and have high ionization lines compared to the star-forming galaxies \citep{zhang2019}. Moreover, it is not affected by reddening, since these two lines are close to each other. An additional advantage of this method is that it can be used up to redhsift z$\sim$1.4 for optical spectra. TBT diagram is able to disentangle the star-forming galaxies from AGNs, but not the composite galaxies. However, 70\% of the composite galaxies fall inside the AGN area \citep{pons2016}. In Figure~\ref{TBTplot}, the TBT diagram is shown for the 138/160 sources that have good quality spectra. The separation line (dotted) is defined as follows:
\begin{equation}
    {\rm{^{0.0}(g-z)=-1.2\times log([NeIII]/[OII])-0.4.}}
\end{equation}
\noindent 126 sources (91\%) fall inside the AGN area (shaded blue), while twelve sources in the star-forming region (shaded green).

The MEx and TBT diagrams have classified seven and twelve of our SED selected AGN, as star-forming galaxies. However, 7/7 and 5/12 of these sources in the MEx and the TBT diagram, respectively, are selected as AGN through other methods (e.g. X-rays, mid-IR, [NeV] emission). Previous studies have also found a small percentage of AGN misclassified as star-forming systems, when using diagnostics based on optical spectroscopy. \citet{juneau2011} and \citep{juneau2014} found 8\% and 20\%, respectively, of X-ray sources lying inside the star-forming regime of the MEx diagram. In our sample, there are 14 X-ray sources and 16 [NeV] emitters that fall inside the spectral coverage of MEx emission lines. 93\% and 100\%, respectively, are also classified as AGN based on the MEx diagram. \citet{trouille2011} found that 3\% of the X-ray sources are classified as star-forming galaxies, using the TBT diagram. In our case, the percentage of the misclassified AGNs is 5\% (2/40) for the [NeV] emitters and 6.5\% (2/31) for the X-ray AGNs. Given the reliability of the X-ray emission ($L_{X}>10^{42} erg s^{-1}$) and/or the significant [NeV] emission line, we assume that sources lying inside the star-forming area in both diagrams but identified as AGNs via other methods to have real AGN activity.

In total, optical spectroscopy confirms that 134/160 ($\sim$84\%) SED selected AGNs present signs of AGN activity. The remaining sources mainly have poor quality spectra, thus the above diagnostics could not be applied.


\subsection{Intrinsic absorption estimation}\label{obscuration}
\subsubsection{Mildly obscured AGNs}
\begin{figure}
     \includegraphics[width=0.48\textwidth]{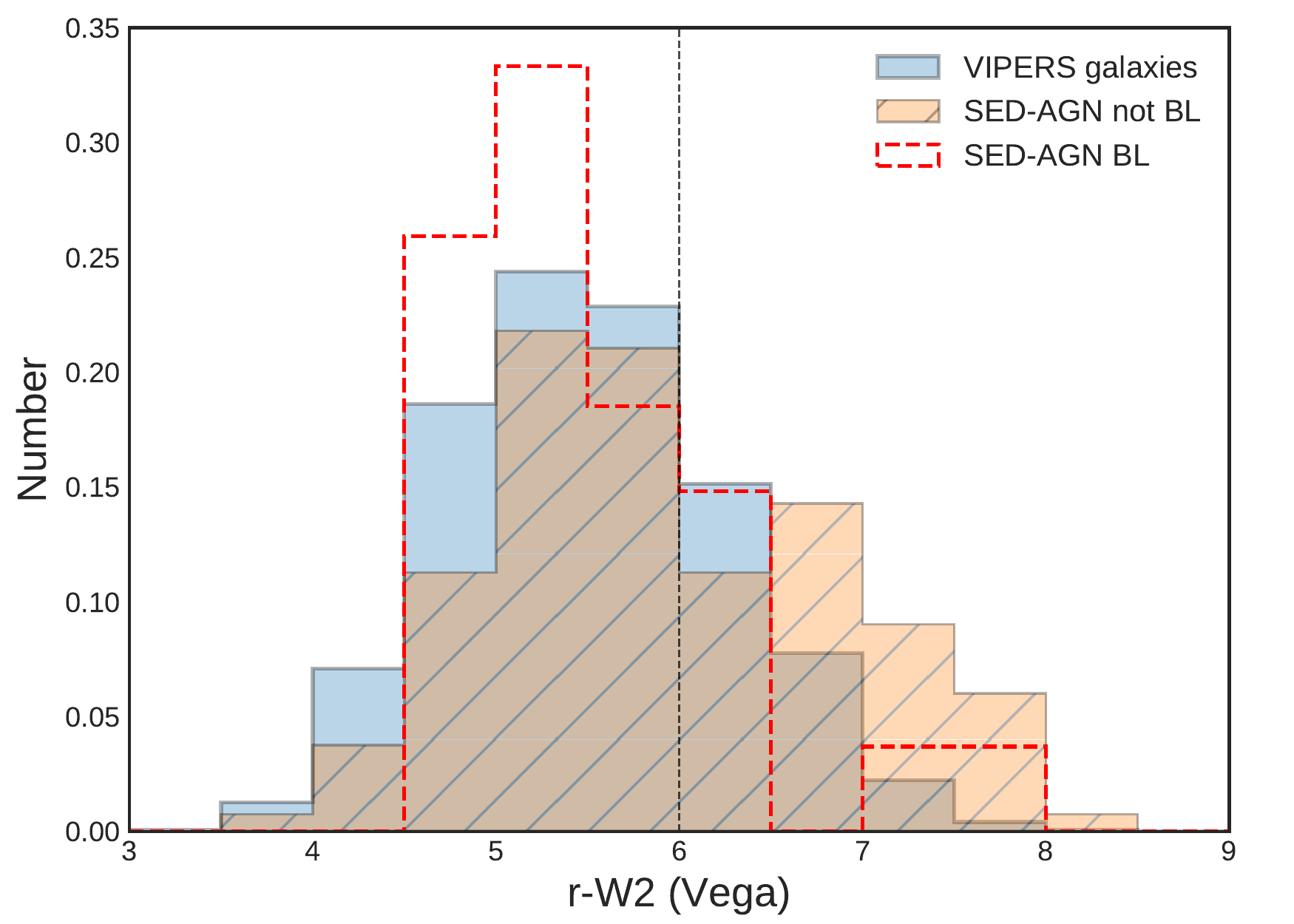}
\caption{The normalized distribution of r--W2 colour for the SED-selected AGNs that present (red-dashed) or not (green-hatched) broad lines in their spectra. For reference, we show the histogram of the VIPERS sample classified as normal galaxies (blue). The vertical dotted line represents the threshold used in \citet{yan2013} to select obscured AGNs.}\label{yanhist}
\end{figure}
SED fitting results revealed that 71\% of the AGNs are obscured ($\psi \leq 70$ corresponding to AGNs of type 1.5 and 2), based on the estimated inclination angle (Section~\ref{4.1}). However, in this section we also explore different diagnostic criteria of obscuration using optical and mid-IR colours or the X-ray to mid-IR relation.

Obscured sources are expected to be unbiased by dust in the mid-IR regime, while being absorbed in the optical bands. \citet{yan2013} used a r--W2>6 mag cut-off along with the Stern criteria (W1--W2>0.8 and W2<15.2 mag) to select type II AGNs. \citet{lamassa2016} used W1 instead of W2 band with a more relaxed threshold (r--W1>4 mag) in a sample of X-ray sources in Stripe 82 field and highlighted the power of this diagnostic to reveal obscured AGNs not detected through the classic W1--W2 colour criterion by \citet{assef2013}. In this work, we used the \citet{yan2013} criteria. To be consistent with previous studies, we converted the CFHTLS r band to r\textsubscript{SDSS} Vega system to calculate the r--W2 colour as defined by \citet{yan2013}. In Figure~\ref{yanhist}, we plot the normalized r--W2 colour distribution for the 160 high-confidence SED-selected AGNs and also the VIPERS sources classified as normal galaxies, for reference. We separate the AGN population based on whether or not broad lines are presented in their spectra. The distribution of SED-AGNs without broad lines occupies redder r--W2 colours compared to the VIPERS galaxies and broad line AGN.

To evaluate in a statistical manner if the two AGN populations come from the same distribution as the VIPERS galaxies and to assess any similarities between these samples, we performed the two-side Kolmogorov-Smirnov (KS) test. The KS test between the VIPERS galaxies and the SED-AGNs without broad lines revealed that there is a $<1.8\times10^{-05}$ chance (D\textsubscript{KS}=0.21) that they are drawn from the same parent population. In the case of the SED-AGNs with broad lines, we obtained a 68\% probability (D\textsubscript{KS}=0.14) for the r--W2 distributions to be representative of the same population. The SED-AGNs without broad lines have redder (r--W2=5.93) colour on average than VIPERS galaxies (r--W2=5.50) or the broad line AGNs (r--W2=5.51), while almost 40\% (55/133) of them have red colours ($\rm{r-W2}\geq6$). On the other hand, broad-line AGNs have blue colours, with the exception of only six sources being optically red. These red, type 1 AGNs have $\rm{E(B-V)}>0.05$, while most of them have $\rm{E(B-V)}=1$ that indicates a large amount of polar dust in these sources \citep{yang2020}.

\begin{figure}
    \includegraphics[width=0.48\textwidth]{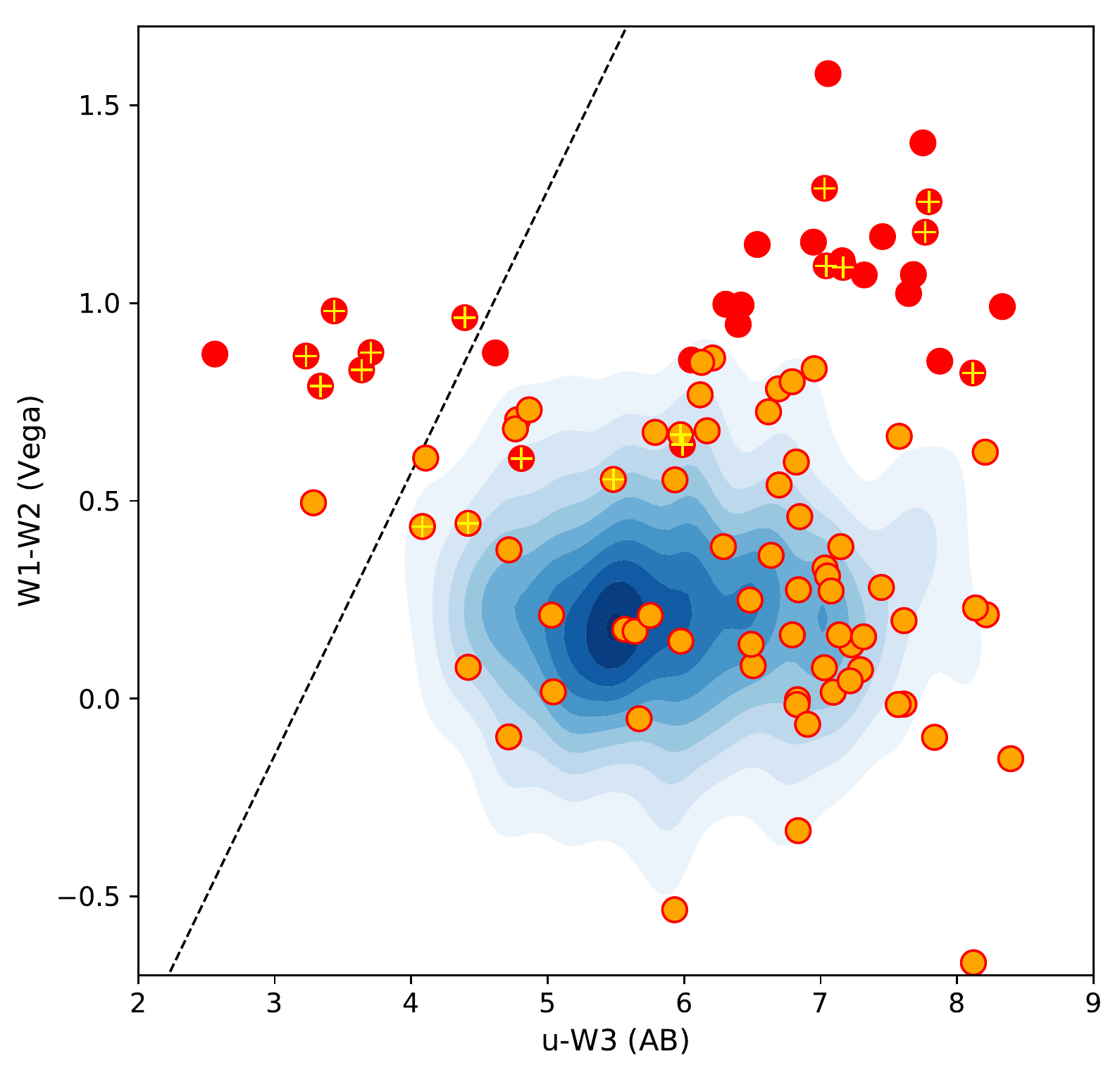}
\caption{Optical-IR colour diagram for the 98 SED selected AGNs (filled circles). \citet{hickox2017} selection relation is presented by the dashed line (see text for more details). Sources on the right side of this line are considered obscured. The density contours present the VIPERS sample, while the red filled circles and the crosses show the mid-IR and X-ray selected AGNs.}\label{hickox}
\end{figure}

\citet{hickox2017} explored the mid-IR colours and the SEDs of a large sample of type I and II quasars selected via SDSS spectroscopy. They showed that simple mid-IR colour cuts could identify the majority of luminous AGNs but may miss the most heavily obscured AGNs. On the other hand, r--W2 colour criteria might be biased against low redshift regimes (z<0.5). Thus, they defined a new optical-IR selection criterion that cleanly separates the unobscured and obscured AGNs: (u--W3[AB])>1.4$\times$(W1--W2[Vega])+3.2. This is more effective as it uses the maximum baseline between optical and IR wavelength range. However, it requires detections (or upper limits) in all four bands u, W1, W2 and W3. In Figure~\ref{hickox}, we plot the W1--W2 versus the u--W3 colours for all the sources in the VIPERS sample requiring detections in all four bands with a minimal signal-to-noise ratio greater than two as in \citet{hickox2017}, resulting in 98 SED selected AGNs. For reference, we over-plot the AGN samples selected through mid-IR (red circles) and X-ray (crosses) selection techniques. More than 90\% of the SED selected AGNs lie in the area characterized by obscuration. As expected all the sources near and on the left side of the line are unobscured broad-line AGNs.


\subsubsection{Highly obscured AGNs}
\begin{figure}
    \includegraphics[width=0.48\textwidth]{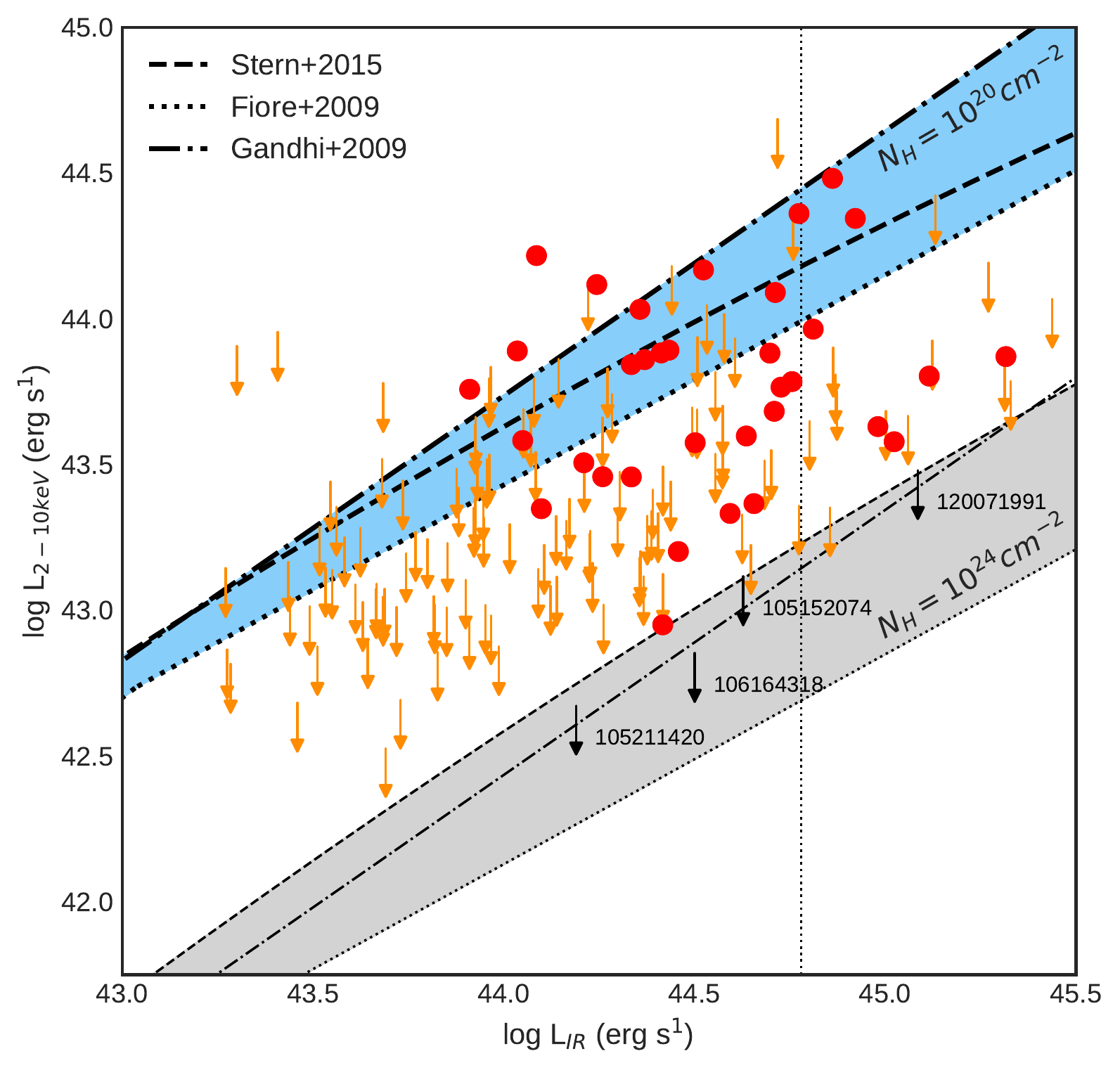}
\caption{The X-ray luminosity in the [2-10 kev] band as a function of the IR luminosity of the SED selected AGNs with (red circles) and without (orange arrows) X-ray detections. The thick lines represent the correlation derived by \citep{stern2015}, \citet{fiore2009} and \citet{gandhi2009} for unabsorbed AGNs. The lower thinner lines are for absorbed ($N_{H}=10^{24}$cm\textsuperscript{-2}) AGNs, assuming X-ray suppression with a factor of 20. The shaded areas correspond to the scatter of the relations derived in the aforementioned studies. The vertical line at L\textsubscript{IR}>6$\times$10\textsuperscript{44} erg s\textsuperscript{-1} corresponds to our threshold for luminous sources. The thick black arrows indicate the most heavily absorbed AGNs according to this plot.}\label{LxLir}
\end{figure}
To search for sources with extreme intrinsic absorption in our sample, we used the relation of their IR and X-ray luminosity. It is well known that the mid-IR luminosity is correlated with the unabsorbed X-ray emission in the AGNs for a wide range of luminosities \citep{lutz2004,gandhi2009,mateos2015,stern2015}. The mid-IR emission provides a good measurement of the AGN luminosity regardless absorption. On the other hand, the X-ray emission is expected to be suppressed at some level with increasing HI column densities \citep{alexander2005, alexander2008}. Thus, the X-ray to mid-IR luminosity ratio constitutes another measurement of absorption. In Figure~\ref{LxLir}, we plot the X-ray luminosity in the hard band as a function of the AGN mid-IR luminosity derived by \texttt{X-CIGALE}. To facilitate comparison with previous studies, we transformed the luminosities from [2-8 keV] band into the [2-10 keV] band using the \texttt{WebPIMMS}\footnote{\url{https://heasarc.gsfc.nasa.gov/docs/software/tools/pimms.html}}~\texttt{v4.8d} software, assuming a photon index of $\Gamma=1.7$ and the Galactic HI column density \rm{$N_{H}=10^{20}$cm\textsuperscript{-2}}. The dashed bold line in the plot represents the relation derived from \citet{stern2015} from an unabsorbed AGN sample distributed over several orders of magnitude. For reference, we plot the relations given by \citet{fiore2009} and \citet{gandhi2009}. The bottom lines indicate the corresponding relation for absorbed sources with \rm{$N_{H}=10^{24}$cm\textsuperscript{-2}}. For the latter, we assumed that the X-ray emission is suppressed by a factor of 20 \citep{lansbury2015}.

The majority of the sources lie below the \citet{stern2015} relation, suggesting that these are absorbed AGNs. Four sources (thick black arrows in Fig.~\ref{LxLir}) lie in the Compton-Thick (CT) regime, that suggests they may have column densities \rm{$N_{H} \geq 10^{24}$cm}\textsuperscript{-2}. Figure~\ref{CTseds} in the Appendix, presents the SEDs of these four sources. The SED fitting results show that in only one source (ID=106164318) the AGN emission is obscured in the optical wavelengths. \cite{Georgantopoulos2011}, argued that even though the majority of CT sources have low L\textsubscript{X}-L\textsubscript{IR} ratio (at least in the local Universe), this does not necessarily imply that all sources with low L\textsubscript{X}-L\textsubscript{IR} ratio are CT. Furthermore, they argued that at higher redshifts, this method alone is not complete or capable to reveal real CT AGNs.

\section{Discussion}\label{discussion}

\begin{figure}
\centering
\begin{tikzpicture}[thick,scale=1, every node/.style={transform shape}]
  \begin{scope}[opacity=0.7]
    \fill[blue!40!white]   ( 90:1.2) circle (2);
    \fill[green!40!white] (210:1.2) circle (1.6);
    \fill[red!40!white]  (330:1.2) circle (1.9);
 \end{scope}
 \def\firstcircle{(90:1.2) circle (1.95)}
 \draw[solid,white!30!blue] \firstcircle node[below] {};
  \def\firstcircle{(210:1.2) circle (1.55)}
 \draw[dotted,black!50!green] \firstcircle node[below] {};
  \def\firstcircle{(330:1.2) circle (1.85)}
 \draw[dashed,white!30!red] \firstcircle node[below] {};

  \node at ( 90:3.65)  [font=\Large][text=black!30!blue]   [align=center]{SED\\160};
  \node at (230:3.5)  [font=\Large][text=black!50!green]  [align=center] {mid-IR\\54};
  \node at (305:3.5)   [font=\Large] [text=black!30!red][align=center] {X-ray\\116};
 
  \node at (90:2) [font=\large] {95(74)};
  \node at (210:1.9) [font=\large] {6};
  \node at (330:2) [font=\large] {80};

  \node at (190:0.2) [font=\large] {15(14)};
  \node at (260:1.3) [font=\large] {2};
  \node at (30:1.3) [font=\large] {19(17)};
  \node at (160:1.2) [font=\large] {31(29)};

\end{tikzpicture}
\caption{Venn diagram of the AGN samples selected through SED decomposition (blue-solid), X-ray detection (red-dashed) and mid-IR criteria (green-dotted). These samples consists of sources that fall inside the VIPERS field, have spectroscopic redshifts from VIPERS in the range [0.5,1.2] and have optical, near-IR and mid-IR counterparts. The numbers inside parenthesis are the confirmed SED selected AGNs through optical spectroscopy.}\label{Venn}

\end{figure}
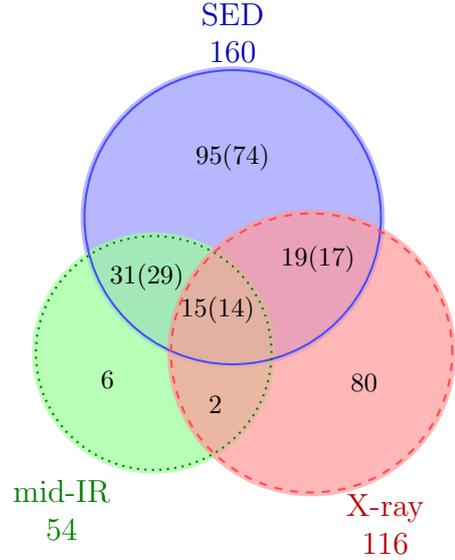
 \begin{figure*}
    \includegraphics[width=1\textwidth]{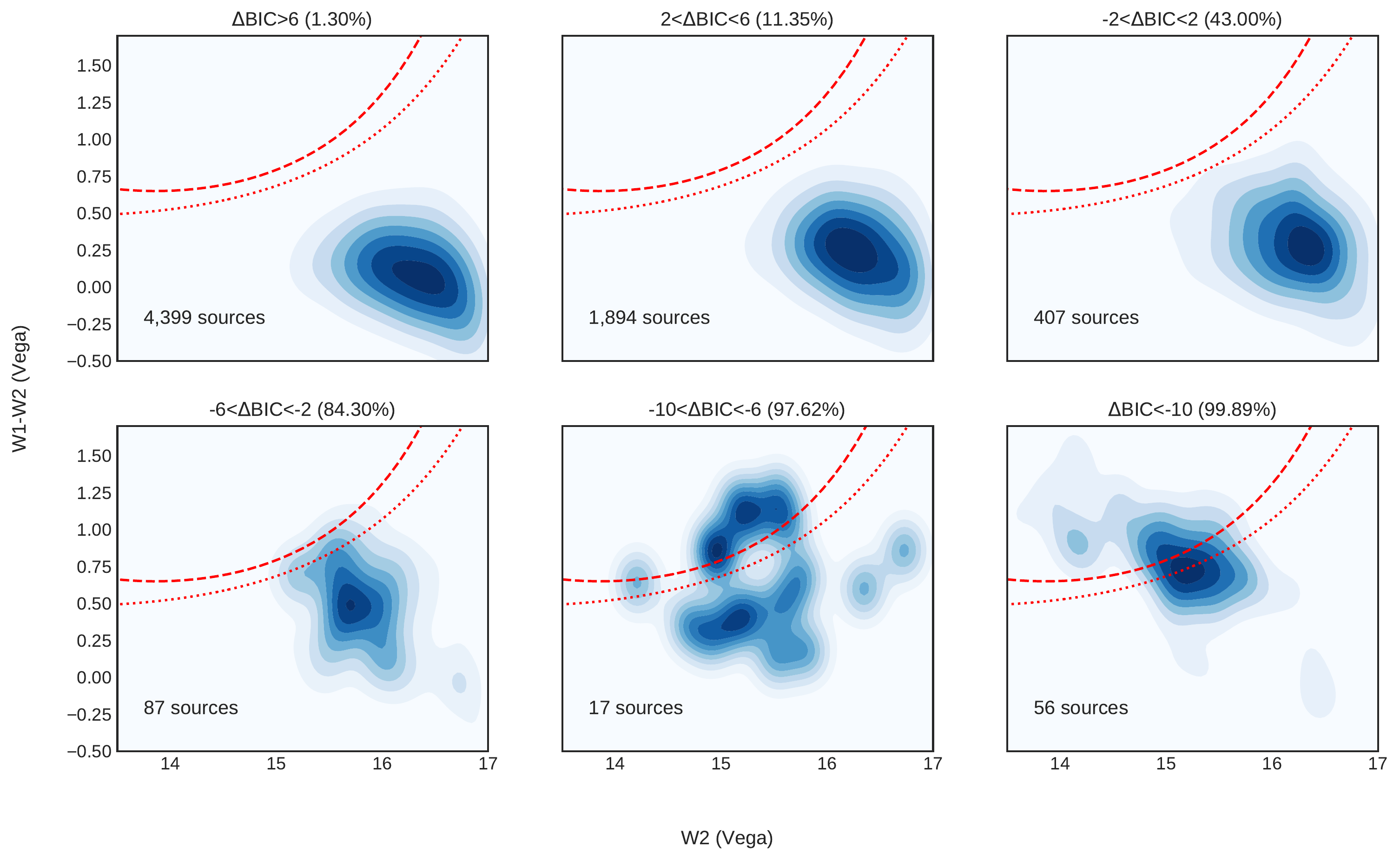}
    \caption{Density plots of the VIPERS sample in the W2, W1--W2 magnitude-colour diagram separated into bins with different $\Delta$BIC values (AGN probabilities). Above each panel, we label the range of $\Delta$BIC and in the parenthesis the average probability of the sources hosting an AGN. The dashed and dotted lines represent the AGN selection criteria defined by \citet{assef2013} with 90\% and 75\% reliability, respectively.}\label{assefplotBIC}
 \end{figure*}

\begin{figure}
    \includegraphics[width=0.50\textwidth]{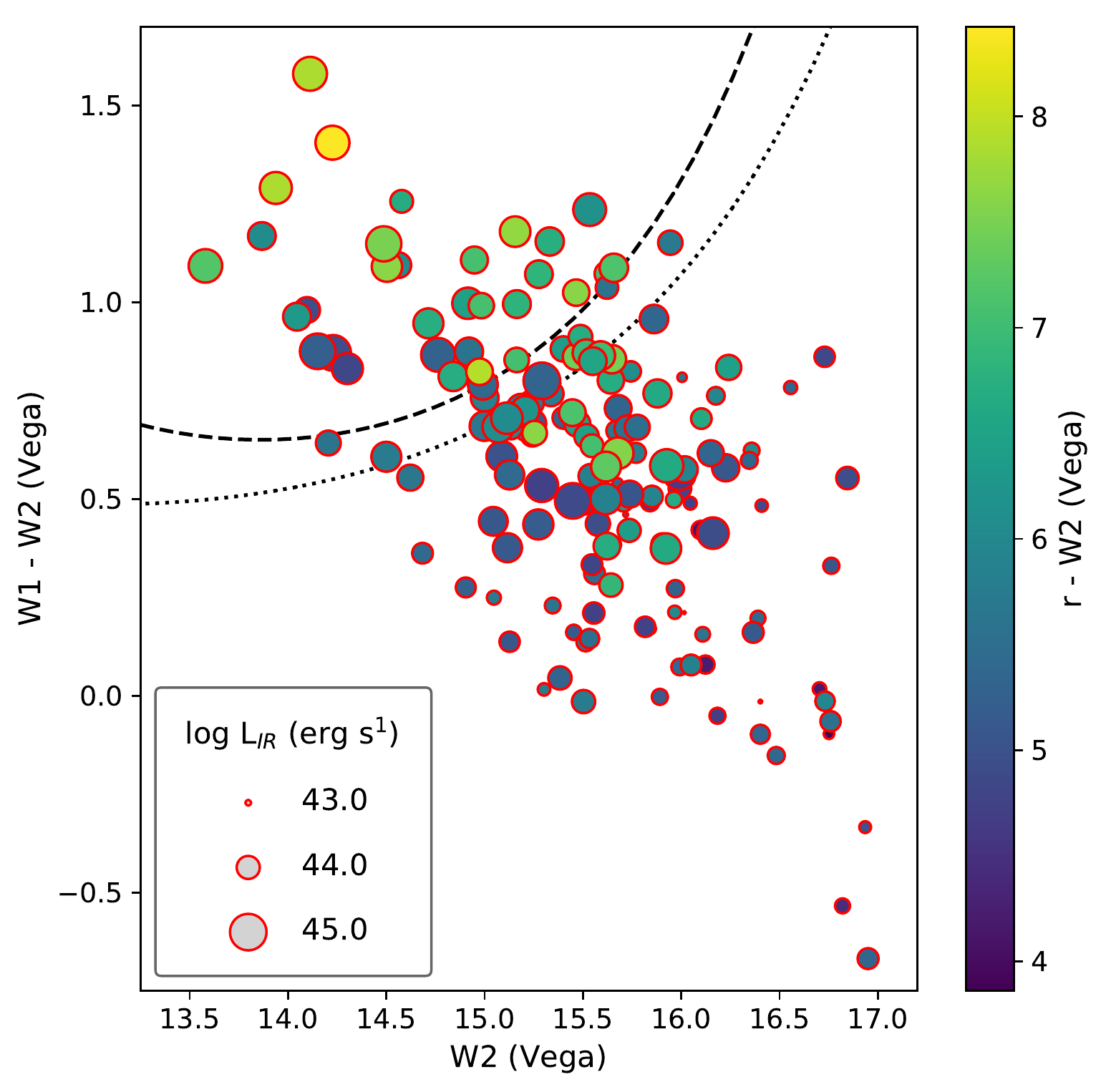}
    \caption{\textit{WISE} magnitude-colour (W2, W1--W2) diagram for the 160 SED selected AGNs colour-coded based on their optical colour (r--W2). The size of the circles corresponds to the IR luminosity as indicated in the legend. The dashed (dotted) line represents the \citet{assef2013} selection threshold with reliability of 90\% (75\%).}\label{assefcolour}
\end{figure}

\begin{figure*}
\begin{center}
   \begin{tabular}{c c}
    \includegraphics[width=0.475\textwidth]{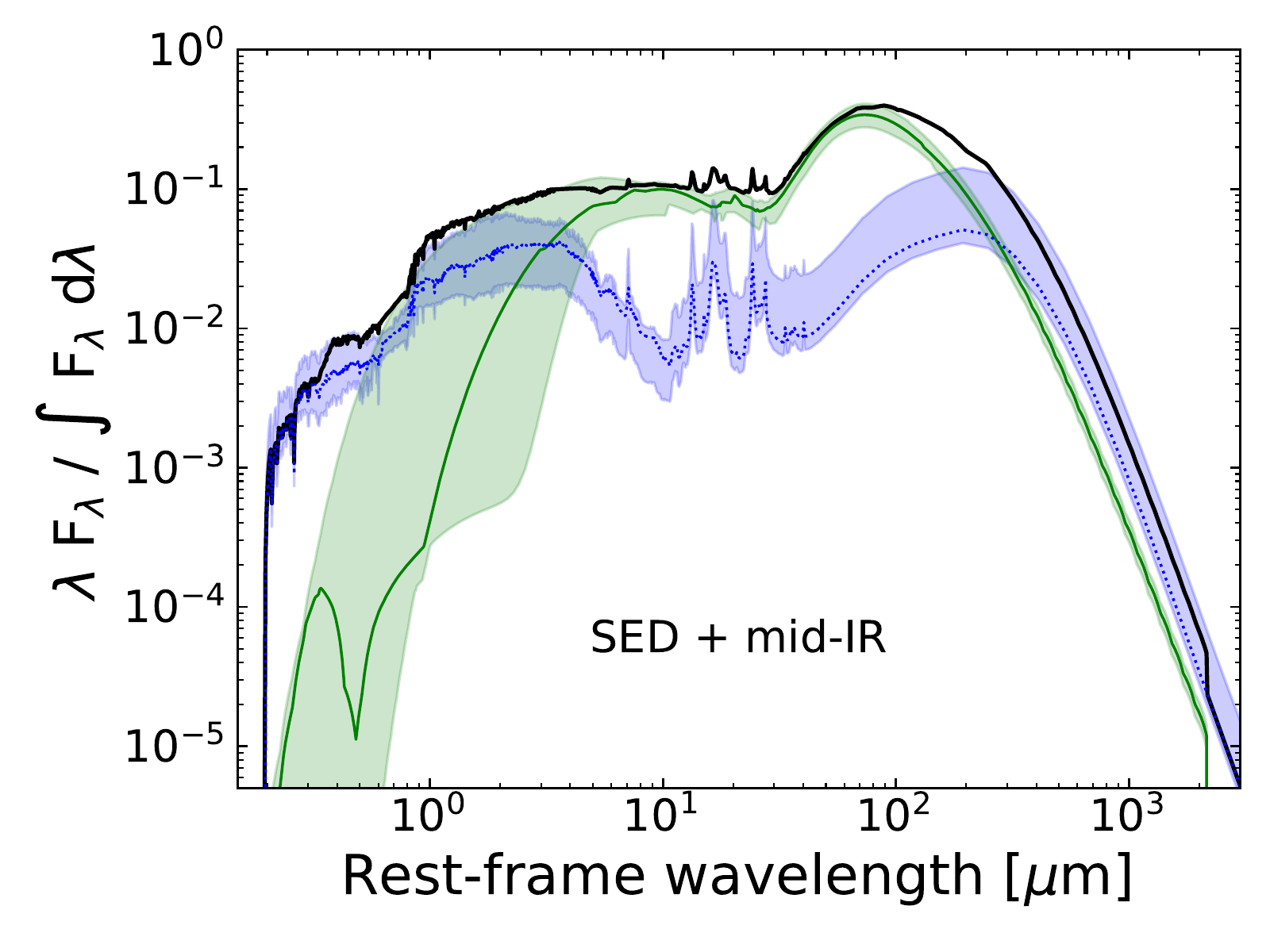} &
    \includegraphics[width=0.475\textwidth]{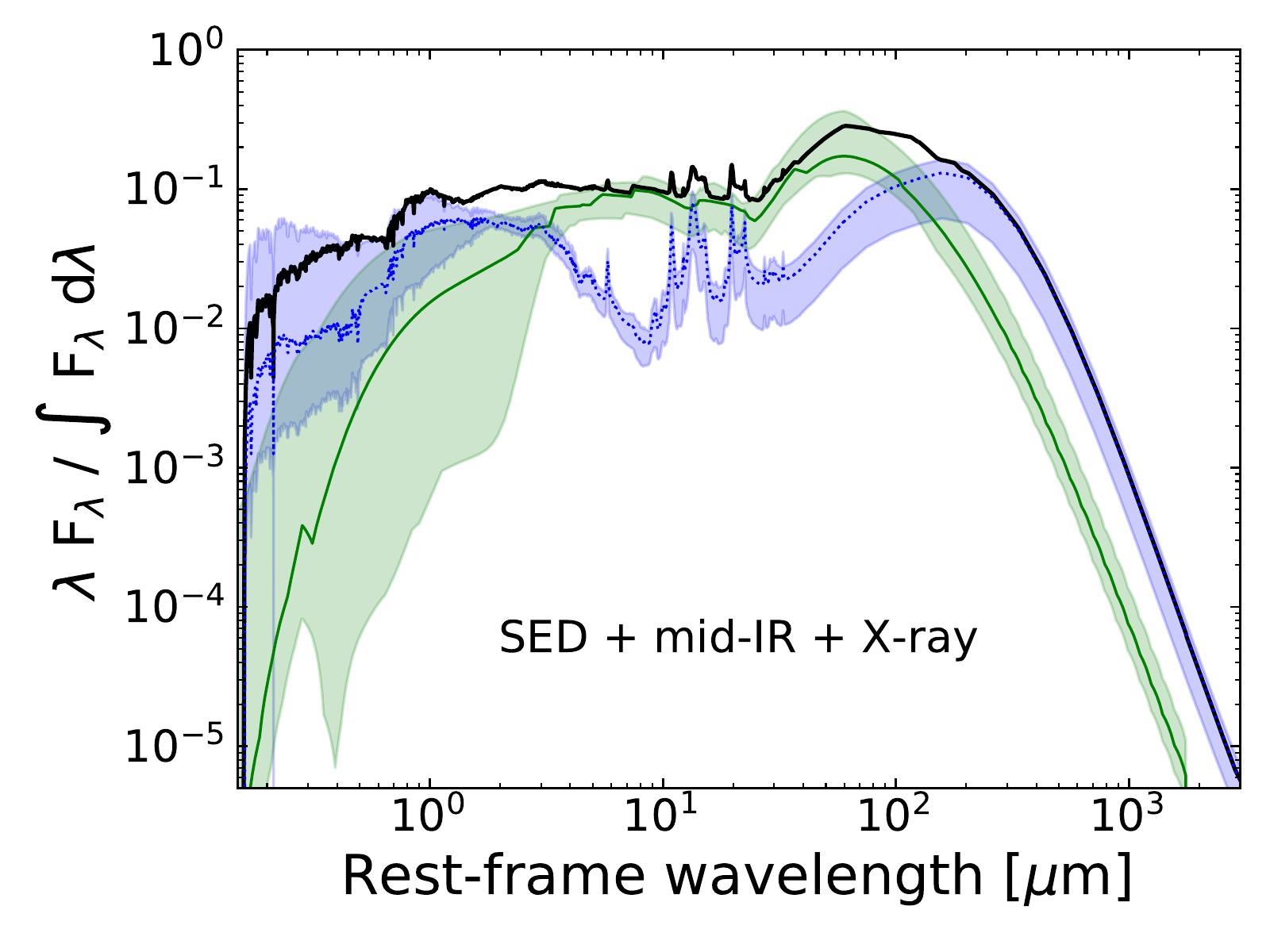} \\
    \includegraphics[width=0.475\textwidth]{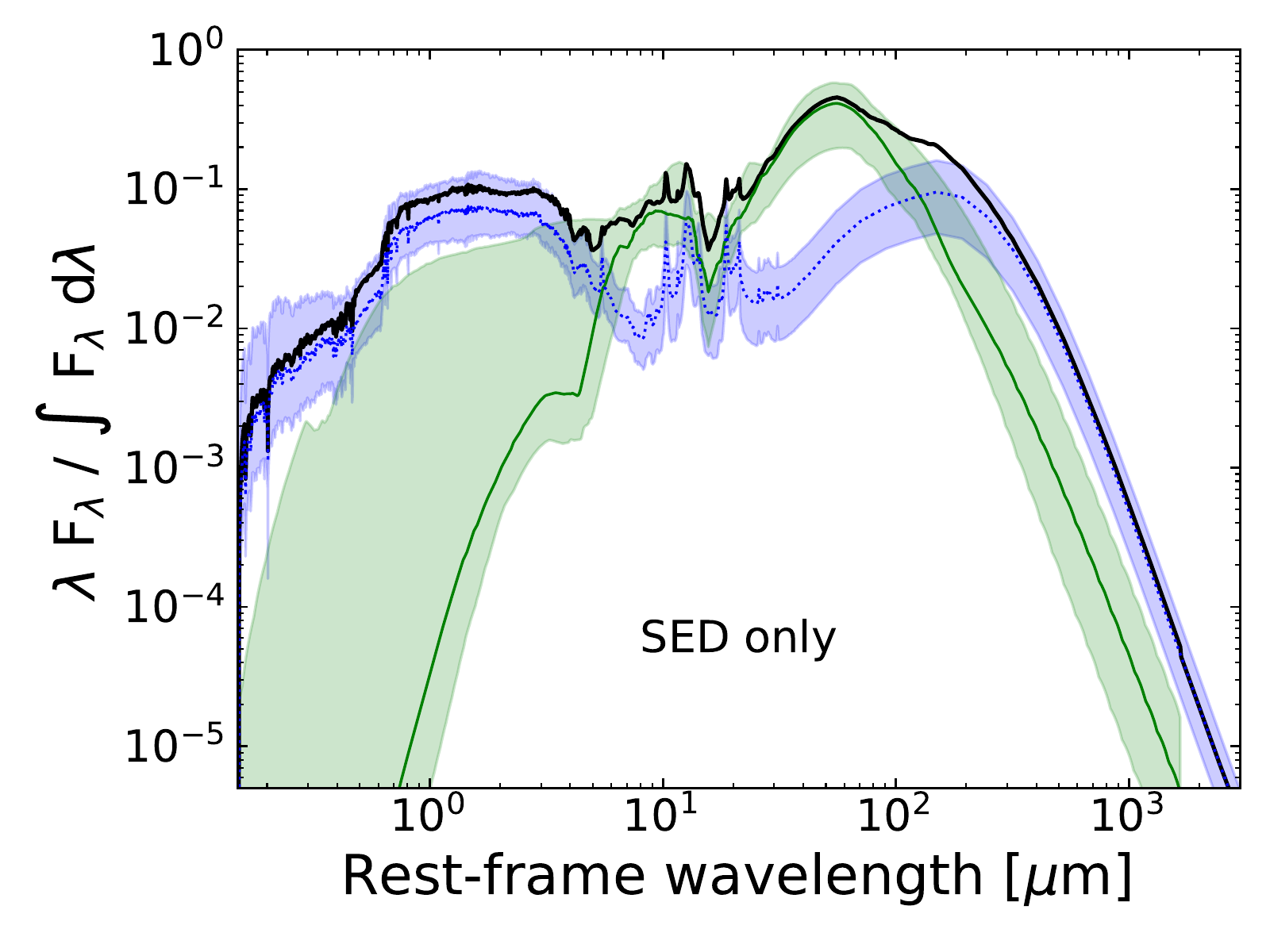}&
    \includegraphics[width=0.475\textwidth]{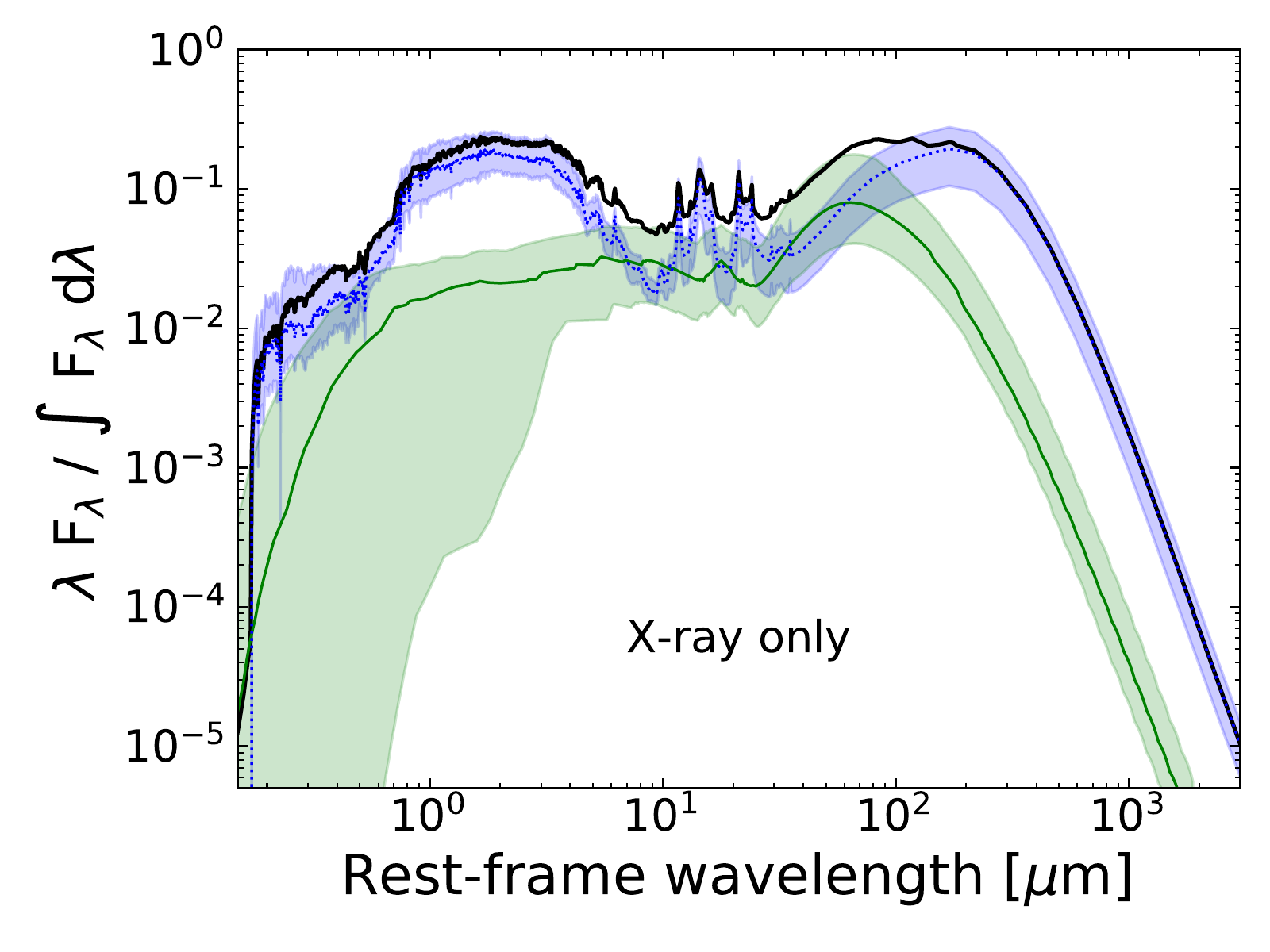}  
    \end{tabular}
    \end{center}
\caption{Stacked SEDs at rest-frame for AGN samples selected through different selection methods as indicated. The median SEDs of the host galaxy emission and the AGN components are plotted with blue dotted and green solid lines, respectively. The black thick line shows the median total flux, while the shaded areas correspond to 15\textsuperscript{th} up to 75\textsuperscript{th} percentiles at each wavelength.}\label{stacked}
\end{figure*}
The identification of AGNs in a specific wavelength regime depends strongly on the physical and observational properties of the sources, such as the luminosity, redshift, absorption and black hole mass. Thus, different selection techniques are more sensitive to different AGN populations and the overlapping among them is affected by the completeness and reliability along with the depth of each survey. To compare the various selection methods used in this work in a uniform manner, we selected the X-ray detected AGNs that have a counterpart in our initial VIPERS catalogue (with optical, near-IR and mid-IR counterparts). The Venn diagram, in Figure~\ref{Venn}, presents the overlap of the SED (160), mid-IR (54) and X-ray (116) selected AGN samples. The numbers in the parenthesis indicate the number of confirmed AGN through optical spectroscopy ([NeV] emission, MEx and TBT diagrams). In total, 139/160 ($\sim$87\%) of the SED selected AGNs revealed by our analysis, are also classified as AGN in at least one of the different AGN selection methods presented in the literature.


Specifically, all the mid-IR selected AGN \citep{mateos2012, stern2012} are classified as AGN via our SED fitting technique. Regarding AGNs selected by the \citet{assef2013} criteria (Section~\ref{mid-IR}), 32 (44) out of 35 (52) sources with a reliability of 90\% (75\%), were also characterised as AGN, by our SED fitting criteria. This corresponds to a percentage of 91.5\% (84.6\%). The mid-IR sources not selected in our analysis have values of $\Delta$BIC lower than the threshold adopted in this work.
To better understand the behaviour of $\Delta$BIC in this colour diagnostic and check if the latter sources are real AGNs, we plotted the populations defined in Table~\ref{tableKassRaft} in the \citet{assef2013} diagram (Fig.~\ref{assefplotBIC}). Starting from the upper left panel to the lower right, $\Delta$BIC is decreasing, while the probability of the sources hosting an AGN is increasing. The normal galaxy population with values $\Delta$BIC>2 (upper left and middle panels) is consistent with the mid-IR colours and all lie outside the wedges of \citet{assef2013}. In cases where it is uncertain which model describes better the observed SEDs (upper right) and the probability is equal for both galaxy and AGN templates, the majority of this population lies still below the lines. Increasing the probability of a source to host an AGN (the observed data fit better models with galaxy and AGN templates), the sources move towards the mid-IR AGN wedges defined by \citet{assef2013} (lower panels).

Despite the relaxed threshold adopted in this work, there is a large population that would not be selected through simple colour-colour cuts. This SED selected population is twice the number of the AGN selected through the \citet{assef2013} criteria. Figure~\ref{assefplotBIC} shows the robustness of the latter diagnostic criteria, but also the importance of selecting AGNs via SED decomposition techniques, as they are able to identify a much larger AGN population. SED fitting may better separate the mid-IR AGN emission from their host galaxy. In Figure~\ref{assefcolour}, we plot the colour-magnitude diagram for the 160 SED selected AGNs colour-coded based on the r--W2 colour. Larger symbols correspond to higher AGN IR luminosities, derived by the SED fitting. It is evident that while the method of \citet{assef2013} selects the brightest and optically reddest AGNs, SED decomposition allows us to identify less luminous AGNs with bluer colours.

Furthermore, among the \citet{assef2013} selected AGNs with reliability 75\% and 90\%, 17 ($\sim$33\%) and 14 ($\sim$40\%) sources have X-ray emission, respectively. \citet{mendez2013} showed that the percentage of mid-IR selected AGNs that have X-ray detections depends strongly on the depths of the surveys and ranges between 45\% to 90\%. When the depth of the IR surveys increases, this fraction decreases. On the other hand, increasing the X-ray depth increases the fraction of mid-IR AGNs that are X-ray detected. For example, \citet{pouliasis2019} found that $\sim$70\% of mid-IR AGNs have X-ray detections in the $\sim$7 Ms \textit{Chandra} Deep Field South (CDFS) image. Our results agree well with those of \citet{mendez2013} for shallow depth X-ray surveys (in this work the average is $\sim$20 ks) and low sensitivity limits in the IR bands. The majority of the X-ray sources lie outside the mid-IR wedges (Fig.~\ref{assef}). This could be due to the fact that these sources may be less luminous in mid-IR or that the relative emission from the host galaxy overpowers that from the AGN. Sources inside the \citet{assef2013} wedge with no X-ray counterparts are expected to be luminous absorbed AGNs. Out of these, 12/37 ($\sim$33\%) have [NeV] emission in their optical spectra while the vast majority have red colours.

Finally, $\sim$21\% of the SED selected AGNs have X-ray emission. On the other hand, we found that 30\% of the X-ray population is selected as AGN via SED decomposition with high significance. As shown in Table~\ref{tableKassRaft}, the majority of the X-ray sources have low AGN probabilities, based on our Bayesian analysis. One possible explanation is that these sources do not have a large enough amount of dust around their black hole to overpower the host galaxy luminosity. Indeed, the average IR AGN luminosity for X-ray sources that are not selected through SED decomposition is almost 0.5 dex lower compared to SED selected AGNs. The SED decomposition is biased towards those objects which are bright in the infrared compared to their hosts, and so this would naturally directly lead to this difference. In order to visualize these results, we plot in Figure~\ref{stacked} the stacked SEDs for AGN samples identified via different combinations of AGN selections methods. AGNs selected through the SED fitting technique, mid-IR colours and are detected in X-rays (upper right panel) are in general AGN-dominated systems. Furthermore, the median AGN emission extends to the optical wavelengths which indicates that these are less obscured sources. In the case of AGN that are SED and mid-IR selected (upper left panel) the AGN component is dominant but the AGNs are more obscured. Similarly, the SED-only selected sample (bottom left panel) consists of obscured AGNs, though with less dominant AGN component. In contrast, the X-ray only selected AGNs (lower right panel) comprise systems that the host galaxy component dominates the SED. The scatter in the AGN emission in those stacked SEDs, especially in the optical regime of the spectrum, is mostly due to the different types of AGN (obscured, unobscured, intermediate type).

Taking into account only sources with high mid-IR luminosity, 7 out of 21 sources ($\sim$34\%) have an X-ray counterpart. The luminosity cut at L\textsubscript{IR}>6$\times$10\textsuperscript{44} erg s\textsuperscript{-1} (vertical line in Fig.~\ref{LxLir}) was defined in \citet{delmoro2016} and corresponds to the X-ray quasar regime according to the L\textsubscript{X}-L\textsubscript{IR} relation. \citet{delmoro2016} found that 70\% of mid-IR luminous AGNs in the redshift range 1<z<3 is detected in the X-rays. This discrepancy comes partially from the different redshift regime and mostly from the depth of their X-ray observations (2 and 4 Ms) \textit{Chandra} X-ray observations compared to the \textit{XMM-Newton} observations used in this study with much lower exposure times ($\sim$20 ks). In \citet{mountrichas2017}, the obscured fraction among a sample of type I AGNs (average redshift equal to 2.3) with luminosities higher than L\textsubscript{IR}>1.6$\times$10\textsuperscript{46} erg s\textsuperscript{-1} was found $\sim$10\%. This indicates a higher number of absorbed sources at higher redshifts and type I AGNs. Finally, 10/21 of these sources are also [NeV] emitters, while the majority of them are classified as optically red (r--W2$\geq$6). Example SEDs of the luminous mid-IR AGNs with and without X-ray detections are shown in Figure~\ref{sedLUMINOUS} in the Appendix~\ref{luminous}.

\section{Summary and Conclusions}\label{summary}

AGN selection based on X-ray emission is by far the most reliable and effective tool to select a large number of AGNs. However, it is biased against the most absorbed ones. SED fitting techniques in the mid-IR regime are able to disentangle the IR emission coming from the torus, which is heated by the central engine from the host galaxy, and identify these populations that X-ray selection misses. In this work, we built and modelled the SEDs of 6,860 sources in the CFHTLS W1 field, using the \texttt{X-CIGALE} software. All sources in this sample have spectroscopic redshifts from the VIPERS survey and have been observed in optical (CFHTLS), near-IR (VHS) and mid-IR (\textit{WISE}) photometric bands. We fitted these SEDs with galaxy only templates and in a second run we combined both galaxy and AGN templates. Using a Bayesian approach, we compared these two fits and selected the objects for which the addition of an AGN component provides a higher statistical confidence level. 

We ended up with 160 sources with high AGN probability. Analysis of their optical spectra revealed 27 broad line AGNs, with 42/114 (37\%) inside the spectral coverage having [NeV] emission. Using the MEx and TBT diagrams, 90\% of our SED selected AGNs, with the corresponding emission lines detected, fall inside the AGN or composite areas in these diagrams. In total, 134/160 (84\%) of our SED selected sources are confirmed AGNs, by these robust diagnostic tools. Our main results can be summarized as follows:

\begin{itemize}

\item To compare our SED decomposition technique with mid-IR selection methods that identify AGN, we applied the selection criteria of \citet{assef2013}, \citet{stern2012} and \citet{mateos2012}. Our analysis revealed that the SED method recovers the mid-IR AGN population with high completeness (92\%). However, we found in addition a significant number of AGNs, twice as high as with mid-IR colour techniques. This population that simple mid-IR colours fail to uncover consists of lower luminosity AGNs with systems that are dominated by the host galaxy emission.

\item Among the X-ray selected AGNs, 34/116 (30\%) sources are selected through our SED analysis method. The remaining X-ray population not selected through SED fitting has host-galaxy dominated systems. On the other hand, 34/160 (21\%) of the SED selected AGNs have X-ray emission. In addition to this, the SED fitting results suggest that the vast majority ($\sim$70\%) of the AGNs have inclination angles of view $\psi \leq 50$ indicating some level of obscuration (type 1.5 and 2). We verify these results using the optical and mid-IR colour selection criteria defined by \citet{yan2013} and \citet{hickox2017} and utilizing the L\textsubscript{X}-L\textsubscript{IR} diagram. Interestingly, only 35\% of the most luminous mid-IR selected AGNs have X-ray counterparts, suggesting strong absorption].

\end{itemize}

We conclude that the different methods (SED decomposition, mid-IR colours, X-ray emission) used to identify AGNs are complementary to each other and equally important to constrain the full picture of the AGN demographics. In particular, SED fitting is able to identify a large number of obscured AGNs that the X-ray surveys miss and the simple mid-IR colour cuts do not select. These are critical for studying the AGN population with high obscuration in host-galaxy dominated systems and it might be the key between the connection of normal galaxies and AGNs.

\section*{Acknowledgments}\label{ackn}

The authors are grateful to the anonymous referee for valuable suggestions that significantly improved the manuscript. E.P. is thankful to Dr. A. Georgakakis for useful discussions and suggestions about this work. E.P. acknowledges financial support by ESA under the HCV programme, contract no. 4000112940. GM acknowledges support by the Agencia Estatal de Investigación, Unidad de Excelencia María de Maeztu, ref. MDM-2017-0765 and by the PROTEAS II project (MIS 5002515), which is implemented under the "Reinforcement of the Research and Innovation Infrastructure" action, funded by the "Competitiveness, Entrepreneurship and Innovation" operational programme (NSRF 2014-2020) and co-financed by Greece and the European Union (European Regional Development Fund). This research has made use of the VizieR catalogue access tool, CDS, Strasbourg, France. The original description of the VizieR service is presented by \citet{ochsenbein2000}. This research has made use of the SIMBAD database \citep{wenger2000}, operated at CDS, Strasbourg, France and, also, of NASA's Astrophysics Data System. This research made use of Astropy, a community-developed core Python package for Astronomy (Astropy Collaboration et al. 2013 \url{http://www.astropy.org}). This publication made use of TOPCAT \citep{taylor2005} for all table manipulations. The plots in this publication were produced using Matplotlib, a Python library for publication quality graphics \citep{hunter2007} and R\footnote{R Core Team (2016). R: A language and environment for statistical computing. R Foundation for Statistical Computing, Vienna, Austria. URL \url{https://www.R-project.org/}.}. Based on observations obtained with MegaPrime/MegaCam, a joint project of CFHT and CEA/DAPNIA, at the Canada-France-Hawaii Telescope (CFHT) which is operated by the National Research Council (NRC) of Canada, the Institut National des Sciences de l'Univers of the Centre National de la Recherche Scientifique (CNRS) of France, and the University of Hawaii. This work is based in part on data products produced at Terapix and the Canadian Astronomy Data Centre as part of the Canada-France-Hawaii Telescope Legacy Survey, a collaborative project of NRC and CNRS.




\bibliographystyle{mnras}
\bibliography{main}


\appendix

\section{Bayesian factor}
To compare models in Bayesian statistics, one needs to calculate the Bayes factor, \textbf{BF}, that is the ratio of the posteriori to a priori complementary probabilities (or the so-called evidence ratio). Let's assume that two models (\textbf{m\textsubscript{1}, m\textsubscript{2}}) have to be compared to a specific data set \textbf{y} and \textbf{p\textsubscript{1}}, \textbf{p\textsubscript{2}} their parameters, respectively. Applying the Bayes theorem, the posteriori probabilities for the two models are:

\begin{equation}
\centering
f(\textbf{m\textsubscript{1}}|\textbf{y})=\frac{f(\textbf{m\textsubscript{1}})f(\textbf{y}|\textbf{m\textsubscript{1}})}{\sum\limits_{i=1}^{2}f(\textbf{m\textsubscript{i}})f(\textbf{y}|\textbf{m\textsubscript{i}})}\quad\text{and}\quad f(\textbf{m\textsubscript{2}}|\textbf{y})=1-f(\textbf{m\textsubscript{1}}|\textbf{y}).
\end{equation}

\noindent Then, we calculate the ratio of the posteriori complementary probabilities  of the two models (posterior odds, $\textbf{PO}=\frac{f(\textbf{m\textsubscript{1}}|\textbf{y})}{f(\textbf{m\textsubscript{2}}|\textbf{y})}$) and also the Bayes factor of model \textbf{m\textsubscript{1}} against model \textbf{m\textsubscript{2}}:

\begin{equation}
\centering
\begin{multlined}
    \textbf{BF}= \frac{f(\textbf{m\textsubscript{1}}|\textbf{y})/f(\textbf{m\textsubscript{2}}|\textbf{y})}{f(\textbf{m\textsubscript{1}})/f(\textbf{m\textsubscript{2}})} \\
     = \frac{[\frac{f(\textbf{m\textsubscript{1}})f(\textbf{y}|\textbf{m\textsubscript{1}})}{\sum\limits_{i=1}^{2}f(\textbf{m\textsubscript{i}})f(\textbf{y}|\textbf{m\textsubscript{i}})}]/[\frac{f(\textbf{m\textsubscript{2}})f(\textbf{y}|\textbf{m\textsubscript{2}})}{\sum\limits_{i=1}^{2}f(\textbf{m\textsubscript{i}})f(\textbf{y}|\textbf{m\textsubscript{i}})}]}{f(\textbf{m\textsubscript{1}})/f(\textbf{m\textsubscript{2}})}=\frac{f(\textbf{y}|\textbf{m\textsubscript{1}})}{f(\textbf{y}|\textbf{m\textsubscript{2}})},
\end{multlined}
\end{equation}

\noindent where $f(\textbf{y}|\textbf{m\textsubscript{i}})$ is the marginal likelihood (or the so-called evidence) of the i model and is calculated by integrating with respect to parameters of each model:
\begin{equation}
\centering
f(\textbf{y}|\textbf{m\textsubscript{i}})=\int  f(\textbf{y}|\textbf{k\textsubscript{i}},\textbf{m\textsubscript{i}})f\textsubscript{i}(\textbf{p\textsubscript{i}})d\textbf{p\textsubscript{i}}, \quad i=1,2.
\end{equation}

The value of the Bayes factor determines the rejection or not of the initial assumption with high values (>1) indicating evidence in favor of that model. In practice, the Bayes factor is a measure of the weight of the information that is included in the data in a favor of one model against an other. It can be used as a relative measurement for the comparison of the two models. \citet{kass1995} gave the possible explanation of the Bayes factor for the comparison between two models that are shown in Table~\ref{tableKassRaft}.

\section{SEDs and properties of the SED selected AGNs}\label{luminous}
\begin{figure*}
   \begin{tabular}{c c}
    \includegraphics[width=0.425\textwidth]{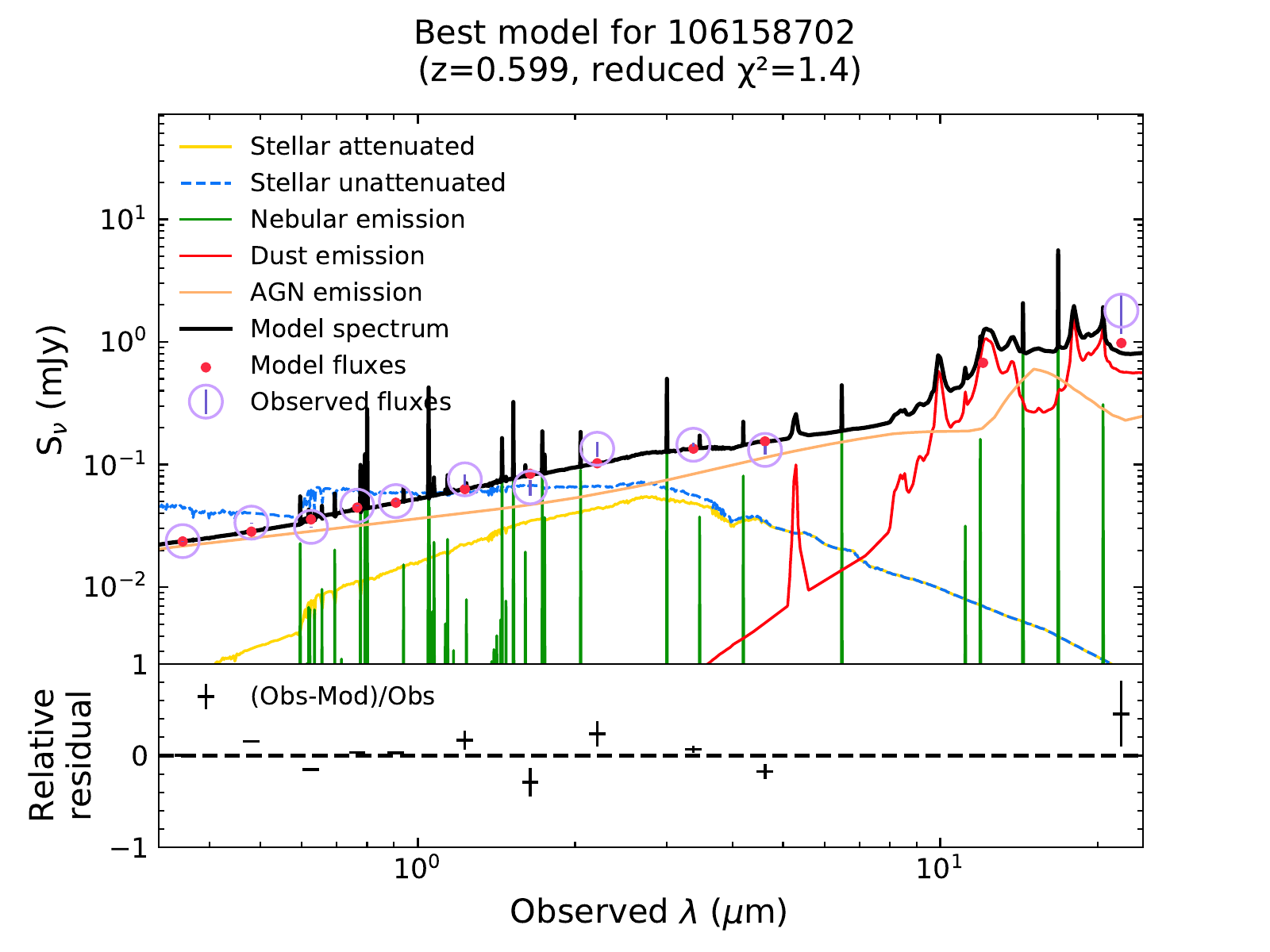} &
    \includegraphics[width=0.425\textwidth]{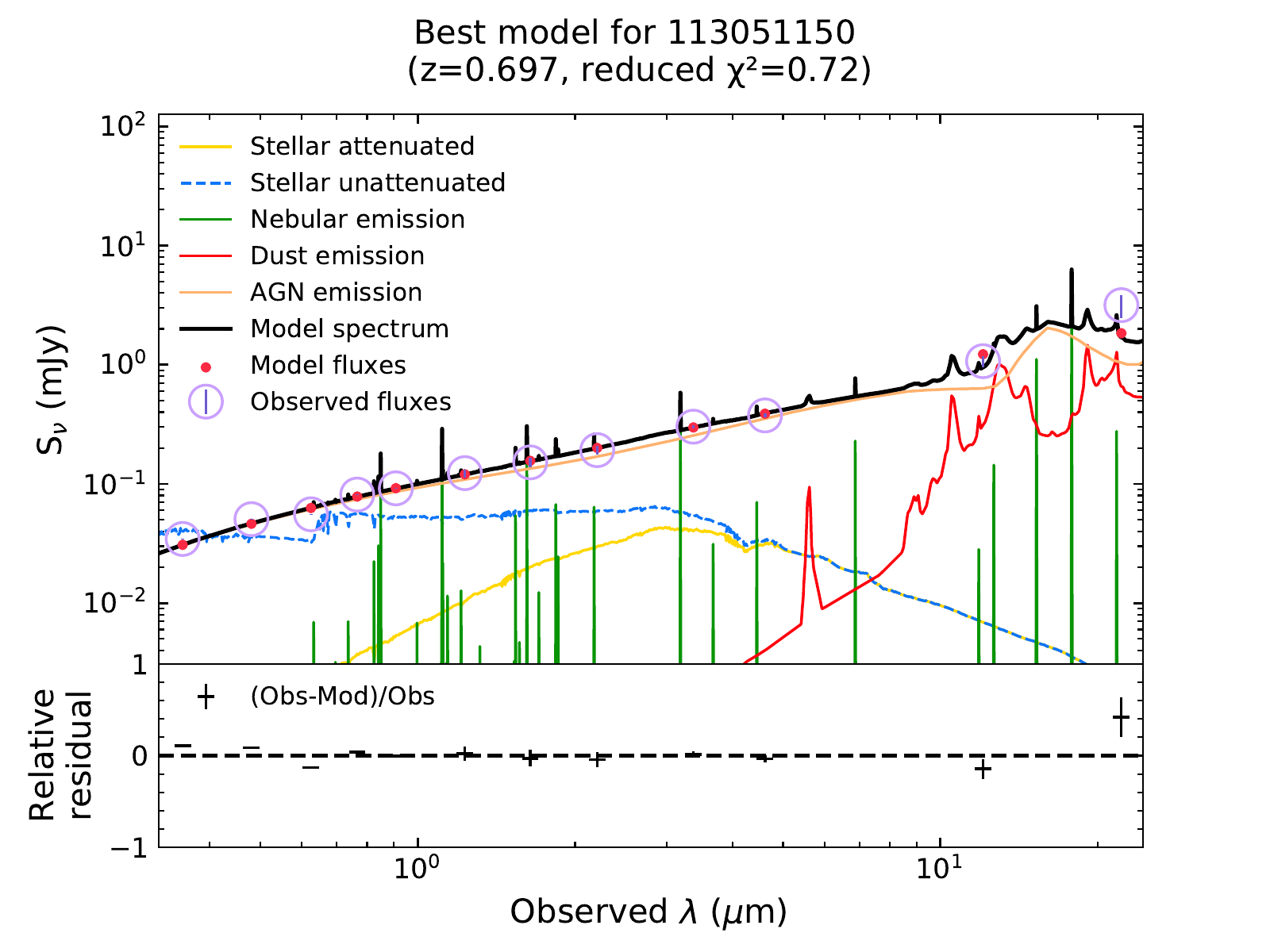} \\
    \includegraphics[width=0.425\textwidth]{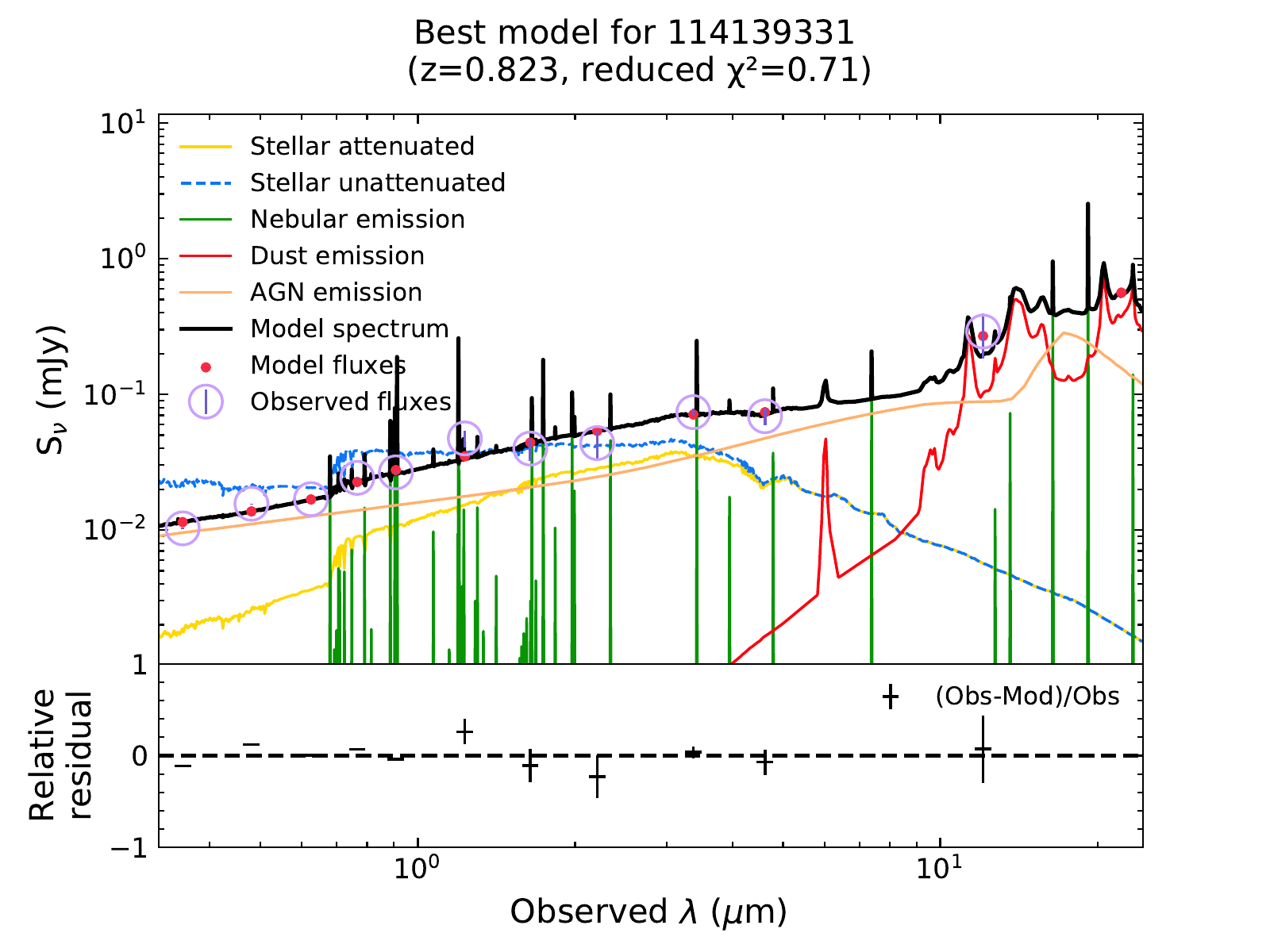}&
    \includegraphics[width=0.425\textwidth]{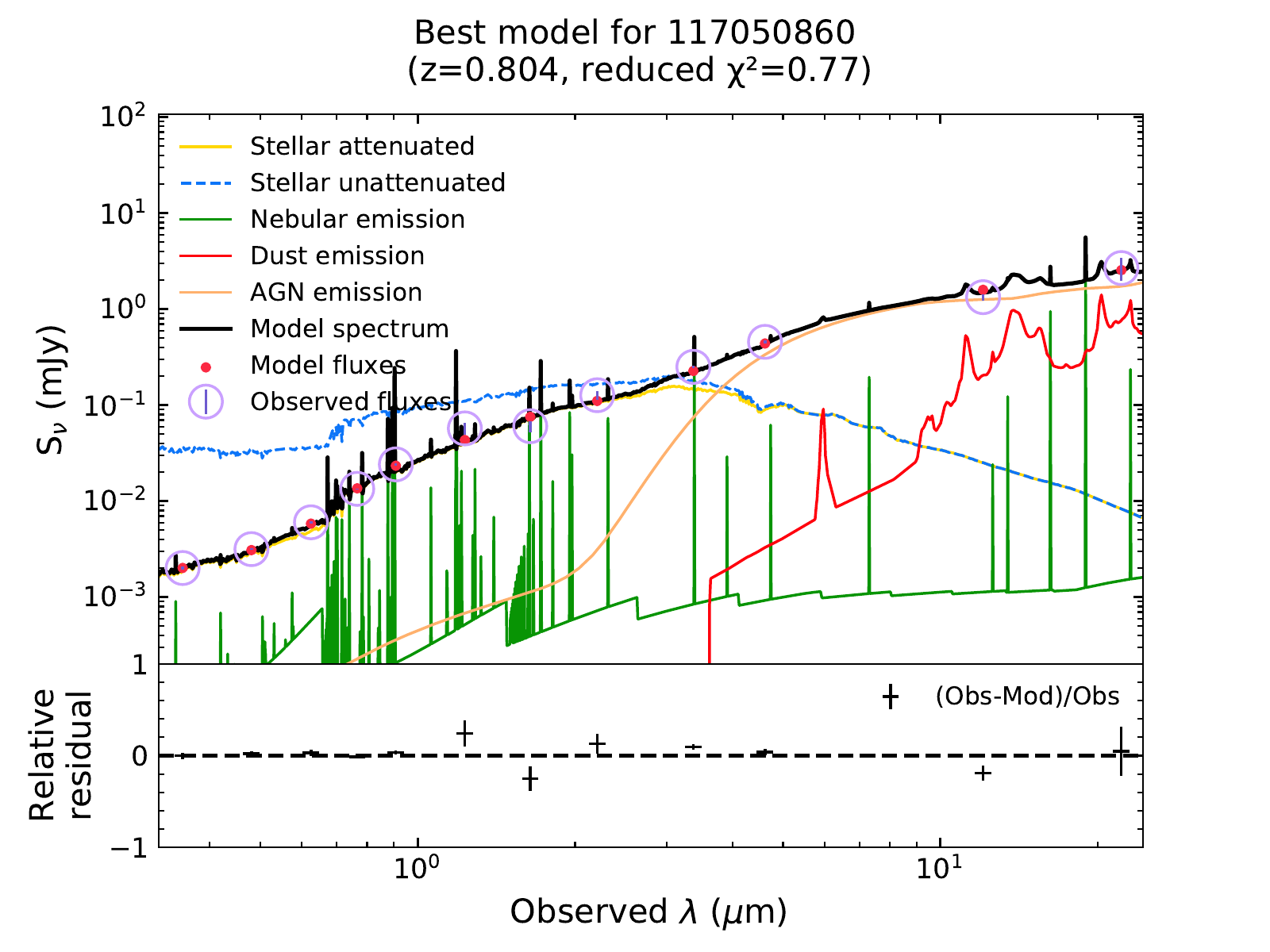} \\
    \includegraphics[width=0.425\textwidth]{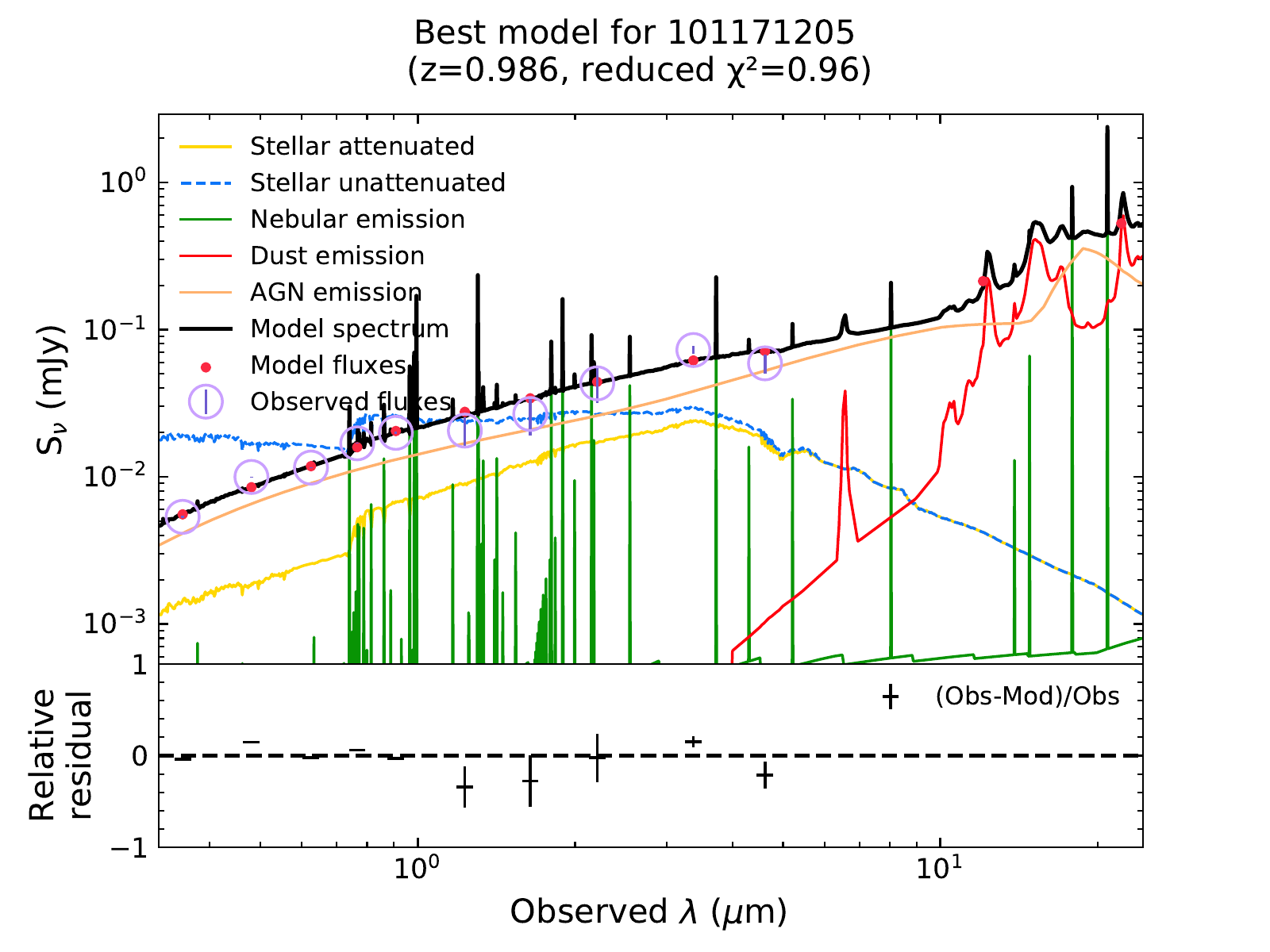} &
    \includegraphics[width=0.425\textwidth]{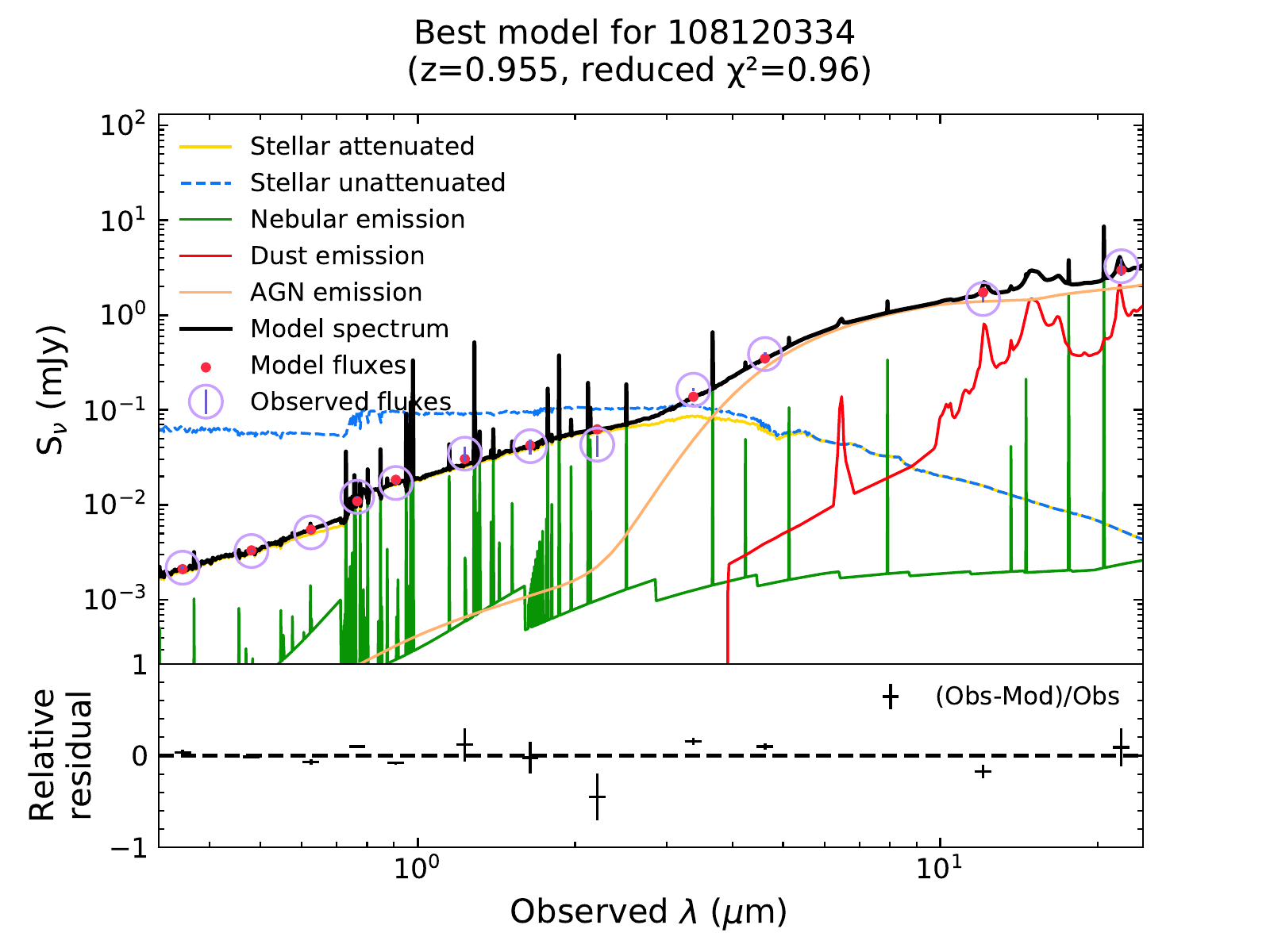}\\
    \includegraphics[width=0.425\textwidth]{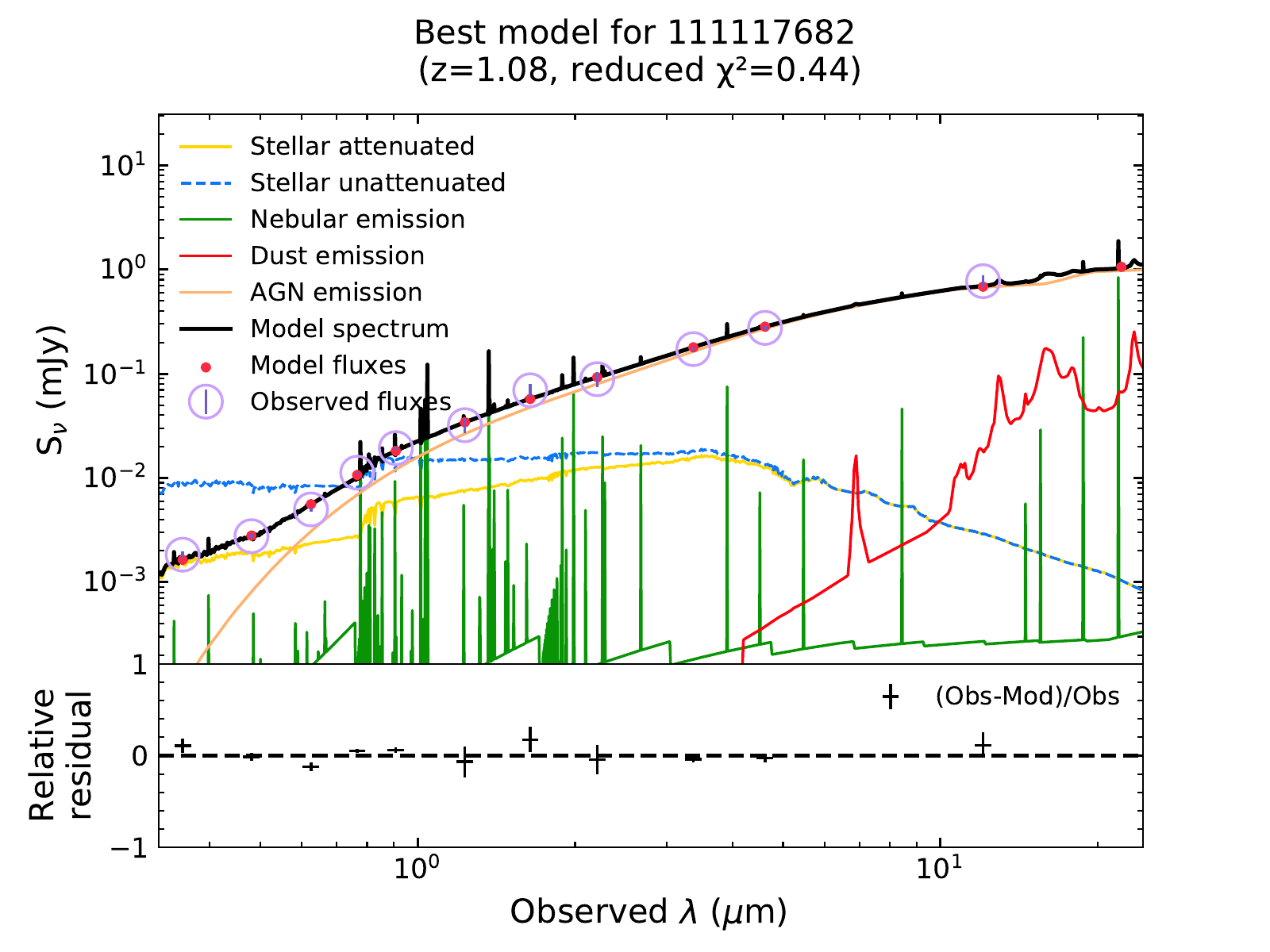}&
    \includegraphics[width=0.425\textwidth]{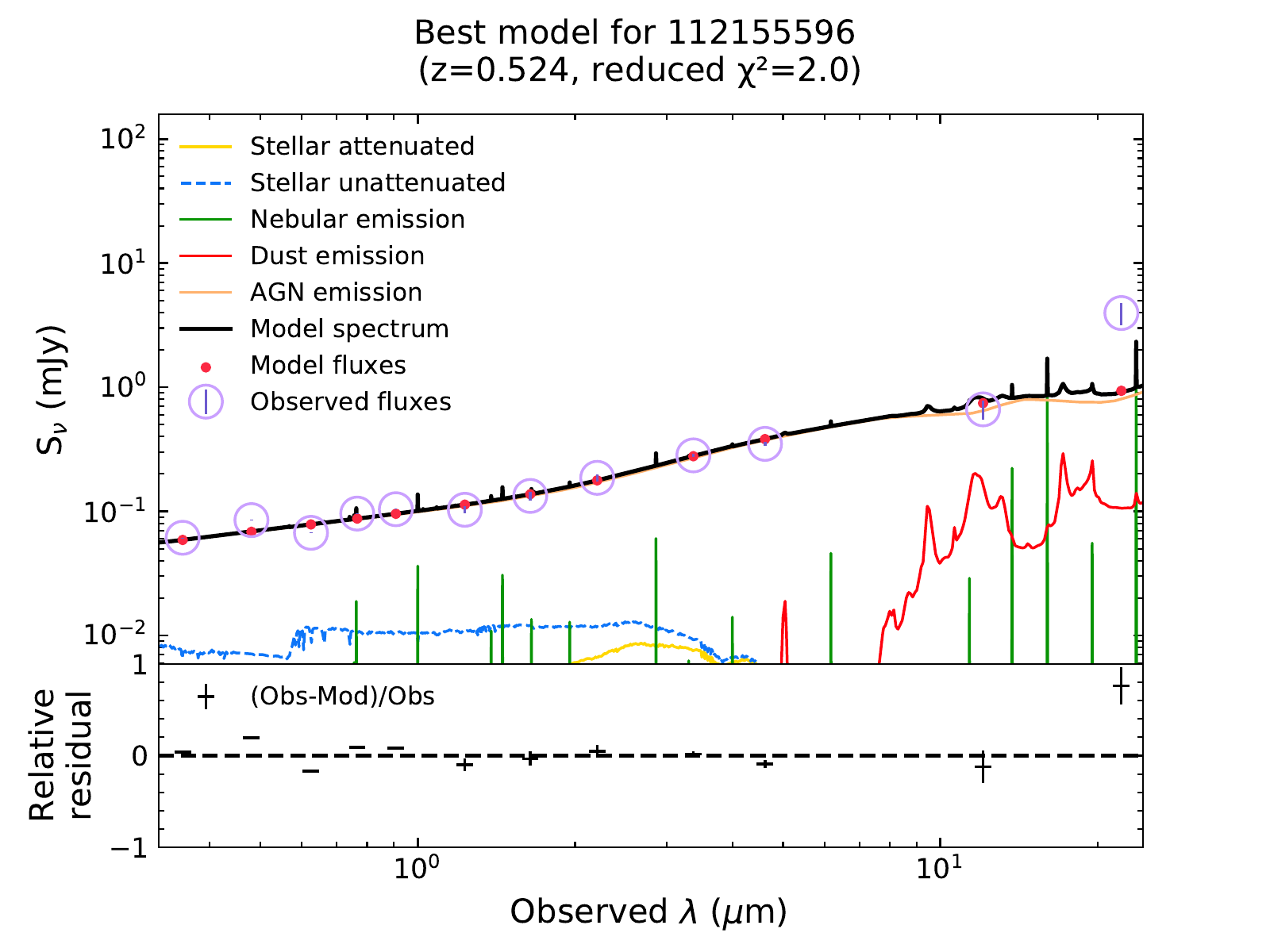}
    \end{tabular}
\caption{Example SEDs of eight luminous AGNs (L\textsubscript{IR}>6$\times$10\textsuperscript{44} erg s\textsuperscript{-1}) with (upper four) and without (lower four) X-ray detections. The dust emission is plotted in red, the AGN component in green, and the attenuated (unattenuated) stellar component is shown with the yellow (blue) solid (dashed) line, while the orange lines shows the nebular emission. The total flux is represented with black colour. Below each SED, we plot the relative residual fluxes versus wavelength.}\label{sedLUMINOUS}
\end{figure*}

\begin{figure*}
   \begin{tabular}{c c}
    \includegraphics[width=0.5\textwidth]{120071991_best_model.pdf} &
    \includegraphics[width=0.5\textwidth]{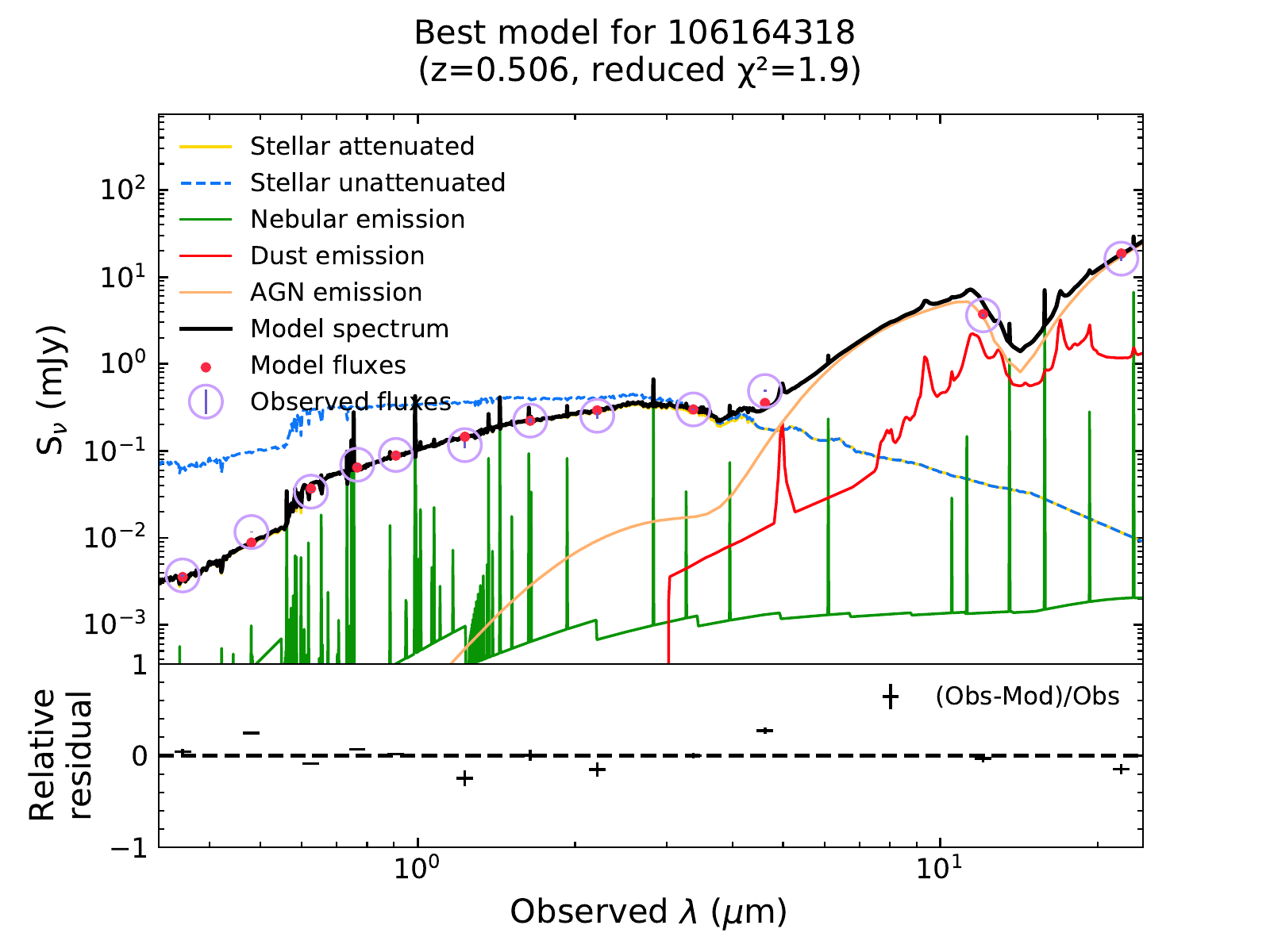} \\
    \includegraphics[width=0.5\textwidth]{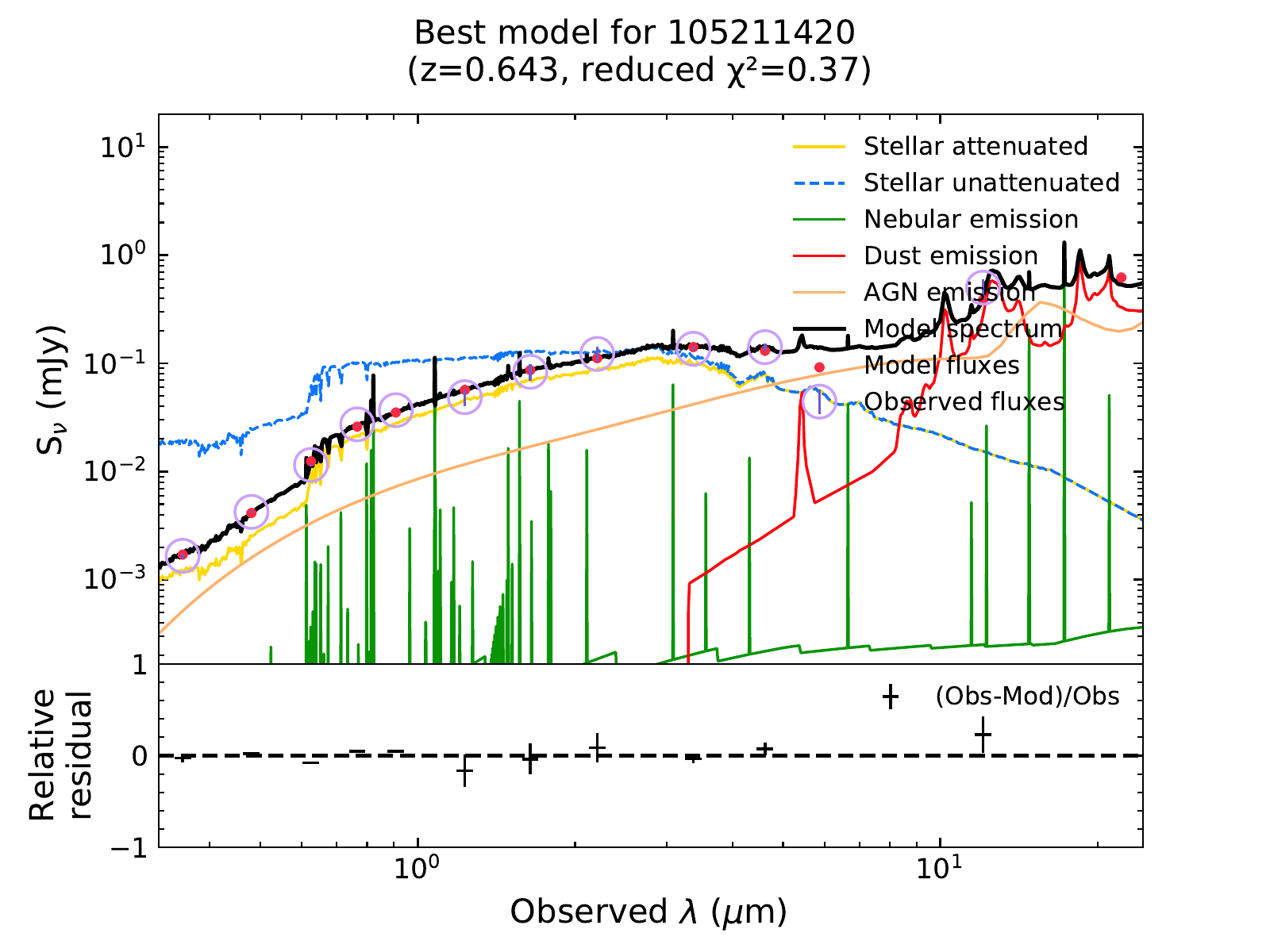}&
    \includegraphics[width=0.5\textwidth]{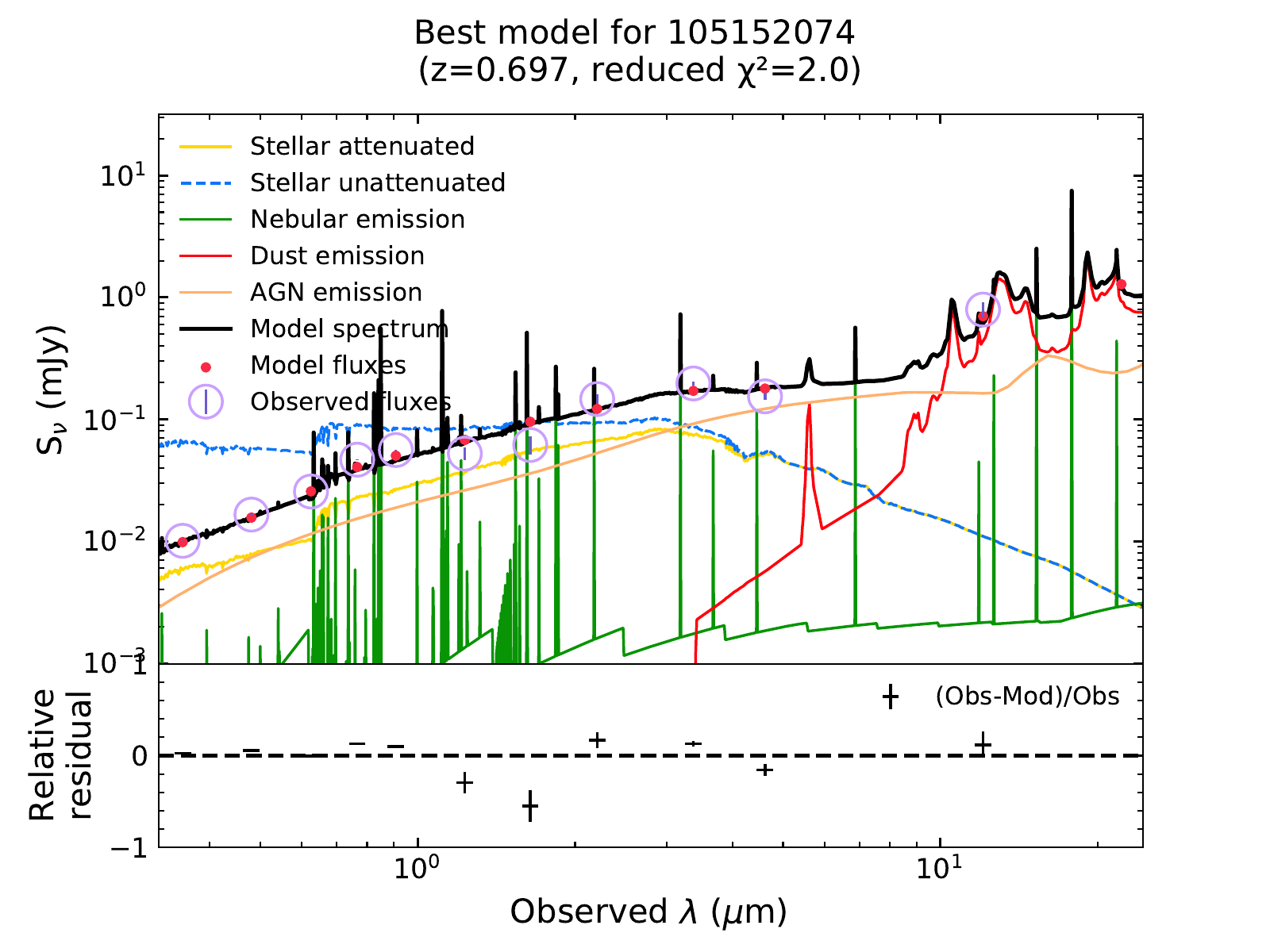} 
    \end{tabular}
\caption{SEDs of the four AGNs with extreme obscuration according to the L\textsubscript{X}--L\textsubscript{IR} relation. The dust emission is plotted in red, the AGN component in green, and the attenuated (unattenuated) stellar component is shown with the yellow (blue) solid (dashed) line, while the orange lines shows the nebular emission. The total flux is represented with black colour. Below each SED, we plot the relative residual fluxes versus wavelength.}\label{CTseds}
\end{figure*}

Figure~\ref{sedLUMINOUS} presents some example SEDs of the 21 luminous AGNs with L\textsubscript{IR}>6$\times$10\textsuperscript{44} erg s\textsuperscript{-1} with and without X-ray detections, while Figure~\ref{CTseds} shows the SEDs for four Compton Thick candidates according to the L\textsubscript{X}-L\textsubscript{IR} diagram. In Table~\ref{tableALL}, we list the observational and physical properties of all the 160 SED selected AGNs.

\begin{table*}
\centering
\caption{Catalogue of the 160 SED selected AGNs.}
\scalebox{0.75}{
\begin{threeparttable}
\begin{tabular}{c c c c c c c c c c c c c} 
\hline
VIPERS   &   RA   &   DEC   &   z   &   L\textsubscript{IR}   &   L\textsubscript{X}   &   mid-IR  &   Broad   &   [NeV]   &   MEx  & TBT   &   SED   &   r--W2\\
 ID & (J2000) & (J2000) &  & (erg s\textsuperscript{-1}) & (erg s\textsuperscript{-1}) & AGN & lines & emission & AGN  & AGN & AGN type & (Vega)  \\ 
(1) & (2) & (3) & (4) & (5) & (6) & (7) & (8) & (9) & (10) & (11) & (12)  & (13)\\ \hline\hline
101128062   &   30.44515   &   -5.94690   &   0.652   &   43.93   &   <43.68   &   0   &   0   &   0   &   0  &  1   &   2.0   &   4.76\\
101143290   &   30.65517   &   -5.87594   &   0.736   &   44.07   &   <43.68   &   0   &   0   &   0   &   0  &  1   &   2.0   &   4.94\\
101145879   &   31.06676   &   -5.85939   &   0.529   &   43.51   &   <42.89   &   0   &   0   &   0   &   0  &  1   &   2.0   &   4.93\\
101166083   &   30.97806   &   -5.76278   &   0.516   &   43.46   &   <42.70   &   0   &   0   &   0   &   0  &  0   &   2.0   &   4.63\\
101167170   &   30.61551   &   -5.75817   &   1.029   &   45.06   &   <43.68   &   0   &   0   &   0   &   0  &  0   &   2.0   &   6.64\\
101171205   &   30.42398   &   -5.74079   &   0.986   &   44.86   &   <43.92   &   0   &   1   &   1   &   0  &  1   &   1.0   &   4.92\\
101175047   &   30.91564   &   -5.72366   &   0.549   &   43.77   &   <43.28   &   0   &   0   &   0   &   0  &  1   &   1.5   &   5.99\\
101183266   &   31.14391   &   -5.68359   &   0.690   &   44.26   &   <43.68   &   0   &   0   &   0   &   0  &  1   &   1.5   &   5.75\\
101184375   &   30.53652   &   -5.68232   &   0.655   &   43.93   &   <43.65   &   0   &   0   &   0   &   0  &  0   &   2.0   &   4.85\\
101199625   &   30.43171   &   -5.60912   &   0.530   &   43.30   &   <43.92   &   0   &   0   &   0   &   1  &  1   &   2.0   &   5.39\\
102121299   &   32.02825   &   -5.96117   &   0.538   &   44.33   &   43.84   &   1   &   0   &   0   &   1  &  1   &   2.0   &   6.02\\
102129812   &   31.40485   &   -5.92148   &   0.817   &   43.95   &   <43.42   &   0   &   0   &   0   &   0  &  0   &   1.5   &   6.63\\
102133752   &   31.33519   &   -5.90048   &   0.537   &   44.04   &   43.89   &   0   &   0   &   0   &   1  &  0   &   1.0   &   5.36\\
102138525   &   31.56050   &   -5.87794   &   0.797   &   44.57   &   <43.60   &   0   &   0   &   0   &   1  &  1   &   1.5   &   6.22\\
102178761   &   31.22399   &   -5.67968   &   0.689   &   43.68   &   <43.54   &   0   &   0   &   1   &   0  &  1   &   2.0   &   4.70\\
102183859   &   31.62379   &   -5.65619   &   0.599   &   43.69   &   <43.09   &   0   &   0   &   0   &   0  &  0   &   2.0   &   5.09\\
103134908   &   32.21963   &   -5.97658   &   0.644   &   44.37   &   <43.13   &   0   &   0   &   0   &   1  &  1   &   1.5   &   4.84\\
103143919   &   32.29891   &   -5.93776   &   0.773   &   44.76   &   43.78   &   1   &   1   &   1   &   0  &  1   &   1.0   &   7.63\\
103151883   &   32.35777   &   -5.90568   &   0.636   &   45.12   &   43.80   &   1   &   1   &   1   &   0  &  0   &   1.0   &   5.29\\
103180263   &   32.25861   &   -5.78499   &   0.644   &   44.71   &   43.68   &   0   &   1   &   0   &   0  &  1   &   1.0   &   5.14\\
103180825   &   32.77869   &   -5.78191   &   0.748   &   44.73   &   43.77   &   0   &   1   &   0   &   0  &  1   &   1.5   &   5.19\\
103197293   &   32.67400   &   -5.71261   &   0.664   &   43.52   &   <43.30   &   0   &   0   &   1   &   0  &  1   &   2.0   &   5.76\\
104170739   &   33.70780   &   -5.90915   &   0.715   &   43.80   &   <43.26   &   0   &   0   &   0   &   1  &  1   &   2.0   &   4.22\\
104236309   &   33.97194   &   -5.65539   &   0.546   &   43.85   &   <43.03   &   0   &   0   &   0   &   0  &  0   &   1.0   &   6.34\\
105139296   &   34.09171   &   -5.95674   &   1.048   &   44.14   &   <43.88   &   0   &   0   &   1   &   0  &  1   &   2.0   &   6.88\\
105145409   &   34.09651   &   -5.93140   &   1.095   &   44.51   &   <43.71   &   0   &   0   &   1   &   0  &  1   &   2.0   &   6.54\\
105152074   &   34.22651   &   -5.90089   &   0.697   &   44.63   &   <43.13 &   0   &   0   &   1   &   1  &  1   &  1.0   &   5.11\\
105189949   &   34.16295   &   -5.72786   &   0.643   &   43.55   &   <43.16   &   0   &   0   &   0   &   1  &  1   &   2.0   &   6.01\\
105201540   &   34.64631   &   -5.67718   &   0.851   &   44.58   &   <43.63   &   0   &   0   &   0   &   0  &  1   &   1.0   &   5.37\\
105211420   &   34.71244   &   -5.62929   &   0.643   &   44.19   &   <42.69   &   0   &   0   &   0   &   1  &  1   &   1.0   &   5.88\\
105216790   &   34.47889   &   -5.60829   &   0.843   &   44.39   &   <43.43   &   1   &   0   &   0   &   1  &  0   &   1.5   &   7.61\\
106158702   &   35.13075   &   -5.87612   &   0.599   &   45.02   &   43.58   &   0   &   1   &   0   &   0  &  1   &   1.5   &   4.72\\
106164318   &   35.79339   &   -5.85072   &   0.506   &   44.50   &   <42.87   &   1   &   0   &   0   &   1  &  1   &   2.0   &   6.06\\
106204876   &   35.60449   &   -5.66945   &   1.078   &   44.44   &   <44.20   &   1   &   0   &   1   &   0  &  0   &   2.0   &   6.94\\
106208862   &   35.82120   &   -5.65156   &   0.802   &   43.88   &   <43.50   &   0   &   0   &   0   &   0  &  1   &   2.0   &   5.95\\
107161301   &   36.85607   &   -5.65162   &   1.077   &   44.57   &   <43.72   &   1   &   0   &   1   &   0  &  1   &   2.0   &   7.15\\
108114624   &   37.12663   &   -5.97348   &   0.553   &   43.55   &   <43.46   &   0   &   0   &   0   &   1  &  1   &   2.0   &   5.36\\
108120334   &   37.46318   &   -5.94224   &   0.955   &   45.12   &   <43.94   &   1   &   0   &   1   &   0  &  1   &   2.0   &   7.86\\
108137534   &   37.02236   &   -5.85604   &   0.778   &   44.70   &   43.88   &   1   &   1   &   1   &   0  &  1   &   2.0   &   5.40\\
108163578   &   37.59208   &   -5.70686   &   0.716   &   44.68   &   <43.53   &   1   &   0   &   0   &   1  &  1   &   1.0   &   5.67\\
108167635   &   37.26583   &   -5.68088   &   1.043   &   44.87   &   <43.77   &   1   &   1   &   1   &   0  &  1   &   1.0   &   6.05\\
108168148   &   37.07837   &   -5.68167   &   0.626   &   43.93   &   <43.56   &   0   &   0   &   0   &   1  &  1   &   1.5   &   6.10\\
109130669   &   38.48938   &   -5.91916   &   0.554   &   43.99   &   <42.89   &   0   &   0   &   0   &   1  &  1   &   2.0   &   4.74\\
109161838   &   38.20769   &   -5.76429   &   1.102   &   44.28   &   <43.76   &   0   &   0   &   1   &   0  &  1   &   2.0   &   6.44\\
109164682   &   38.43727   &   -5.74709   &   0.825   &   44.42   &   <43.14   &   0   &   0   &   0   &   0  &  0   &   1.0   &   6.70\\
109179237   &   38.46824   &   -5.67911   &   0.593   &   43.44   &   <43.06   &   0   &   0   &   0   &   1  &  1   &   2.0   &   3.86\\
110043474   &   31.13770   &   -5.43120   &   0.666   &   44.55   &   <43.55   &   1   &   0   &   1   &   1  &  1   &   1.0   &   5.63\\
110054196   &   31.15407   &   -5.37707   &   0.596   &   44.77   &   <43.38   &   0   &   1   &   0   &   0  &  1   &   1.0   &   5.02\\
110066563   &   30.80605   &   -5.31692   &   0.696   &   43.82   &   <43.04   &   0   &   0   &   0   &   0  &  1   &   2.0   &   5.78\\
110072899   &   30.28438   &   -5.28715   &   0.776   &   44.51   &   <43.95   &   1   &   0   &   1   &   1  &  1   &   1.5   &   6.81\\
110078074   &   30.76310   &   -5.26278   &   0.512   &   43.66   &   <43.09   &   0   &   0   &   0   &   0  &  1   &   2.0   &   5.64\\
110100629   &   30.69404   &   -5.15765   &   0.859   &   44.09   &   44.22   &   0   &   1   &   0   &   0  &  1   &   1.0   &   6.01\\
110133664   &   30.77885   &   -5.00982   &   0.507   &   43.67   &   <43.11   &   0   &   0   &   0   &   1  &  1   &   2.0   &   6.06\\
110180344   &   30.94828   &   -4.79023   &   0.617   &   44.02   &   <43.31   &   0   &   0   &   0   &   1  &  1   &   1.5   &   5.92\\
110202029   &   30.44836   &   -4.68994   &   0.558   &   43.96   &   <43.55   &   0   &   0   &   0   &   0  &  1   &   2.0   &   5.46\\
110204490   &   30.46040   &   -4.67859   &   0.764   &   44.08   &   <43.81   &   1   &   0   &   1   &   1  &  1   &   2.0   &   5.60\\
111063425   &   31.62215   &   -5.31458   &   0.613   &   43.91   &   <42.98   &   0   &   0   &   0   &   0  &  1   &   1.5   &   5.69\\
111076897   &   31.61439   &   -5.24968   &   0.963   &   44.37   &   43.86   &   0   &   1   &   1   &   1  &  1   &   1.0   &   5.59\\
111108509   &   31.47468   &   -5.10136   &   1.135   &   44.53   &   <44.06   &   0   &   0   &   0   &   0  &  0   &   2.0   &   6.62\\
111117682   &   31.51571   &   -5.05702   &   1.084   &   45.27   &   <44.21   &   1   &   0   &   1   &   0  &  1   &   1.5   &   7.51\\
111120886   &   31.32135   &   -5.04079   &   0.516   &   43.90   &   <43.12   &   0   &   0   &   0   &   0  &  1   &   2.0   &   5.26\\
111153701   &   31.84755   &   -4.88503   &   0.704   &   44.30   &   <43.37   &   1   &   0   &   1   &   1  &  1   &   1.5   &   6.38\\
111157163   &   31.98383   &   -4.86751   &   0.557   &   43.73   &   <42.71   &   0   &   0   &   0   &   1  &  1   &   2.0   &   5.36\\
111161925   &   31.35488   &   -4.84550   &   1.093   &   44.70   &   <43.57   &   0   &   0   &   1   &   0  &  0   &   2.0   &   7.30\\
111192162   &   31.88525   &   -4.70298   &   0.500   &   44.42   &   42.92   &   0   &   0   &   0   &   0  &  0   &   1.5   &   5.33\\
112032836   &   32.35279   &   -5.47099   &   0.649   &   44.21   &   <43.52   &   0   &   0   &   0   &   0  &  1   &   1.0   &   4.95\\
112071341   &   32.94422   &   -5.28638   &   0.789   &   44.49   &   <43.71   &   1   &   0   &   1   &   1  &  1   &   1.5   &   6.83\\
112089875   &   32.39088   &   -5.19864   &   0.620   &   43.74   &   <43.21   &   0   &   0   &   0   &   0  &  1   &   2.0   &   5.27\\
112100049   &   33.02364   &   -5.15279   &   1.035   &   44.87   &   <43.83   &   0   &   0   &   0   &   0  &  0   &   2.0   &   7.58\\
112105956   &   32.70154   &   -5.12298   &   0.565   &   43.95   &   <43.33   &   0   &   0   &   0   &   0  &  0   &   1.0   &   4.80\\
112130497   &   32.44284   &   -5.01467   &   0.983   &   44.58   &   <44.03   &   0   &   0   &   1   &   0  &  1   &   2.0   &   7.47\\
112141684   &   33.06693   &   -4.96233   &   0.983   &   44.77   &   44.36   &   1   &   0   &   0   &   0  &  0   &   2.0   &   7.69\\
112155596   &   32.76740   &   -4.89028   &   0.524   &   45.33   &   <43.80   &   1   &   1   &   0   &   0  &  1   &   1.5   &   4.94\\
112163489   &   33.00277   &   -4.85407   &   0.754   &   44.40   &   <43.35   &   1   &   0   &   0   &   1  &  1   &   1.5   &   7.42\\
112167778   &   32.26360   &   -4.83555   &   0.843   &   44.86   &   <43.37   &   1   &   0   &   0   &   1  &  1   &   1.5   &   6.47\\
112169655   &   32.31060   &   -4.82707   &   0.655   &   44.23   &   <43.18   &   1   &   0   &   0   &   0  &  1   &   1.0   &   7.05\\
112173291   &   33.05381   &   -4.81045   &   0.507   &   43.44   &   <43.18   &   0   &   0   &   0   &   1  &  1   &   2.0   &   5.41\\
112196243   &   32.45579   &   -4.70356   &   0.701   &   43.85   &   <43.25   &   0   &   0   &   0   &   0  &  1   &   2.0   &   5.35\\
113037156   &   33.96294   &   -5.45440   &   0.641   &   44.36   &   <43.22   &   0   &   0   &   0   &   1  &  0   &   1.0   &   6.48\\
113044521   &   33.32085   &   -5.42048   &   1.014   &   44.86   &   44.48   &   0   &   1   &   1   &   0  &  1   &   1.0   &   6.09\\
\hline
\end{tabular}
\begin{tablenotes}
\item \textbf{Note.} -- (1): Identifier. (2): Right ascension. (3): Declination. (4): Redshift. (5): Infrared AGN luminosity. (6): X-ray [2-10 keV] luminosity. The '<' symbol represent the upper limit. (7): "1" ("0") if AGN is (is not) selected through mid-IR colours. (8): "1" ("0") for existence (absence) of broad lines. (9), (10), (11): "1" ("0") if AGN is (is not) selected via [NeV] emission, MEx and TBT diagrams, respectively. (12): AGN type derived from the SED fitting results ($\psi$ value). (13): r--W2 colour \citep{yan2013}.
\end{tablenotes}
\end{threeparttable}}
\label{tableALL}
\end{table*}

\begin{table*}
\centering
\contcaption{}
\scalebox{0.75}{
\begin{threeparttable}
\begin{tabular}{c c c c c c c c c c c c c} 
\hline
VIPERS   &   RA   &   DEC   &   z   &   L\textsubscript{IR}   &   L\textsubscript{X}   &   mid-IR  &   Broad   &   [NeV]   &   MEx  & TBT   &   SED   &   r--W2\\
 ID & (J2000) & (J2000) &  & (erg s\textsuperscript{-1}) & (erg s\textsuperscript{-1}) & AGN & lines & emission & AGN  & AGN & AGN type & (Vega)  \\ 
(1) & (2) & (3) & (4) & (5) & (6) & (7) & (8) & (9) & (10) & (11) & (12)  & (13)\\ \hline\hline
113051150   &   33.34912   &   -5.38764   &   0.697   &   45.32   &   43.87   &   1   &   1   &   0   &   0  &  1   &   1.5   &   5.23\\
113064992   &   33.13933   &   -5.32305   &   0.540   &   43.27   &   <43.16   &   0   &   0   &   0   &   1  &  1   &   2.0   &   5.05\\
113092199   &   33.45236   &   -5.19618   &   0.585   &   43.95   &   <43.03   &   0   &   0   &   0   &   1  &  1   &   2.0   &   5.42\\
113110546   &   33.99779   &   -5.11639   &   0.500   &   44.10   &   43.35   &   1   &   0   &   0   &   1  &  1   &   2.0   &   6.66\\
113169191   &   33.07584   &   -4.85186   &   0.742   &   43.53   &   <43.16   &   0   &   0   &   0   &   0  &  0   &   2.0   &   4.73\\
113180911   &   33.82514   &   -4.80045   &   0.624   &   43.68   &   <43.06   &   0   &   0   &   0   &   0  &  0   &   2.0   &   5.18\\
113206860   &   33.08376   &   -4.67540   &   0.807   &   43.97   &   <43.85   &   0   &   0   &   0   &   0  &  1   &   2.0   &   5.37\\
114073711   &   34.67060   &   -5.31094   &   0.659   &   44.23   &   <43.29   &   0   &   0   &   0   &   1  &  1   &   2.0   &   6.44\\
114082043   &   34.38863   &   -5.26415   &   0.599   &   44.36   &   44.03   &   0   &   0   &   0   &   0  &  0   &   1.0   &   5.66\\
114108550   &   34.74651   &   -5.16858   &   0.753   &   44.65   &   <43.24   &   1   &   0   &   0   &   1  &  1   &   1.0   &   6.70\\
114131874   &   34.58537   &   -5.07403   &   0.649   &   44.52   &   44.17   &   1   &   1   &   0   &   0  &  1   &   1.5   &   6.32\\
114139331   &   34.32735   &   -5.04438   &   0.823   &   44.98   &   43.63   &   0   &   1   &   1   &   0  &  1   &   1.0   &   4.68\\
114161732   &   34.28235   &   -4.95660   &   1.092   &   44.61   &   <43.95   &   0   &   0   &   0   &   0  &  1   &   1.0   &   6.98\\
115027291   &   35.78914   &   -5.49197   &   0.626   &   44.26   &   43.46   &   1   &   0   &   0   &   1  &  1   &   2.0   &   5.59\\
115115472   &   35.66146   &   -5.06443   &   0.907   &   44.05   &   43.58   &   0   &   1   &   1   &   0  &  1   &   1.0   &   7.31\\
115122976   &   35.65010   &   -5.02796   &   0.845   &   44.24   &   44.12   &   0   &   0   &   1   &   1  &  1   &   2.0   &   7.63\\
115135448   &   35.82728   &   -4.97046   &   0.828   &   44.41   &   43.88   &   1   &   0   &   0   &   1  &  1   &   1.0   &   7.91\\
116014737   &   36.78589   &   -5.55085   &   0.553   &   44.16   &   <43.32   &   0   &   0   &   0   &   1  &  1   &   1.0   &   6.18\\
116032161   &   36.57659   &   -5.45495   &   0.770   &   43.68   &   <43.80   &   0   &   0   &   1   &   1  &  1   &   2.0   &   5.65\\
116042241   &   36.60305   &   -5.40163   &   1.000   &   44.76   &   <44.39   &   0   &   0   &   1   &   0  &  1   &   2.0   &   6.63\\
116046207   &   36.75194   &   -5.38182   &   0.870   &   45.31   &   <43.87   &   0   &   1   &   1   &   0  &  1   &   1.5   &   4.88\\
116085907   &   36.02205   &   -5.18092   &   0.649   &   43.69   &   <42.54   &   0   &   0   &   0   &   0  &  1   &   1.0   &   6.67\\
116099542   &   36.73734   &   -5.11584   &   0.704   &   43.58   &   <43.27   &   0   &   0   &   1   &   1  &  1   &   2.0   &   5.74\\
116150199   &   36.37517   &   -4.87492   &   0.629   &   43.72   &   <43.03   &   0   &   0   &   0   &   0  &  1   &   2.0   &   5.43\\
116167222   &   36.39511   &   -4.79289   &   0.854   &   44.17   &   <43.40   &   1   &   0   &   1   &   1  &  1   &   2.0   &   6.68\\
116192311   &   35.95629   &   -4.66974   &   0.622   &   44.21   &   43.51   &   0   &   0   &   0   &   1  &  1   &   2.0   &   5.84\\
117010534   &   37.79089   &   -5.57627   &   0.986   &   45.13   &   <44.44   &   1   &   0   &   0   &   0  &  1   &   1.5   &   8.43\\
117027325   &   37.62852   &   -5.49188   &   0.784   &   43.88   &   <43.44   &   0   &   0   &   0   &   0  &  1   &   2.0   &   5.54\\
117037264   &   37.08046   &   -5.44453   &   0.786   &   44.44   &   <43.46   &   1   &   0   &   0   &   1  &  1   &   2.0   &   7.08\\
117050860   &   36.96234   &   -5.37541   &   0.804   &   44.92   &   44.34   &   1   &   0   &   1   &   1  &  1   &   2.0   &   7.86\\
117062279   &   37.26845   &   -5.31627   &   0.779   &   43.96   &   <43.54   &   0   &   0   &   0   &   0  &  1   &   2.0   &   5.85\\
117066130   &   37.84448   &   -5.29393   &   0.586   &   43.92   &   <43.37   &   0   &   0   &   0   &   1  &  1   &   2.0   &   5.09\\
117110515   &   37.15691   &   -5.09081   &   0.698   &   43.49   &   <43.03   &   0   &   0   &   0   &   0  &  1   &   2.0   &   5.04\\
117135140   &   37.13011   &   -4.97454   &   0.933   &   44.05   &   <43.71   &   0   &   0   &   0   &   0  &  0   &   2.0   &   4.99\\
117173795   &   37.68207   &   -4.76728   &   0.630   &   43.56   &   <43.37   &   0   &   0   &   0   &   0  &  1   &   2.0   &   4.17\\
117177187   &   37.35116   &   -4.75558   &   0.892   &   44.30   &   <43.49   &   1   &   0   &   0   &   0  &  1   &   2.0   &   7.08\\
117178247   &   37.36978   &   -4.74554   &   0.610   &   44.66   &   43.37   &   0   &   0   &   0   &   1  &  1   &   1.0   &   4.87\\
117186794   &   37.30807   &   -4.70517   &   0.662   &   43.91   &   43.76   &   0   &   1   &   0   &   0  &  1   &   1.5   &   5.81\\
118042646   &   38.32293   &   -5.41092   &   0.708   &   44.63   &   <43.34   &   0   &   1   &   1   &   0  &  1   &   1.0   &   5.79\\
118086325   &   38.44772   &   -5.18750   &   0.638   &   44.22   &   <44.14   &   1   &   0   &   0   &   1  &  1   &   1.0   &   5.72\\
118086892   &   38.49890   &   -5.18475   &   0.516   &   43.41   &   <43.97   &   0   &   0   &   0   &   0  &  0   &   2.0   &   5.61\\
118086923   &   38.28170   &   -5.18458   &   0.731   &   44.12   &   <43.10   &   0   &   0   &   1   &   1  &  1   &   1.0   &   6.38\\
118087346   &   38.60147   &   -5.18252   &   0.803   &   44.55   &   <43.83   &   1   &   0   &   0   &   0  &  0   &   1.5   &   6.74\\
118128010   &   38.48948   &   -4.98375   &   0.646   &   43.74   &   <43.46   &   0   &   0   &   0   &   1  &  1   &   2.0   &   5.38\\
118161115   &   38.01033   &   -4.80912   &   0.543   &   43.28   &   <42.83   &   0   &   0   &   0   &   1  &  1   &   2.0   &   5.13\\
118187397   &   38.10186   &   -4.67423   &   0.763   &   44.22   &   <43.28   &   1   &   0   &   0   &   1  &  1   &   2.0   &   7.06\\
119009735   &   30.37515   &   -4.64988   &   0.894   &   44.72   &   <44.70   &   1   &   0   &   1   &   0  &  1   &   2.0   &   6.72\\
119034376   &   30.75731   &   -4.53195   &   0.939   &   44.08   &   <43.56   &   0   &   0   &   0   &   0  &  1   &   2.0   &   6.54\\
119061358   &   30.48839   &   -4.39574   &   0.549   &   43.96   &   <43.81   &   1   &   0   &   0   &   0  &  0   &   1.5   &   7.23\\
119088613   &   31.14702   &   -4.26421   &   0.578   &   43.83   &   <42.87   &   0   &   0   &   0   &   0  &  1   &   2.0   &   4.61\\
119091830   &   30.98208   &   -4.25297   &   0.735   &   44.14   &   <43.34   &   0   &   0   &   0   &   0  &  0   &   2.0   &   5.74\\
119095831   &   30.75257   &   -4.22947   &   0.525   &   43.64   &   <42.92   &   0   &   0   &   0   &   0  &  0   &   2.0   &   5.17\\
120040724   &   31.90079   &   -4.52882   &   0.971   &   44.27   &   <43.85   &   0   &   1   &   0   &   0  &  1   &   1.0   &   5.52\\
120071991   &   31.39916   &   -4.39738   &   0.741   &   45.08   &   <43.50   &   1   &   0   &   0   &   1  &  1   &   1.0   &   7.19\\
120103089   &   31.84380   &   -4.27291   &   0.707   &   44.46   &   43.20   &   0   &   1   &   0   &   0  &  1   &   1.5   &   5.10\\
120114522   &   32.09921   &   -4.22791   &   0.745   &   44.39   &   <43.36   &   1   &   0   &   1   &   1  &  1   &   1.0   &   6.69\\
121034054   &   32.41920   &   -4.54024   &   0.591   &   43.62   &   <43.30   &   0   &   0   &   0   &   0  &  0   &   2.0   &   5.44\\
121043880   &   32.91610   &   -4.48952   &   0.851   &   44.38   &   <43.34   &   0   &   0   &   0   &   0  &  1   &   1.5   &   5.95\\
121046919   &   32.42821   &   -4.47505   &   0.533   &   44.81   &   43.96   &   1   &   1   &   0   &   0  &  1   &   1.5   &   4.82\\
121062481   &   32.42717   &   -4.40335   &   0.842   &   44.42   &   <43.51   &   0   &   0   &   0   &   0  &  0   &   1.5   &   7.13\\
121099761   &   32.53067   &   -4.24116   &   0.822   &   43.93   &   <43.40   &   0   &   0   &   1   &   1  &  1   &   2.0   &   5.56\\
121102071   &   32.87646   &   -4.23080   &   0.609   &   44.33   &   43.46   &   1   &   1   &   0   &   0  &  1   &   1.0   &   4.95\\
122022800   &   33.37111   &   -4.57965   &   0.674   &   44.64   &   43.60   &   0   &   1   &   1   &   0  &  1   &   1.0   &   5.43\\
122042685   &   33.50952   &   -4.48348   &   0.514   &   43.82   &   <43.06   &   0   &   0   &   0   &   1  &  1   &   1.0   &   4.12\\
122043807   &   33.72362   &   -4.47911   &   0.722   &   44.80   &   <43.67   &   0   &   0   &   0   &   0  &  1   &   2.0   &   5.86\\
122052858   &   33.11464   &   -4.43304   &   1.149   &   45.44   &   <44.08   &   1   &   1   &   0   &   0  &  0   &   1.0   &   5.31\\
122100077   &   33.17477   &   -4.21485   &   0.669   &   44.50   &   43.57   &   1   &   0   &   1   &   1  &  1   &   1.0   &   5.87\\
123021037   &   34.20095   &   -4.59600   &   0.502   &   43.27   &   <42.88   &   0   &   0   &   0   &   1  &  1   &   2.0   &   4.46\\
123057675   &   34.49237   &   -4.42283   &   0.646   &   44.43   &   43.89   &   0   &   0   &   1   &   1  &  1   &   1.0   &   5.08\\
123098574   &   34.78088   &   -4.24129   &   0.660   &   43.97   &   <43.00   &   0   &   0   &   0   &   0  &  1   &   2.0   &   5.14\\
124035576   &   35.84055   &   -4.52366   &   0.619   &   44.09   &   <43.16   &   0   &   0   &   0   &   0  &  0   &   1.5   &   4.82\\
124072245   &   35.24707   &   -4.36030   &   0.812   &   44.36   &   <43.20   &   0   &   0   &   0   &   0  &  0   &   1.0   &   5.36\\
124075097   &   35.57515   &   -4.34718   &   0.837   &   44.11   &   <43.24   &   0   &   0   &   0   &   0  &  1   &   1.0   &   7.06\\
124109737   &   35.53904   &   -4.19366   &   0.616   &   44.14   &   <43.13   &   0   &   0   &   0   &   0  &  1   &   2.0   &   5.32\\
125022958   &   36.15360   &   -4.58911   &   0.639   &   43.63   &   <43.04   &   0   &   0   &   0   &   0  &  0   &   2.0   &   4.44\\
125026840   &   36.42350   &   -4.57150   &   0.611   &   44.26   &   <43.04   &   0   &   1   &   0   &   0  &  1   &   1.5   &   5.78\\
127028630   &   38.34431   &   -4.55794   &   0.578   &   44.59   &   43.33   &   0   &   0   &   0   &   1  &  1   &   1.0   &   5.09\\
127061363   &   38.40941   &   -4.39736   &   0.753   &   44.71   &   44.09   &   1   &   0   &   0   &   1  &  1   &   1.0   &   5.76\\
127103639   &   38.46575   &   -4.21191   &   1.150   &   45.00   &   <43.70   &   1   &   0   &   1   &   0  &  0   &   1.5   &   6.16\\
127111323   &   38.32294   &   -4.17705   &   0.658   &   43.61   &   <43.11   &  0   &   0   &   0   &   0  &  1   &   2.0   &   5.63 \\
\hline
\end{tabular}
\begin{tablenotes}
\item \textbf{Note.} -- (1): Identifier. (2): Right ascension. (3): Declination. (4): Redshift. (5): Infrared AGN luminosity. (6): X-ray [2-10 keV] luminosity. The '<' symbol represent the upper limit. (7): "1" ("0") if AGN is (is not) selected through mid-IR colours. (8): "1" ("0") for existence (absence) of broad lines. (9), (10), (11): "1" ("0") if AGN is (is not) selected via [NeV] emission, MEx and TBT diagrams, respectively. (12): AGN type derived from the SED fitting results ($\psi$ value). (13): r--W2 colour \citep{yan2013}.
\end{tablenotes}
\end{threeparttable}}
\label{tableALL}
\end{table*}

\section{Dependence of the method's reliability on the available photometry}\label{farIR}
In our analysis, we constructed SEDs using optical, near-IR and mid-IR photometry. In this section, we examine the effect of adding ancillary data on the reliability of our method. In particular, we test if the omission of (i) near-IR photometry, (ii) the longest mid-IR photometric bands and (iii) the addition of far-IR data, affects the AGN selection method. Toward this end, we followed the procedure described in Section~\ref{methods} to calculate the $\Delta$BIC values for all samples with different number of photometric bands. Then, we compared these values assuming that real AGNs are the SED-AGNs derived from the sample with the highest number of bands used.

Our analysis revealed, that the absence of near-IR data results in a high number of false-positive AGN identifications. For relaxed thresholds ($\Delta$BIC<-2), even though the completeness is high (62\%), the false-positive rate reaches up to 65\% overestimating the number of the selected AGNs. This is probably due to the fact that in these bands the AGN and galaxy SEDs are almost equal in fluxes. Thus, we chose to include near-IR bands in our analysis to better constraint the SED fitting and increase the reliability of our method. Furthermore, we tested if the absence of both the W3 and W4 bands affects the reliability of the AGN selection method. Comparing the SED selected samples, we concluded that the absence of W3 and W4 bands resulted in 16\% of false-positive AGN candidates. W3 and W4 photometric bands are available for 86\% of the galaxies in the VIPERS sample. Thus, we expect $\sim2\%$ of our SED selected AGN to be misclassified.

We also examined whether the addition of far-IR data affects the AGN selection method. For that purpose, we used the \textit{Herschel} Multi-tiered Extragalactic Survey (HerMES) data \citep{oliver2012} that overlap with the VIPERS field. HerMES used PACS \citep{poglitsch2010} and SPIRE \citep{griffin2010} photometric data from the ESA \textit{Herschel Space Observatory} \citep{pilbratt2010}. We cross-matched the VIPERS sample with the far-IR sources as described in Section~\ref{finalsample}. For the 174 sources with \textit{Herschel} data, we run the \texttt{X-CIGALE} code with and without AGN templates. We, then, compared the SED selected AGNs and we found no statistical differences between the results. Thus, in our analysis, we do not require far-IR photometry. This does not affect the reliability of our method, while it allows us to apply the methodology in a significantly larger galaxy sample.

\bsp	
\label{lastpage}
\end{document}